\documentclass[aps,prd,10pt,superscriptaddress,nofootinbib,preprintnumbers]{revtex4-2}
\usepackage{amsmath,amsfonts,amssymb}
\usepackage{graphicx}
\usepackage{dcolumn}
\usepackage[mathlines]{lineno}
\usepackage[usenames,dvipsnames]{color}
\usepackage{booktabs}
\usepackage{adjustbox}
\usepackage{physics}
\usepackage{hyperref}

\begin{document}

\title{Neutrino mass, scalar dark matter, and collider signatures in a radiative doublet–triplet model}

\author{Isaac Bamwidhi}
\email{isaac.bamwidhi@uaeu.ac.ae, isaac.bamwidhi@cern.ch}
\affiliation{Department of Physics, United Arab Emirates University, Al-Ain, UAE}

\author{Mohamed Belfkir}
\email{mohamed.belfkir@udst.edu.qa, mohamed.belfkir@cern.ch}
\affiliation{College of General Education, University of Doha for Science and Technology, Qatar}

\author{Mohamed Amin Loualidi}
\email{ma.loualidi@uaeu.ac.ae}
\affiliation{Department of Physics, United Arab Emirates University, Al-Ain, UAE}

\author{Salah Nasri}
\email{snasri@uaeu.ac.ae, salah.nasri@cern.ch}
\affiliation{Department of Physics, United Arab Emirates University, Al-Ain, UAE}

\begin{abstract}
We explore a radiative neutrino mass model based on the topology \texttt{T4-3-i},  which is a one-loop completion of the Weinberg operator.  We focus on its \texttt{T4-3-i-C1} realization,  containing an inert scalar doublet, a real scalar triplet, a Majorana fermion triplet, and a Dirac singlet fermion. An exact $Z_2$ symmetry forbids the tree-level type-III term and stabilizes the lightest $Z_2$-odd particle, leaving the one-loop topology as the leading source of neutrino mass. The model accomodate a neutral {\it CP}-even scalar dark matter candidate and predicts a rank two neutrino mass matrix, and thus one neutrino is massless. A combined analysis incorporating theoretical  and experimental constraints reveals viable solutions to both neutrino mass orderings, with the  inverted ordering predicting larger values of $m_{\beta\beta}$, $m_\beta$, and $\sum_i m_i$, thereby enhancing its testability through future beta-decay, neutrinoless double beta decay, and cosmological probes. Within the same viable parameter region, the observed  dark matter relic abundance is accounted for by annihilation and coannihilation processes, while the electroweak triplet sector gives rise to  prompt, displaced, or long-lived particle signatures, offering  complementary collider probes  to dark matter and low-energy searches.
\end{abstract}


\maketitle
\section{Introduction}
\label{intro}
Neutrino oscillations and cosmological observation of non-baryonic dark matter (DM) provide two compelling indications of physics beyond the Standard Model (SM). Oscillation data requires at least two massive neutrinos~\cite{Super-Kamiokande:1998kpq,SNO:2002tuh} and global analyses have substantially improved the precision with which the mixing angles and mass-squared differences are known~\cite{Esteban:2024eli,deSalas:2020pgw,Capozzi:2021fjo}, yet the mechanism responsible for the small neutrino mass scale remains unresolved. Meanwhile, the DM relic abundance is accurately measured~\cite{Planck:2018vyg}, while its particle nature remains unknown. For broader reviews of particle DM, see Refs.~\cite{Bergstrom:2000pn,Bertone:2004pz,Cirelli:2024ssz}. These open questions motivate extensions of the SM in which neutrino mass generation and a viable DM candidate originate from a common sector. With the SM field content, the leading effective operator generating Majorana neutrino masses is the dimension five Weinberg operator $LLHH/\Lambda$ after electroweak symmetry breaking (EWSB)\cite{Weinberg:1979sa}. Its radiative ultraviolet completions are particularly appealing because loop suppression, often supplemented by small couplings or symmetry protected lepton number (LN) violating parameters, can naturally accommodate the observed neutrino mass scale while allowing new states near the TeV scale~\cite{Zee:1980ai,Tao:1996vb,Ma:2006km}. Systematic classifications and studies of one-loop realizations were presented in Refs.~\cite{Bonnet:2012kz,Law:2013dya,Arbelaez:2022ejo}; see Ref.~\cite{Cai:2017jrq} for a comprehensive review. \\

At one loop, only finite neutrino mass amplitudes correspond to genuinely independent radiative mechanisms. Ultraviolet-divergent amplitudes require counterterms associated with tree-level seesaw realizations of the Weinberg operator and therefore represent loop corrections to those mechanisms rather than new sources of neutrino mass~\cite{Bonnet:2012kz}. The finite, irreducible one-loop topologies are labeled \texttt{T1-i}, \texttt{T1-ii}, \texttt{T1-iii}, and \texttt{T3}. Three additional finite topologies—\texttt{T4-1-ii}, \texttt{T4-2-i}, and \texttt{T4-3-i} can be viewed as one-loop extensions of tree-level seesaw mechanisms. However, only \texttt{T4-2-i} and \texttt{T4-3-i} can generate the leading neutrino mass contribution, provided that the loop contains Majorana fermions, the relevant couplings conserve LN, and a discrete symmetry forbids the scalar fields from acquiring vacuum expectation values (VEVs). Thus, the viable leading-order one-loop structures are \texttt{T1-i}, \texttt{T1-ii}, \texttt{T1-iii}, \texttt{T3}, \texttt{T4-2-i}, and \texttt{T4-3-i}~\cite{Bonnet:2012kz}.
The \texttt{T1} and \texttt{T3} classes encompass many familiar radiative neutrino mass models. In particular, the original scotogenic model~\cite{Ma:2006km} corresponds to a \texttt{T3} realization~\cite{Restrepo:2013aga}. In scotogenic-type constructions, an exact and unbroken dark symmetry, usually a $\mathbb{Z}_2$, can simultaneously forbid the tree-level neutrino mass contribution and stabilize the lightest symmetry-odd state. If this state is electrically neutral and phenomenologically viable, it can constitute a DM candidate belonging to the same sector that radiatively generates neutrino mass~\cite{Restrepo:2013aga,Law:2013saa,Hirsch:2013ola}. The requirement that the dark symmetry remain unbroken is itself a nontrivial consistency condition, as emphasized in Ref.~\cite{Merle:2016scw}. For a recent review of scotogenic and related dark sector realizations of neutrino mass, see Ref.~\cite{Avila:2025qsc}. \\

The two  finite \texttt{T4} structures require special care because they 
generically coexist with lower-order contributions~\cite{Bonnet:2012kz}. The \texttt{T4-2-i} structure is a radiative extension of the type-II seesaw involving an electroweak scalar triplet $\Delta$, two inert scalar doublets, and Majorana singlet fermions. The $HH\Delta^\dagger$ interaction required by the topology, together with an allowed $LL\Delta$ coupling, reproduces the tree-level type-II seesaw, while an allowed $\overline{L}\widetilde{H}N$ coupling also opens a type-I contribution. The first explicit realization avoided both terms through a
$D_4\times Z_3\times Z_5\times Z_2$ flavor symmetry and additional flavon fields~\cite{Loualidi:2020jlj}. A more economical supersymmetric modular-$A_4$ construction subsequently used modular weights to suppress the tree-level operators without introducing fields beyond those required by the topology~\cite{Kashav:2022kpk}. Alternative realizations have since been proposed using non-invertible $Z_7$ Tambara--Yamagami selection rules~\cite{Kashav:2026jjg} and, more recently, a non-holomorphic $T'$ modular symmetry, in which even- and odd-weight polyharmonic Maa{\ss} forms forbid the type-I and type-II operators while a residual $Z_2$ symmetry stabilizes the lightest dark sector state~\cite{Loualidi:2026pld}. By contrast, \texttt{T4-3-i} is a one-loop extension of the type-I/III seesaw: if the fermion coupled to the external $LH$ legs is a Majorana singlet or triplet, the corresponding tree-level type-I or type-III contribution is generated. A radiative linear-seesaw model based on $U(1)_{B-L}$ was previously shown to reduce to the \texttt{T4-3-i} structure after symmetry breaking \cite{Wang:2015saa}. More recently, a complete field-theoretic realization of the type-I branch was constructed, its viable electroweak assignments were classified, and the minimal singlet-doublet realization \texttt{T4-3-i-B1} was studied in detail in Ref. \cite{Belfkir:2026slp}. In that realization, the fermion coupled to $LH$ is a Dirac singlet $N$, LN violation is carried by a separate $Z_2$-odd Majorana singlet $\psi$, and both scalar and fermionic DM scenarios were investigated. \\

In this work, we study the \texttt{T4-3-i-C1} realization of the \texttt{T4-3-i} topology identified in Ref.\cite{Belfkir:2026slp}. Its $Z_2$-odd sector consists of an inert scalar doublet $\phi\sim 2^S_1$, a real scalar triplet $\eta\sim 3^S_0$, and a Majorana fermion triplet $\psi\sim 3^F_0$, while the fermion coupled to $LH$ is the Dirac singlet $N$. This assignment removes the tree-level type-I contribution, and the exact $Z_2$ symmetry forbids the $\overline{L}\widetilde{H}\psi$ interaction that would generate a type-III seesaw term. LN violation is instead carried by the Majorana mass of $\psi$, while the same $Z_2$ symmetry keeps the new scalars inert and stabilizes the lightest odd state. Possible ultraviolet origins of the dark parity are discussed in Refs~\cite{Merle:2015gea,Leite:2019grf,deBoer:2023phz,Garnica:2024wur,VanDong:2023xmf,Lavoura:2012cv}. Compared with the singlet-doublet realization of Ref.~\cite{Belfkir:2026slp}, the triplet fields lead to a richer electroweak phenomenology, including two charged scalar mass eigenstates, additional neutral and charged triplet components, gauge-mediated fermion production, and $Z_2$-odd cascade decays. The model therefore links radiative neutrino mass, scalar DM, compressed electroweak spectra, and charged-state collider signatures~\cite{Brdar:2013iea,Ma:2008cu,Schmidt:2012yg,Ibarra:2016dlb,Avila:2019hhv,Karan:2023adm,Arbelaez:2024lcr,Kahlhoefer:2017dnp,Boveia:2018yeb,Dutta:2017lny,Belyaev:2016lok,Diaz:2016udz,Choubey:2017yyn,Baumholzer:2019twf,Lozano:2025tst}. We focus on the scalar-DM regime, taking {\it CP}-even neutral state $H_1^0$ to be the lightest $Z_2$-odd particle, and perform independent scans for normal ordering (NO) and inverted ordering (IO) of the neutrino mass spectrum requiring  perturbativity, vacuum stability, perturbative unitarity, while imposing constraints from neutrino oscillation data,  electroweak precision tests, charged lepton flavor violation, the Higgs diphoton rate, the relic abundance, and direct-detection. Both orderings remain viable, with IO yielding a slightly better overall fit to the combined constraints and predicts larger values of $m_{\beta\beta}$, $m_\beta$, and $\sum_i m_i$, thereby enabling complementary tests of  the model. We also map the viable scalar DM parameter space and examine charged-state lifetimes and production at current and future colliders. \\

The paper is organized as follows. In Sec.~\ref{sec:model}, we present the model and discuss its fermionic and scalar sectors. Section~\ref{sec:constraints} describes the theoretical and experimental constraints, together with the decay and collider formalism. The relic abundance and direct-detection phenomenology are discussed in Sec.~\ref{sec:dark_matter}. Section~\ref{results} presents the statistical framework, numerical results, and benchmark collider phenomenology. We summarize our conclusions in Sec.~\ref{sec:concl}.
\section{Model}
\label{sec:model}
\subsection{Yukawa interactions in topology \texttt{T4-3-i}}
\begin{figure}[!htbp]
    \centering
    \includegraphics[scale=0.4]{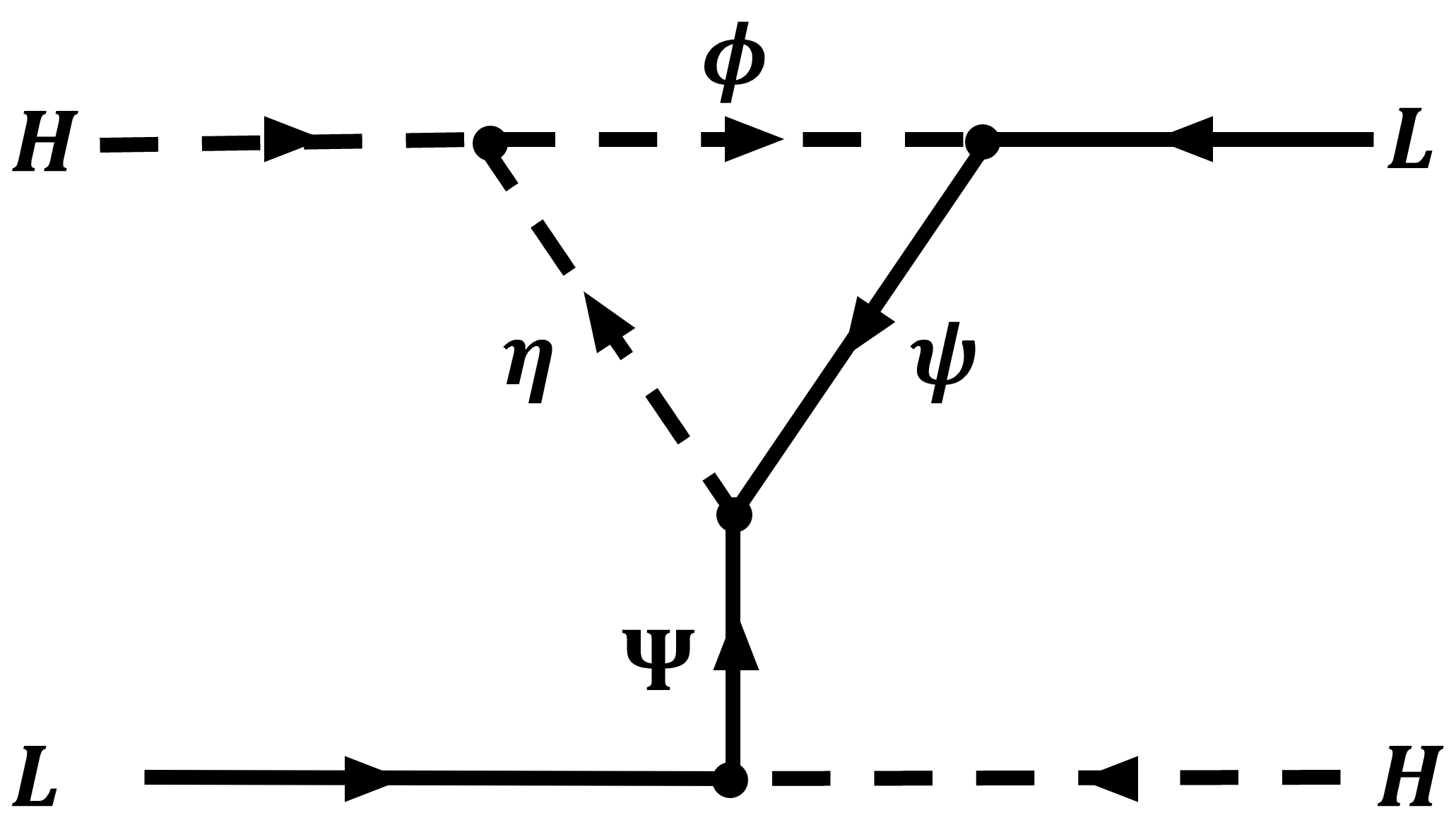}
    \hspace{2.5cm}
    \includegraphics[scale=0.4]{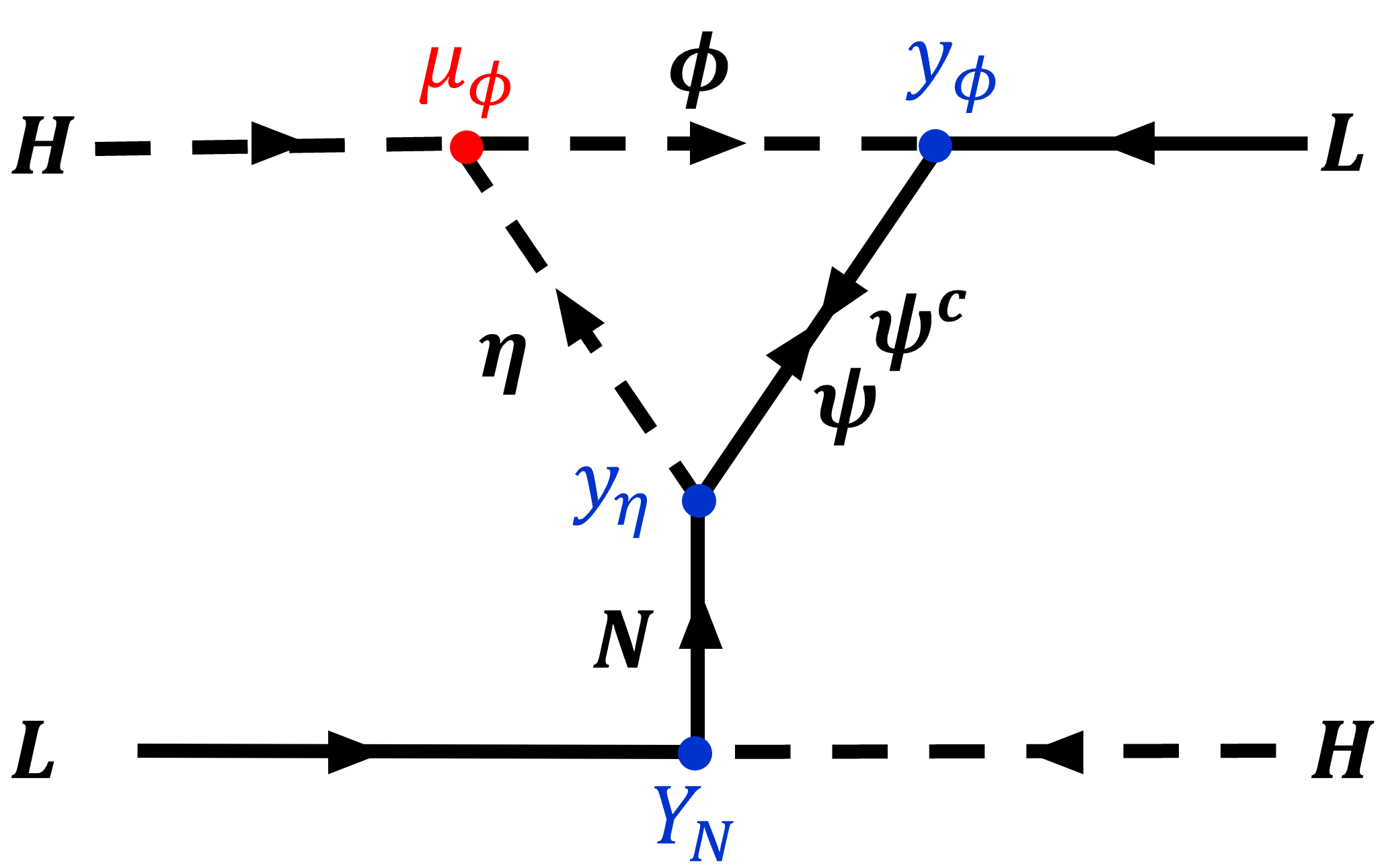}
    \caption{Left: Topology \texttt{T4-3-i} accompanied by the tree level type-I or type-III seesaw contributions when $\Psi$ is a Majorana fermion which transforms under $SU(2)_{\mathrm{L}}$ as $\Psi \sim 1$ or $\Psi \sim 3$, respectively. Right: Genuine \texttt{T4-3-i} topology obtained by promoting $\Psi$ to a Dirac spinor $N$, $\psi$ to a Majorana fermion, while imposing LN conservation for all the couplings.}
    \label{f1}
\end{figure}

The dimension-5 Weinberg operator $O_5 = LLHH/\Lambda$ can be generated at the one-loop level via scalar and fermion exchanges, corresponding to topology \texttt{T4-3-i} depicted in the left panel of Fig.~\ref{f1}. The $SU(2)_L$ representation of the fermion $\Psi$, which couples to the usual SM lepton and Higgs doublets $LH$, determines the nature of the accompanying tree-level contribution: an $SU(2)_L$ Majorana singlet $\Psi$ induces a dominant type-I seesaw at tree level, whereas an $SU(2)_L$ Majorana triplet $\Psi$ gives rise to a dominant tree-level type-III seesaw.
In addition to the choice of $\Psi$, various realizations of this topology can be constructed by considering different $SU(2)_L$ representations for the particles running in the loop. In the case where $\Psi$ is an $SU(2)_L$ singlet, the possible gauge-invariant realizations are listed in Table~\ref{t1}. A complete list of models for both singlet and triplet $\Psi$ can be found in Ref.~\cite{Bonnet:2012kz}.
\begin{table}[htbp]
    \centering
    \begin{tabular}{|c|c|c|c|c|}
         \hline
    Models & \texttt{T4-3-i-A}  & \texttt{T4-3-i-B} & \texttt{T4-3-i-C} & \texttt{T4-3-i-D}  \\
         \hline \hline
    $\phi$ & $1_{1+\alpha}^S$ & $2_{1+\alpha}^S$ & $2_{1+\alpha}^S$ & $3_{1+\alpha}^S$  \\
         \hline
    $\eta$ & $2_{\alpha}^S$ & $1_{\alpha}^S$ & $3_{\alpha}^S$ & $2_{\alpha}^S$  \\
         \hline
    $\psi$ & $2_{\alpha}^F$ & $1_{\alpha}^F$ & $3_{\alpha}^F$ & $2_{\alpha}^F$   \\
         \hline
    \end{tabular}
    \caption{Field content for the sub-models of topology \texttt{T4-3-i}.}
    \label{t1}
\end{table}
Moreover, for each of these four realizations, several distinct models can be obtained depending on the value of the index $\alpha$ that labels the $SU(2)_L$ representations. This index fixes the corresponding hypercharges of the particles propagating in the loop, thereby defining the specific model. Model \texttt{T4-3-i-B} with $\alpha = 0$ has been studied in detail in Ref.\cite{Belfkir:2026slp}. While model \texttt{T4-3-i-A} has not previously been employed in a field-theoretic framework, its scalar sector is expected to be identical to that of model \texttt{T4-3-i-B}, with the roles of $\phi$ and $\eta$ interchanged. In contrast, the fermionic sector is anticipated to differ between the two models. Specifically, in \texttt{T4-3-i-A} the fermion $\psi$ transforms as an $SU(2)_L$ doublet, whereas in \texttt{T4-3-i-B} it is an $SU(2)_L$ singlet. Models \texttt{T4-3-i-C} and \texttt{T4-3-i-D} are the only realizations of this topology that contain scalar and fermion triplets. In this work, we focus on model \texttt{T4-3-i-C1} with $\alpha = 0$, which represents one of the simplest and most minimal realizations of this topology and allows for both scalar and fermionic DM candidates. As discussed in Ref.\cite{Belfkir:2026slp}, for $\alpha = 0, \pm 1, \pm 2, \pm 3$ only three viable sub-models of topology \texttt{T4-3-i-C} are permitted. For our case of interest with $\alpha = 0$, the fields running in the loop are given by
\begin{equation}
\phi = \begin{pmatrix} \phi^{+} \\ \phi^{0} \end{pmatrix} \sim 2_1^S, \quad
\eta = \begin{pmatrix} \eta^+ \\ \eta^0 \\ \eta^- \end{pmatrix} \sim 3_{0}^S, \quad
\psi = \begin{pmatrix} \psi^+ \\ \psi^0 \\ \psi^- \end{pmatrix} \sim 3_{0}^F.
\label{m1}
\end{equation}
The other viable realizations correspond to $\alpha = \pm 2$. These scenarios are phenomenologically richer, as they contain doubly charged scalar and fermionic states; however, their detailed study is beyond the scope of the present work and is deferred to future investigations. In contrast, the cases $\alpha = \pm 1$ and $\alpha = \pm 3$ are discarded since they lead to particles with fractional electric charges.

To eliminate the tree-level type-I seesaw contribution, we follow the approach in Ref.~\cite{Bonnet:2012kz} and redefine the properties of the new fields as follows:  
\begin{itemize}
    \item To avoid a Majorana mass term for $\Psi$ and the resulting tree-level type-I seesaw, we promote $\Psi$ to a Dirac spinor, denoted as $N = \Psi_{\mathrm{L}} + \Psi_{\mathrm{R}}$.
    \item The field $\psi$ is promoted to a Majorana fermion, with all couplings conserving lepton number, while the Majorana mass term for $\psi$ is allowed to break lepton number. This setup forbids a Dirac mass term $\Bar{L}\Tilde{H}\psi$.
    \item To keep the loop in topology \texttt{T4-3-i} closed, we impose an exact $Z_2$ symmetry, where $\phi$, $\eta$, and $\psi$ are $Z_2$-odd, while all other fields are $Z_2$-even, preventing $\phi$ and $\eta$ from acquiring VEVs.
\end{itemize}

After implementing these modifications, the Feynman diagram shown in the right panel of Fig.~\ref{f1} is obtained. From this diagram, the most general renormalisable interactions that are invariant under the gauge symmetry and the imposed $Z_2$ symmetry can be written as
\begin{eqnarray}
\label{eq:yukawa}
\mathcal{L} &=& \Big[ (Y_N)_\beta^{~\alpha}\overline{N^\beta} P_{\mathrm{L}} L_\alpha H 
+ (y_\eta)^{a\alpha}\overline{\psi_a^c} P_{\mathrm{L}} N_\alpha \eta^\ast 
+ (y_\phi)_a^{~\alpha}\overline{\psi^a} P_{\mathrm{L}} L_\alpha \phi 
+ \mu_\phi\phi^\ast \eta H 
+ \mathrm{h.c.} \Big] \nonumber \\
&+& M_N\overline{N}N + \left[\frac{1}{2} M_\psi \mathrm{Tr}(\overline{\psi} \psi^c + \mathrm{h.c.}) \right] 
+ m_\phi^2\phi^\ast\phi + m_\eta^2\eta^\ast\eta.
\end{eqnarray}
Here $L_\alpha$ and $H$ denote the SM lepton and Higgs doublets, respectively, while $N$ is a Dirac singlet and $\psi$ is a Majorana $SU(2)_L$ triplet. The Yukawa matrices $Y_N$, $y_\eta$, and $y_\phi$ encode the flavor structure of the interactions, and $\mu_\phi$ is a trilinear scalar coupling that softly connects the two scalar multiplets $\phi$ and $\eta$ with the SM Higgs. After EWSB, the Higgs field develops a VEV $\upsilon_H$, which plays a crucial role in generating the neutrino mass. As a consequence of these interactions, a Majorana neutrino mass is generated radiatively through the one-loop diagram shown in the right panel of Fig.~\ref{f1}. The resulting neutrino mass matrix takes the form
\begin{equation}
\label{eq:neutrino_mass}
m_\nu^{\alpha\beta} = -\mu_\phi \langle H^0 \rangle^2
\left[
(Y_N^T)_\gamma^{~\alpha} (M_N^{-1})^\gamma (y_\eta)^{\gamma a} 
M_{\psi a} (y_\phi)_a^{~\beta} + (y_\phi^T)_a^{~\alpha} M_{\psi a} (y_\eta^{T})^{a\gamma} 
(M_N^{-1})^\gamma (Y_N)_\gamma^{~\beta}
\right] I_3(m_\phi^2,m_\eta^2,M_\psi^2).
\end{equation}
The two terms inside the square brackets correspond to the two possible orientations of the fermion flow in the loop diagram. Their sum ensures that the resulting neutrino mass matrix is symmetric in flavor space, as required for Majorana neutrinos. $I_3$ represents a dimensionful loop function arising from the momentum integration over the internal propagators of the scalar fields $\phi$, $\eta$, and the fermion $\psi$. Explicitly, it is given by
\begin{equation}
I_3(m_\phi^2,m_\eta^2,M_\psi^2) = -\frac{1}{(4\pi)^2}
\left[
\frac{m_\phi^2 \ln\left(\frac{M_\psi^2}{m_\phi^2}\right)}
{(m_\phi^2-m_\eta^2)(m_\phi^2-M_\psi^2)} + \frac{m_\eta^2 \ln\left(\frac{M_\psi^2}{m_\eta^2}\right)} {(m_\eta^2-m_\phi^2)(m_\eta^2-M_\psi^2)}
\right].
\end{equation}
This loop function encodes the dependence of the neutrino mass on the masses of the particles running in the loop. The neutrino mass vanishes when the LN breaking or scalar mixing ingredients are removed, for example in the limits $\mu_\phi \to 0$, $M_\psi \to 0$ in the LN restoring theory, or vanishing Yukawa combinations. In general, the neutrino mass is suppressed by the loop factor $1/(16\pi^2)$, the heavy mediator masses, the trilinear coupling $\mu_\phi$, and the flavor structure of the Yukawa products, thereby naturally explaining the smallness of neutrino masses within this radiative framework.

\subsection{Scalar sector}
\textcolor{red}{T}he most general gauge and $Z_{2}$-invariant, and renormalizable scalar potential for the model \texttt{T4-3-i-C1} is given by 
\begin{eqnarray}
\mathcal{V} &=& -\mu_H^2 H^{\dagger} H + m_{\phi}^2\phi^{\dagger}\phi + \frac{M_{\eta}^2}{2} \mathrm{Tr}[ \eta^{2}] + \mu_{\phi}\left(H^T~i\tau_2~\eta~\tilde{\phi} + \mathrm{h.c.}\right) + \lambda_1 \left( H^{\dagger} H \right)^2 + \lambda_2\left( \phi^{\dagger }\phi \right)^2 + \lambda_{3} [\mathrm{Tr}[\eta^2]]^2 \nonumber
\\
&& + \kappa_1 \left(H^{\dagger} H \right) \left( \phi^{\dagger }\phi \right) + \kappa_{2} \left( H^{\dagger}\phi \right) \left( \phi^{\dagger} H \right) +\frac{\kappa_{3}}{2} \left[ \left( H^{\dagger}\phi \right)^{2} + \mathrm{h.c.} \right] + \frac{\kappa_{4}}{2}\left( H^{\dagger} H \right) \mathrm{Tr}[\eta^2] + \frac{\kappa_{5}}{2} \left( \phi^{\dagger}\phi \right) \mathrm{Tr}[\eta^2] 
\label{scal.pot}
\end{eqnarray}
After EWSB, the {\it CP}-even neutral component of $H$ obtains a VEV $\upsilon_H = 246~\mathrm{GeV}$, while the $Z_{2}$ symmetry ensures that the neutral components of $\phi$ and $\eta$ do not acquire VEVs. These scalar fields can be parameterized as follows 
\begin{equation}
H=\left( 
\begin{array}{c}
G^+ \\ 
\frac{1}{\sqrt{2}}(\upsilon_H + h_1 + iG^0)
\end{array}
\right),\quad \phi = \left( 
\begin{array}{c}
\phi_2^+ \\ 
\frac{1}{\sqrt{2}}(h_2 + i \omega_2)
\end{array}
\right), \quad \eta = \left(
\begin{array}{c}
     \eta^+  \\
     \eta^0  \\
     \eta^-
\end{array} \right) = 
\begin{pmatrix}
\frac{1}{\sqrt 2} \eta^0 & \eta^+ \\
\eta^- & -\frac{1}{\sqrt 2} \eta^0
\end{pmatrix}
\label{part.repre}
\end{equation}
To prevent {\it CP} violation, we assume all parameters in $\mathcal{V}$ are real, ensuring explicit invariance under the standard {\it CP} transformation $\left\{ H\rightarrow H^{\dagger },\phi \rightarrow \phi^{\dagger},\eta \rightarrow \eta^{\ast }\right\} $. After EWSB, we are
left with three {\it CP}-even scalars $\left\{h_1,H_1^0,H_2^0\right\}$, one {\it CP}-odd scalar $\omega_2$ and two charged scalars $H_1^{\pm}$ and $H_2^{\pm}$. The term of $\mu_{\phi}$ in Eq.~\ref{scal.pot} allows for a mixing between the {\it CP}-even neutral fields $h_2$ and $\eta^0$, and also the mixing of the charged fields $\phi^\pm$ and $\eta^\pm$ given as follows 
\begin{equation}
\left( 
\begin{array}{c}
H_1^0 \\ 
H_2^0
\end{array}
\right) =\left( 
\begin{array}{cc}
c_{\theta_{N}} & s_{\theta_{N}} \\ 
-s_{\theta_{N}} & c_{\theta_{N}}
\end{array}
\right) \left( 
\begin{array}{c}
h_2 \\ 
\eta^0
\end{array}
\right),\quad \left( 
\begin{array}{c}
H_1^{\pm} \\ 
H_2^{\pm}
\end{array}
\right) =\left( 
\begin{array}{cc}
c_{\theta_{C}} & s_{\theta_{C}} \\ 
-s_{\theta_{C}} & c_{\theta_{C}}
\end{array}
\right) \left( 
\begin{array}{c}
\phi_2^{\pm} \\ 
\eta^{\pm}
\end{array}
\right) 
\label{eq:states}
\end{equation}
where we have used the shorthand $c_{\theta_x}=\cos \theta_x$ and $s_{\theta_x}=\sin \theta_x$ with $\theta_{N}$ and $\theta_{C}$ representing the mixing angles that diagonalize the {\it CP}-even and the charged scalar mass matrices, respectively. The minimization of the scalar potential along the $Z_2$-even SM direction gives $\mu_H^2=\lambda_1 \upsilon_H^2$, and $m_{h_1}^2=2\lambda_1 \upsilon_H^2$. Thus, the SM-like scalar $h_1$ does not mix with the $Z_2$-odd states, and the neutral inert spectrum separates into a {\it CP}-even sector $(h_2,\eta^0)$ and the {\it CP}-odd field $\omega_2$. In the {\it CP}-even neutral basis $(h_2,\eta^0)$, the squared-mass matrix can be written as
\begin{equation}
    \label{eq:neutral_mass_matrix}
    \mathcal{M}_N^2 =
    \begin{pmatrix}
        A_N & B_N \\
        B_N & D_N
    \end{pmatrix},
    \qquad
    \begin{aligned} \quad \text{with} \quad
        A_N = m_\phi^2+\frac12(\kappa_1+\kappa_2+\kappa_3) \upsilon_H^2, \quad
        B_N = \frac{\mu_\phi \upsilon_H}{\sqrt2}, \quad
        D_N = M_\eta^2+\frac12\kappa_4 \upsilon_H^2.
    \end{aligned}
\end{equation}
The orthogonal rotation in Eq.~\eqref{eq:states} diagonalizes this matrix according to
$R_N \mathcal{M}_N^2 R_N^T = \mathrm{diag}\left(m_{H_1^0}^2,m_{H_2^0}^2\right)$.  With the convention $m_{H_1^0}<m_{H_2^0}$, the eigenvalues and mixing angle satisfy
\begin{align}
    m_{H_{1,2}^0}^2 = \frac12\left[ A_N+D_N \mp  \sqrt{(A_N-D_N)^2+4B_N^2}  \right],
    \qquad
    \tan 2\theta_{N} =  \frac{2B_N}{A_N-D_N},
    \label{eq:neutral_scalar_angle}
\end{align}
up to the usual quadrant choice fixed by the signs of $B_N$ and the ordered eigenstates. The parameter $\mu_\phi$ is therefore the source of doublet-triplet mixing, whereas $\kappa_3$ controls the splitting between the {\it CP}-even and {\it CP}-odd neutral doublet components before mixing. Since the real triplet contains no independent {\it CP}-odd neutral scalar, $\omega_2$ is already a mass eigenstate. Its mass is
\begin{equation}
    m_{\omega_2}^2 = m_\phi^2+\frac12(\kappa_1+\kappa_2-\kappa_3) \upsilon_H^2.
    \label{eq:omega_mass}
\end{equation}
The charged fields $(\phi_2^\pm,\eta^\pm)$ are mixed by the same trilinear operator.  In the scalar potential convention of Eq.~\eqref{scal.pot}, their quadratic terms are described by
\begin{equation}
    \label{eq:charged_mass_matrix}
    \mathcal{M}_C^2 =
    \begin{pmatrix}
        A_C & B_C\\
        B_C & D_C
    \end{pmatrix},
    \quad \text{with} \quad
    \begin{aligned} 
        A_C = m_\phi^2+\frac12\kappa_1v^2, \quad
        B_C = \frac{\mu_\phi v}{\sqrt2}, \quad
        D_C = M_\eta^2+\frac12\kappa_4v^2.
    \end{aligned}
\end{equation}
The charged rotation in Eq.~\eqref{eq:states} gives $R_C \mathcal{M}_C^2 R_C^T  =  \mathrm{diag}\left(m_{H_1^\pm}^2,m_{H_2^\pm}^2\right)$. The corresponding eigenvalues and mixing angle are obtained from Eq.~\eqref{eq:neutral_scalar_angle} by replacing $A_N,B_N,D_N$ with $A_C,B_C,D_C$.  In this minimal potential limit the neutral and charged matrices are correlated because they share $B_N=B_C$ and the same triplet-like diagonal entry.

For our numerical analysis, we use a physical input basis rather than the original set of quadratic parameters in Eq.~\eqref{scal.pot}.  In the scalar sector this basis is chosen as
\begin{equation}
    \label{eq:scalar_input_basis}
    \mathcal{P}_{\mathrm{scalar}}^{\mathrm{in}} =
    \left\{ m_{H_1^0},m_{H_2^0},m_{\omega_2}, m_{H_1^\pm},m_{H_2^\pm}, \theta_{\mathrm{N}},\theta_{\mathrm{C}}, \lambda_2, \lambda_3, \kappa_1,\kappa_2,\kappa_4,\kappa_5
    \right\}.
\end{equation}
Here $m_{h_1}$ and $\upsilon_H$ fix $\lambda_1$ and $\mu_H^2$, while the neutral sector quantities $\kappa_3$, $\mu_\phi$, $m_\phi^2$, and $M_\eta^2$ are dependent parameters reconstructed from the physical neutral masses and $\theta_{N}$. Using the input basis in Eq.~\eqref{eq:scalar_input_basis}, these parameters are obtained by inverting the {\it CP}-even matrix together with Eq.~\eqref{eq:omega_mass}.  With the rotation convention above, one finds
\begin{align}
    \label{eq:neutral_physical_map}
    \kappa_3 &=  \frac{c_{\theta_{N}}^2m_{H_1^0}^2+s_{\theta_{N}}^2m_{H_2^0}^2-m_{\omega_2}^2}{v^2},
    &
    \mu_\phi &= \frac{\sqrt2 c_{\theta_{N}}s_{\theta_{N}}\left(m_{H_1^0}^2-m_{H_2^0}^2\right)}{v},
    \nonumber\\
    m_\phi^2 &= m_{\omega_2}^2-\frac12(\kappa_1+\kappa_2-\kappa_3)v^2,
    &
    M_\eta^2 &= s_{\theta_{N}}^2m_{H_1^0}^2+c_{\theta_{N}}^2m_{H_2^0}^2-\frac12\kappa_4v^2.
\end{align}
Similarly, the physical charged parameters define the gauge basis entries through
\begin{align}
    \label{eq:charged_physical_map}
    A_C = c_{\theta_{C}}^2m_{H_1^\pm}^2+s_{\theta_{C}}^2m_{H_2^\pm}^2,
    \qquad
    B_C = c_{\theta_{C}}s_{\theta_{C}}\left(m_{H_1^\pm}^2-m_{H_2^\pm}^2\right),
    \qquad
    D_C = s_{\theta_{C}}^2m_{H_1^\pm}^2+c_{\theta_{C}}^2m_{H_2^\pm}^2.
\end{align}

\section{Theoretical and experimental constraints}
\label{sec:constraints}
\subsection{Perturbativity and Vacuum stability}
\label{perturb_vacu_stab}
The theoretical consistency of the model requires perturbative couplings, a scalar potential that is bounded from below (BFB), and scalar-scattering amplitudes that satisfy perturbative unitarity.
\newline
\textbf{Perturbativity:} As a conventional tree-level perturbativity criterion, we require the quartic and Yukawa couplings to satisfy
\begin{equation}
    |\lambda_i|,~|\kappa_i| < 4\pi,
    \qquad
    |Y_N|,~|y_\eta|,~|y_\phi| < \sqrt{4\pi}.
\end{equation}
\newline
\textbf{Boundedness from below:} At large field values, the quartic part of the scalar potential dominates. The absence of runaway directions therefore requires the quartic potential to remain positive along all directions in field space. For multi-scalar potentials, the corresponding BFB conditions can be derived using copositivity criteria~\cite{Kannike:2012pe}. A classic review of electroweak vacuum stability can be found in Ref.~\cite{Sher:1988mj}. For the scalar potential in Eq.~\eqref{scal.pot}, the terms proportional to $\kappa_2$ and $\kappa_3$ depend on the relative $\mathrm{SU}(2)$ orientation and phase of the two scalar doublets. We parameterize this dependence as
\begin{equation}
    H^\dagger\phi =  \sqrt{(H^\dagger H)(\phi^\dagger\phi)}\, \rho e^{i\theta},
    \qquad  0\leq\rho\leq1.
\end{equation}
Minimizing the quartic potential with respect to $\theta$ and $\rho$ gives the effective $H$--$\phi$ coupling
\begin{equation}
    \label{eqn:123}
    \kappa_{H\phi}^{\min}
    =
    \begin{cases}
        \kappa_1,
        & \kappa_2-|\kappa_3|\geq 0,\\[2mm]
        \kappa_1+\kappa_2-|\kappa_3|,
        & \kappa_2-|\kappa_3|<0.
    \end{cases}
\end{equation}
The resulting pairwise BFB conditions are
\begin{equation}
    \lambda_i>0
    \qquad \text{for } i=1,2,3, \quad \kappa_{H\phi}^{\min} + 2\sqrt{\lambda_1\lambda_2}>0, \quad \kappa_4+4\sqrt{\lambda_1\lambda_3}>0, \quad
    \kappa_5+4\sqrt{\lambda_2\lambda_3}>0.
\label{eq:bounded_conditions}
\end{equation}
Together with the corresponding three-field copositivity condition, these inequalities ensure that the quartic potential is positive in all asymptotic directions of field space.
\newline
\textbf{Perturbative unitarity:} Perturbative unitarity constrains combinations of the quartic couplings through scalar $2\to2$ scattering. In the high-energy limit, the scattering amplitudes are determined by the quartic interactions, and the resulting coupled-channel matrix can be expressed in terms of its eigenvalues $\Lambda_i$. The complete set of eigenvalues is given in Appendix~\ref{sec:unitarity}. With the normalization adopted here, they must satisfy$\left|\Lambda_i(\lambda_j,\kappa_k)\right|<8\pi$ for all $i$~\cite{Lee:1977eg,Goodsell:2018tti}.

\subsection{Electroweak precision observables}
New electroweak states modify the gauge boson vacuum polarizations through loop effects. These corrections are conveniently parameterized by the oblique parameters $S$, $T$ and $U$ introduced by Peskin and Takeuchi~\cite{Peskin:1991sw}. Their evaluation in extended scalar sectors is studied in Ref.~\cite{Grimus:2007if}, while the precision constraints adopted here are summarized by the Particle Data Group~\cite{ParticleDataGroup:2024cfk}. In models where new physics mainly affects the gauge boson propagators, these parameters provide a model independent constraint on the spectrum of new particles using precision electroweak data.  In the present model, the additional $Z_2$-odd fields consist of the inert scalar states $H_1^\pm$, $H_2^\pm$, $H_1^0$, $H_2^0$ and $\omega_2$, together with a fermion triplet $\psi$. Since $Z_2$ symmetry forbids VEVs for the inert scalars, there are no tree-level corrections to the parameter $\rho \equiv m_W^2/(m_Z^2 \cos^2\theta_W)
$. Consequently, the leading electroweak precision constraints arise at one loop. We use the standard definitions of the oblique parameters in terms of the transverse gauge boson self-energies $\Pi_{VV'}(q^2)$~\cite{Peskin:1991sw,Grimus:2007if}. We work in the $U=0$ limit and include the scalar contributions from $H^\pm_1$, $H^\pm_2$, $H^0_1$, $H^0_2$, and $\omega_2$ evaluated in the mass basis, while the fermion triplet is taken to be mass-degenerate in its charged and neutral components, and thus its oblique contribution is neglected. 

The parameter $T$ is particularly sensitive to custodial symmetry breaking and therefore depends strongly on the mass splittings between the charged and neutral inert scalars. It is related to the deviation of $\rho$ from unity through $\Delta\rho \equiv \rho - 1 = \alpha T$. For a generic pair of scalar masses, $m_a$ and $m_b$, the contributions involve the loop function
\begin{equation}
    \label{eqn:obl_3}
    F\left(m_a^2,m_b^2\right) = \frac{m_a^2+m_b^2}{2} - \frac{m_a^2 m_b^2}{m_a^2-m_b^2}
    \ln{\left(\frac{m_a^2}{m_b^2}\right)},
\end{equation}
which vanishes for $m_a = m_b$, showing explicitly that $\Delta T$ is controlled by custodial-symmetry-breaking mass splittings. Here, the scalar contribution to the $T$ parameter can be written explicitly in the mass basis as
\begin{equation}
    \begin{aligned}
        \Delta T_{\text{scalar}} = \frac{1}{16\pi s_W^2 m_W^2} \Bigg[ & \left(c_{\theta_{C}} c_{\theta_{N}}+2s_{\theta_{C}} s_{\theta_{N}}\right)^2 F\left(m_{H_1^\pm}^2,m_{H_1^0}^2\right)
         + \left(c_{\theta_{C}} s_{\theta_{N}}-2s_{\theta_{C}} c_{\theta_{N}}\right)^2 F\left(m_{H_1^\pm}^2,m_{H_2^0}^2\right) \\
         & + \left(2c_{\theta_{C}} s_{\theta_{N}}-s_{\theta_{C}} c_{\theta_{N}}\right)^2 F\left(m_{H_2^\pm}^2,m_{H_1^0}^2\right)
         + \left(2c_{\theta_{C}} c_{\theta_{N}}+s_{\theta_{C}} s_{\theta_{N}}\right)^2 F\left(m_{H_2^\pm}^2,m_{H_2^0}^2\right) \\
         & + c_{\theta_{C}}^2 F\left(m_{H_1^\pm}^2,m_{\omega_2}^2\right) + s_{\theta_{C}}^2 F\left(m_{H_2^\pm}^2,m_{\omega_2}^2\right)
         - c_{\theta_{N}}^2 F\left(m_{H_1^0}^2,m_{\omega_2}^2\right) - s_{\theta_{N}}^2 F\left(m_{H_2^0}^2,m_{\omega_2}^2\right) \\
         & - 2c_{\theta_{C}}^2 s_{\theta_{C}}^2\left(3c_W^2 -s_W^2\right) F\left(m_{H_1^\pm}^2,m_{H_2^\pm}^2\right) \Bigg],
    \end{aligned}
    \label{eq:obl_4}
\end{equation}
where $s_X \equiv \sin\theta_X$ and $c_X \equiv \cos\theta_X$ for $X = C,N$. The $S$ parameter probes the momentum dependence of the neutral gauge boson self-energies and is typically less sensitive to small mass splittings. It is given by 
\begin{eqnarray}
        \Delta S_{\text{scalar}} &= \frac{1}{12\pi} \Bigg[ \left(s_{\theta_{C}}^2+\frac{1}{2} c_{\theta_{C}}^2-s_W^2\right)^2 G\left(m_{H_1^\pm}^2,m_{H_1^\pm}^2,m_Z^2\right)
         + \left(c_{\theta_{C}}^2+\frac{1}{2} s_{\theta_{C}}^2-s_W^2\right)^2 G\left(m_{H_2^\pm}^2,m_{H_2^\pm}^2,m_Z^2\right) \nonumber \\
        & + 2c_{\theta_{C}}^2 s_{\theta_{C}}^2\left(\frac{3}{2}-2s_W^2\right)^2 G\left(m_{H_1^\pm}^2,m_{H_2^\pm}^2,m_Z^2\right) 
         + \frac{c_{\theta_{N}}^2}{4} G\left(m_{H_1^0}^2,m_{\omega_2}^2,m_Z^2\right) + \frac{s_{\theta_{N}}^2}{4} G\left(m_{H_2^0}^2,m_{\omega_2}^2,m_Z^2\right) \Bigg].
    \label{eq:s}
\end{eqnarray}
Here, $G(x,y,z)$ denotes the scalar two-point loop function entering the momentum-dependent part of the neutral gauge-boson self-energies; its explicit definition is given in Ref.~\cite{Grimus:2007if}. In the present analysis, the scalar sector provides the new-physics contributions to the oblique parameters, and we therefore identify
\begin{equation}
\Delta S \equiv \Delta S_{\mathrm{scalar}},
\qquad
\Delta T \equiv \Delta T_{\mathrm{scalar}}.
\end{equation}
In the limit $\Delta U=0$, we adopt the electroweak global-fit inputs~\cite{ParticleDataGroup:2024cfk}
\begin{equation}
\label{eqn}
\bigl(\Delta S_0,\Delta T_0\bigr) = \bigl(-0.05,~0.00\bigr),
\qquad
\bigl(\sigma_S,\sigma_T\bigr) = \bigl(0.07,~0.06\bigr),
\end{equation}
where $\Delta S_0$ and $\Delta T_0$ denote the best-fit values and $\sigma_S$ and $\sigma_T$ their respective standard deviations. The correlation between $\Delta S$ and $\Delta T$ is incorporated in the likelihood analysis described in Sec.~\ref{subsec:numerical_scan}.
\subsection{Higgs to diphoton branching ratio}
The decay of the SM Higgs boson into two photons is a loop-induced process dominated by $W^\pm$-boson and charged fermion loops~\cite{Djouadi:2005gi}. It is therefore particularly sensitive to the presence of new charged particles, including charged scalars~\cite{Carena:2012xa,Picek:2012ei}. In the present model, the scalar sector contains the charged states $H_i^\pm$ arising from the mixing of the fields $\phi^\pm$ and $\eta^\pm$ defined in Eq.~\eqref{eq:states}. These states couple to the SM Higgs boson through the scalar potential in Eq.~\eqref{scal.pot}, and contribute to the decay $h \to \gamma\gamma$ at the one-loop level. After EWSB, the scalar potential generates interaction terms of the form
\begin{equation}
    \label{eq:hcharged-gaugebasis}
    \kappa_1 \upsilon_H h \phi^+ \phi^- + \kappa_4 \upsilon_H h\eta^+ \eta^-+
    \frac{\mu_\phi}{\sqrt{2}} h \left(\phi^+ \eta^- + \phi^- \eta^+ \right).
\end{equation}
These terms induce the couplings between the Higgs boson and the charged scalar fields in the gauge basis. Rotating to the charged-scalar mass eigenstate basis, only the diagonal Higgs-charged-scalar couplings contribute to this process, and they are given by
\begin{equation}
    \label{eq:hcharged-diagonal}
    \begin{aligned}
        g_{hH_1^+H_1^-} = \kappa_1 \upsilon_H c_{\theta_{C}}^2 + \kappa_4 \upsilon_H s_{\theta_{C}}^2+ \sqrt{2} \mu_\phi s_{\theta_{C}} c_{\theta_{C}}, \qquad g_{hH_2^+H_2^-} = \kappa_1 \upsilon_H s_{\theta_{C}}^2+\kappa_4 \upsilon_H c_{\theta_{C}}^2- \sqrt{2} \mu_\phi s_{\theta_{C}} c_{\theta_{C}}.
    \end{aligned}
\end{equation}
The partial decay width of $h \to \gamma \gamma$ is defined as
\begin{equation}
    \Gamma(h \to \gamma\gamma) =
    \frac{G_{\mathrm{F}}\alpha_{\mathrm{em}}^2m_{\mathrm{h}}^3}{128\sqrt{2}\pi^3}
    \left| \mathcal{A}_W + \mathcal{A}_f + \mathcal{A}_{H^\pm} \right|^2,
    \label{eq:diphoton}
\end{equation}
where $\mathcal{A}_W$, $\mathcal{A}_f$, and $\mathcal{A}_{H^\pm}$ denote the contributions from the $W$-boson, charged fermions, and the new charged scalar loops, respectively. The SM contributions are given by
\begin{equation}
    \label{eq:AWAf}
    \mathcal{A}_W = \mathcal{A}_1(\tau_W), \qquad
    \mathcal{A}_f = \sum_f N_c^f Q_f^2 \mathcal{A}_{1/2}(\tau_f),\qquad
    \tau_j=\frac{m_{h}^2}{4m_j^2}.
\end{equation}
We define the charged-scalar amplitude coefficients as 
\begin{equation}
    \begin{aligned}
        \kappa_{H_i^\pm}
        \equiv \frac{\upsilon_H g_{hH_i^+H_i^-}}
        {2m_{H_i^\pm}^{2}},
        \qquad i=1,2, \qquad \text{with } \quad
        \mathcal{A}_{H^\pm} =\sum_{i=1,2}\kappa_{H_i^\pm}\mathcal{A}_0(\tau_i).
    \end{aligned}
\label{eq:kappa_Hpm}
\end{equation}
The loop functions for spin-1, spin-1/2 and spin-0 particles are given respectively by
\begin{equation}
        \mathcal{A}_1(\tau) = -[2\tau^2+3\tau+3(2\tau-1)f(\tau)]\tau^{-2}, \quad
        \mathcal{A}_{1/2}(\tau) = 2[\tau+(\tau-1)f(\tau)]\tau^{-2}, \quad
        \mathcal{A}_0(\tau) = -[\tau-f(\tau)]\tau^{-2},
    \label{eq:loop-functions}
\end{equation}
with
\begin{equation}
    f(\tau) =
    \begin{cases}
        \arcsin^2(\sqrt{\tau}), & \tau \le 1, \\
        -\dfrac{1}{4}\left[\ln\left(\dfrac{1+\sqrt{1-\tau^{-1}}}{1-\sqrt{1-\tau^{-1}}}\right) - i\pi \right]^2, & \tau > 1,
    \end{cases}
    \label{eq:floop}
\end{equation}
Assuming that the new charged scalars do not induce a sizable modification of the total Higgs width, $\Gamma_\mathrm{h}^{\mathrm{tot}}\simeq \Gamma_{\mathrm{h},{\mathrm{SM}}}^{\mathrm{tot}}$, the constraint from the diphoton branching ratio can be expressed in terms of
\begin{equation}
    \label{eq:mugamgam}
    R_{\gamma\gamma}\equiv
\frac{{\mathrm{BR}}(h\to\gamma\gamma)_{\mathrm{model}}}
     {{\mathrm{BR}}(h\to\gamma\gamma)_{\mathrm{SM}}}
    \simeq
    \frac{\Gamma(h\to\gamma\gamma)_{\mathrm{model}}}
     {\Gamma(h\to\gamma\gamma)_{\mathrm{SM}}}
\end{equation}  

Using Eq.~\eqref{eq:diphoton}, the ratio becomes 
\begin{equation}
    \label{eq:diphoton2}
    R_{\gamma\gamma}
    =
    \frac{|A_W+A_f+A_{H^\pm}|^2}
         {|A_W+A_f|^2}
    =
    \left|
    1+
    \frac{
    \displaystyle
    \sum_{i=1,2}
    \frac{g_{hH_i^+H_i^-}v}{2m_{H_i^\pm}^2}
    A_0(\tau_i)}
    {A_W+A_f}
    \right|^2.
\end{equation}

The experimental input is imposed on the normalized branching ratio, 
\begin{equation}
    R_{\gamma\gamma}^{\mathrm{exp}}
    =
    \frac{{\mathrm{BR}}(h\to\gamma\gamma)_{\mathrm{exp}}}
         {{\mathrm{BR}}(h\to\gamma\gamma)_{\mathrm{SM}}},
\end{equation}
The charged scalar contribution can interfere constructively or destructively with the SM $W$-boson and charged-fermion amplitudes, leading to an enhancement or suppression of $R_{\gamma\gamma}$~\cite{Aiko:2023nqj}. Therefore, the measured value of $R_{\gamma\gamma}$ constrains $m_{H_i^\pm}$, $\theta_C$, and the scalar couplings entering $g_{hH_i^+H_i^-}$. The ATLAS combination reports the Higgs-boson branching fractions assuming SM production cross sections. For the diphoton channel, the measured ratio is
\begin{equation}
    R_{\gamma\gamma}^{\mathrm{ATLAS}}
    \equiv
    \frac{{\mathrm{Br}}(h\to\gamma\gamma)_{\mathrm{ATLAS}}}{{\mathrm{Br}}(h\to\gamma\gamma)_{\mathrm{SM}}} = 1.0883^{+0.0952}_{-0.0895}.
    \label{eq:Rgg_ATLAS_input}
\end{equation}
\subsection{Charged lepton flavor violation processes}
\label{sec:clfv}
Charged lepton flavor violation (cLFV) arises from the flavor structure of the Yukawa interaction in Eq.~\eqref{eq:yukawa}. Since the exact $Z_2$ symmetry forbids tree-level cLFV amplitudes, the leading contributions are generated at one loop and are controlled by products of two $y_\phi$ couplings. The relevant loop states are the neutral and charged $Z_2$-odd scalars $H_i^0$, $H_i^\pm$, the {\it CP}-odd state $\omega_2$, and the fermionic triplet components. In our numerical analysis we include the radiative decays $\ell_\alpha \to \ell_\beta\gamma$, the three-body decays $\ell_\alpha\to3\ell_\beta$, and coherent $\mu-e$ conversion in nuclei. These observables provide complementary probes of the flavor structure because they depend on dipole, non-dipole, and four-fermion operators~\cite{Ilakovac:1994kj,Kuno:1999jp,Calibbi:2017uvl}.
 The radiative decays $\ell_\alpha \to \ell_\beta \gamma$ are described by the effective dipole amplitude
\begin{equation}
    \label{eq:muegamma}
    \mathcal{M}(\ell_\alpha \to \ell_\beta \gamma) =
    i e m_\alpha~\overline{u}_{\beta}(p-q)~\sigma^{\mu\nu} q_\nu
    \left(A_{R} P_{R} + A_{L} P_{L}\right)
    u_{\alpha}(p)~\epsilon_\mu^*(q),
\end{equation}
where $A_{L}$ and $A_{R}$ are the left- and right-handed dipole form factors. For an on-shell photon, the transition is fully described by these dipole operators, and the corresponding branching ratio is
\begin{equation}
    \mathrm{BR}(\ell_\alpha \to \ell_\beta \gamma) = \frac{48\pi^3\alpha_{\mathrm{em}}}{G_{F}^2}
    \left(
    |A_{L}|^2 + |A_{R}|^2
    \right)
    \mathrm{BR}(\ell_\alpha \to \ell_\beta \nu_\alpha \bar{\nu}_\beta).
    \label{eq:br1}
\end{equation}
For the present field content, the dipole amplitude receives contributions from the $(\psi^0,H_i^\pm)$, $(\psi^\pm,H_i^0)$, and $(\psi^\pm,\omega_2)$ loop topologies~\cite{Toma:2013zsa}. In the mass basis, the right-handed form factor is
{\small
\begin{equation}
    \begin{aligned}
        A_{R} &= -\frac{1}{32\pi^2} \sum_a (y_\phi)_{\beta a}(y_\phi)^{*}_{\alpha a} \Biggl[ \frac{c_{\theta_{N}}^2}{m_{H_1^0}^2} f_{F}(x_{H_1^0}^a) + \frac{s_{\theta_{N}}^2}{m_{H_2^0}^2} f_{F}(x_{H_2^0}^a) + \frac{1}{m_{\omega_2}^2} f_{F}(x_{\omega_2}^a) + \frac{c_{\theta_{C}}^2}{m_{H_1^\pm}^2} f_{S}(x_{H_1^\pm}^a) + \frac{s_{\theta_{C}}^2}{m_{H_2^\pm}^2} f_{S}(x_{H_2^\pm}^a) \Biggr],
    \end{aligned}
    \label{eq:AR_lgamma}
\end{equation}
}
where $x_X^a\equiv M_{\psi_a}^2/m_X^2$. The opposite-chirality form factor is suppressed by the charged-lepton mass ratio and is given by $A_{\mathrm L}=(m_\beta/m_\alpha)A_{\mathrm R}$. The loop functions are given by 
\begin{equation}
    \label{eq:dipole_loop_functions}
    f_{F}(x) = \frac{2+3x-6x^2+x^3+6x\ln x}{6(1-x)^4}, \quad
    f_{S}(x) = \frac{1-6x+3x^2+2x^3-6x^2\ln x}{6(1-x)^4}.
\end{equation}

The three-body decay $\ell_\alpha \to 3\ell_\beta$ receives contributions from off-shell photon penguins, $Z$-penguin diagrams, and box diagrams. It is therefore sensitive not only to the dipole form factor entering $\ell_\alpha\to\ell_\beta\gamma$, but also to non-dipole vector and four-lepton contact operators. Neglecting the mass-suppressed opposite-chirality contribution, we set $A_D=A_R$. The photon-penguin and box contributions to the branching ratio are then given by~\cite{Vicente:2014wga}
\begin{equation}
    \begin{aligned}
        \mathrm{BR}(\ell_\alpha \to 3\ell_\beta) &= \frac{3(4\pi)^2\alpha_{\mathrm{em}}^2}{8G_{\mathrm{F}}^2}
        \Bigg[
        \underbrace{|A_{ND}|^2}_{\mathrm{non-dipole}} + \underbrace{|A_{D}|^2
        \left( \frac{16}{3}\ln\frac{m_\alpha}{m_\beta} - \frac{22}{3} \right)}_{\mathrm{dipole}}
        \\
        &\quad
        + \underbrace{\frac{1}{6}|B|^2}_{\mathrm{box}} + \underbrace{\left(
        -2A_{ND}A_{D}^* +\frac{1}{3}A_{ND}B^*
        -\frac{2}{3}A_{D}B^* +\mathrm{h.c.} \right)}_{\mathrm{interference}}
        \Bigg]
        \times
        \mathrm{BR}(\ell_\alpha \to \ell_\beta \nu_\alpha \bar{\nu}_\beta).
    \end{aligned}
    \label{eq:br_l3l}
\end{equation}
The corresponding non-dipole photon-penguin coefficient is
\begin{equation}
        A_{ND} = \frac{1}{32\pi^2} \sum_a (y_\phi)_{\beta a}(y_\phi)^{*}_{\alpha a}
        \Bigg[
        \frac{c_{\theta_{N}}^2}{m_{H_1^0}^2} g_{F}(x_{H_1^0}^a)
        + \frac{s_{\theta_{N}}^2}{m_{H_2^0}^2}~g_{F}(x_{H_2^0}^a)
        +
        \frac{1}{m_{\omega_2}^2}~g_{F}(x_{\omega_2}^a)
        + \frac{c_{\theta_{C}}^2}{m_{H_1^\pm}^2}~g_{S}(x_{H_1^\pm}^a) +
        \frac{s_{\theta_{C}}^2}{m_{H_2^\pm}^2}~ g_{S}(x_{H_2^\pm}^a)
        \Bigg],
\end{equation}
where the loop functions are in this case given by
\begin{equation}
    g_{S}(x)=\frac{2-9x+18x^2-11x^3+6x^3\ln x}{6(1-x)^4},
    \qquad
    g_{F}(x)=\frac{2x^3-9x^2+18x-11-6\ln x}{6(1-x)^4}.
\end{equation}
The box coefficient $B$ arises from genuine four-point one-loop diagrams involving the exchange of the fermionic states $\psi^0,\psi^\pm$ and the scalar mass eigenstates $H_i^\pm$, $H_i^0$, and $\omega_2$~\cite{Ilakovac:1994kj}. Defining $B = B^{(\psi^0,H^\pm)} + B^{(\psi^\pm,S^0)}$, we have
\begin{equation}
    \begin{aligned}
        e^2 B^{(\psi^0,H^\pm)}
        &=
        \frac{1}{(4\pi)^2}
        \sum_{a,b}\sum_{i,j=1}^{2}
        \frac{r_i^2 r_j^2}{4}
        \Bigg[
        \frac{1}{2}
        (y_\phi)_{\beta a}(y_\phi)^*_{\alpha a}
        (y_\phi)_{\beta b}(y_\phi)^*_{\beta b}
        \mathcal I_4\left(M_{\psi_a^0}^2,M_{\psi_b^0}^2,m_{H_i^\pm}^2,m_{H_j^\pm}^2\right)
        \\
        &\hspace{3.7cm}
        +
        M_{\psi_a^0}M_{\psi_b^0}
        (y_\phi)_{\beta a}(y_\phi)_{\beta b}
        (y_\phi)^*_{\alpha a}(y_\phi)^*_{\beta b}
        \mathcal J_4\left(M_{\psi_a^0}^2,M_{\psi_b^0}^2,m_{H_i^\pm}^2,m_{H_j^\pm}^2\right)
        \Bigg],
        \\[2mm]
        e^2 B^{(\psi^\pm,S^0)}
        &=
        \frac{1}{(4\pi)^2}
        \sum_{a,b}\sum_{r,s=1}^{3}
        \frac{u_r^2 u_s^2}{4}
        \Bigg[
        \frac12
        (y_\phi)_{\beta a}(y_\phi)^*_{\alpha a}
        (y_\phi)_{\beta b}(y_\phi)^*_{\beta b}
        \mathcal I_4\left(M_{\psi_a^\pm}^2,M_{\psi_b^\pm}^2,m_{S_r^0}^2,m_{S_s^0}^2\right)
        \\
        &\hspace{3.7cm}
        +
        M_{\psi_a^\pm}M_{\psi_b^\pm}
        (y_\phi)_{\beta a}(y_\phi)_{\beta b}
        (y_\phi)^*_{\alpha a}(y_\phi)^*_{\beta b}         
        \mathcal J_4\left(M_{\psi_a^\pm}^2,M_{\psi_b^\pm}^2,m_{S_r^0}^2,m_{S_s^0}^2\right)
        \Bigg],
    \end{aligned}
\end{equation}
Here $S_r^0=\{H_1^0,H_2^0,\omega_2\}$, $r_i=(c_{\theta_{C}},-s_{\theta_{C}})$ and
$u_r=(c_{\theta_{N}},-s_{\theta_{N}},1/\sqrt2)$. The factor $1/2$ accounts
for identical fermions in the loop. The four-point functions are
\begin{equation}
    \mathcal J_4(a,b,c,d) = \sum_{x\in\{a,b,c,d\}} \frac{x\ln x}{\prod_{y\neq x}(x-y)},
    \qquad
    \mathcal I_4(a,b,c,d) = \sum_{x\in\{a,b,c,d\}}
    \frac{x^2\ln x}{\prod_{y\neq x}(x-y)}.
\end{equation}
The three-body decay also receives a genuine $Z$-penguin contribution. Neglecting terms suppressed by external charged lepton masses, the induced flavor changing vertex can be written as
\begin{equation}
    \label{eq:Zpenguin_vertex}
    \Gamma_Z^\mu(\ell_\alpha\to\ell_\beta) = \gamma^\mu P_{L}\mathcal F_Z.
\end{equation}
After integrating out the $Z$ boson, we define the reduced four-lepton coefficients
\begin{equation}
    F_{LL}^{Z}=g_{L}^\ell \mathcal{F_Z}/(e^2m_Z^2) \quad \text{and} \quad
    F_{LR}^{Z}=g_{R}^\ell \mathcal{F_Z}/(e^2m_Z^2),
\end{equation}
where
$g_{L}^\ell=e(-1/2 + s_W^2)/(s_Wc_W)$ and $g_{R}^\ell=e~s_W^2/(s_Wc_W)$. The first chiral index in $F_{LL}^{Z}$ and $F_{LR}^{Z}$ refers to the flavor-changing current $\bar{\ell}_\beta\gamma^\mu P_{\mathrm{L}}\ell_\alpha$, while the second one refers to the chirality of the final-state $\bar{\ell}_\beta\gamma_\mu P_{{L,R}}\ell_\beta$ current. The $Z$-penguin contribution amounts to adding 
\begin{equation}
    \Delta_Z = \frac13\left(2|F_{LL}^{Z}|^2+|F_{LR}^{Z}|^2\right)
\end{equation} 
inside the square bracket of Eq.~\eqref{eq:br_l3l}. The loop-induced form factor can be decomposed as $\mathcal F_Z = \mathcal F_Z^{(\psi^0,H^\pm)} + \mathcal F_Z^{(\psi^\pm,S^0)}$ where
\begin{equation}
    \begin{aligned}
        \mathcal F_Z^{(\psi^0,H^\pm)} = -\frac{1}{16\pi^2} \sum_a\sum_{i,j=1}^{2}
        g_{\beta a}^{C_j}~g_{\alpha a}^{C_i*}
        \Big[
        &
        2 g_{ZH_i^-H_j^+}~C_{24}(M_{\psi_a},m_{H_i^\pm},m_{H_j^\pm}) + \delta_{ij}g_{\mathrm{L}}^\ell~B_1(M_{\psi_a},m_{H_i^\pm})
        \Big],
        \\
        \mathcal F_Z^{(\psi^\pm,S^0)}
        =
        -\frac{1}{16\pi^2}
        \sum_a\sum_{r,s=1}^{3}
        g_{\beta a}^{N_s}
        g_{\alpha a}^{N_r*}
        \Bigg[
        &
        \delta_{rs}
        \Bigg\{
        g_{Z\psi^\pm}
        \left[
        2 C_{24}(m_{S_r^0},M_{\psi_a},M_{\psi_a})
        +\frac12
        +M_{\psi_a}^2
        C_0(m_{S_r^0},M_{\psi_a},M_{\psi_a})
        \right]
        \\
        &+ g_{\mathrm{L}}^\ell B_1(M_{\psi_a},m_{S_r^0})
        \Bigg\}
        +
        2 g_{ZS_r^0S_s^0}
        C_{24}(M_{\psi_a},m_{S_r^0},m_{S_s^0})
    \end{aligned}
\end{equation}
We express the loop amplitudes in terms of the standard Passarino--Veltman functions~\cite{Passarino:1978jh,Denner:1991kt}. In the low-energy cLFV observables considered here, the external charged lepton momenta are much smaller than the heavy mediator masses entering the loop. We therefore work in the leading term of the small-momentum expansion, evaluating the loop functions at vanishing external momenta
\begin{equation}
    B_1(M,m)\equiv B_1(0;M^2,m^2),
    \qquad
    C_0(a,b,c)\equiv C_0(0,0,0;a^2,b^2,c^2),
    \qquad
    C_{24}(a,b,c)\equiv C_{24}(0,0,0;a^2,b^2,c^2).
\end{equation}
The mass basis Yukawa couplings entering these expressions are
$g_{\alpha a}^{C_1}=ic_{\theta_{C}}(y_\phi)_{\alpha a}/\sqrt2$ and
$g_{\alpha a}^{C_2}=-is_{\theta_{C}}(y_\phi)_{\alpha a}/\sqrt2$ for the charged
scalars, while
$g_{\alpha a}^{N_1}=ic_{\theta_{N}}(y_\phi)_{\alpha a}/\sqrt2$,
$g_{\alpha a}^{N_2}=-is_{\theta_{N}}(y_\phi)_{\alpha a}/\sqrt2$ and
$g_{\alpha a}^{N_3}=-(y_\phi)_{\alpha a}/\sqrt2$ for the neutral scalars.
The $Z$-boson couplings follow from the scalar and fermion kinetic terms after rotation to the mass basis. Adopting the vertex convention $i g_{ZXY}(p_X-p_Y)^\mu$, the charged-scalar couplings are
\begin{equation}
g_{Z H_i^- H_j^+} = -\frac{e}{s_Wc_W} \left(\frac{1}{2}-s_W^2\right) R^C_{i1}R^C_{j1}
+ \frac{e}{s_Wc_W} \left(1-s_W^2\right)
R^C_{i2}R^C_{j2},
\end{equation}
where $R^C$ is the charged scalar rotation matrix defined in Eq.~\Ref{eq:states}. In particular,
\begin{equation}
g_{Z H_1^- H_2^+}
=
g_{Z H_2^- H_1^+}
=
\frac{e}{s_Wc_W}
\left(\frac{3}{2}-2s_W^2\right)c_{\theta_{C}}s_{\theta_{C}},
\end{equation}
which is the same coupling combination that appears in the mixed charged scalar contribution to Eq.~\eqref{eq:s}. For the $Z$ couplings to the neutral scalars and charged-fermion, they are given respectively by
\begin{eqnarray}
g_{Z S_r^0 S_s^0} = \frac{e}{2s_Wc_W}
\left( X_r\delta_{s3}-X_s\delta_{r3} \right),
\qquad
g_{Z\psi^\pm} = \frac{e}{s_Wc_W}\left(1-s_W^2\right) = \frac{ec_W}{s_W},
\qquad \text{with} \qquad
X=(c_{\theta_{N}},-s_{\theta_{N}},0).
\end{eqnarray}

Coherent $\mu-e$ conversion provides an additional probe of the $\mu\to e$ flavor changing transition. We evaluate the conversion rate for aluminium, using the same $y_\phi$-dominance approximation as for the radiative and three-body CLFV modes. The short-distance contributions retained are the off-shell photon penguin and the $Z$-penguin, while scalar quark operators are neglected. Following the standard notation of
Ref.~\cite{Vicente:2014wga}, the conversion rate is
\begin{equation}
    \label{eq:mueconv}
    \begin{aligned}
        \mathrm{CR}(\mu-e, \mathrm{Nucleus}) 
        &= \frac{p_e E_e m_\mu^3 G_{F}^2 \alpha_{\mathrm{em}}^3 Z_{\mathrm{eff}}^4 F_p^2}{8\pi^2 Z \Gamma_{\mathrm{capt}}} \quad \times \Biggl\{ \left| (Z+N)(g_{LV}^{(0)}+g_{{LS}}^{(0)}) + (Z-N)(g_{LV}^{(1)}+g_{LS}^{(1)}) \right|^2 \\
        &\quad + \left| (Z+N)(g_{RV}^{(0)}+g_{RS}^{(0)}) + (Z-N)(g_{RV}^{(1)}+g_{RS}^{(1)}) \right|^2 \Biggr\}.
    \end{aligned}
\end{equation}
The right-handed vector coefficient is suppressed by the external charged lepton masses, and we take $g_{RV}^{(0,1)} \simeq 0$. The left-handed quark-level coefficients are
\begin{equation}
    g_{LV}^{(q)} = \frac{\sqrt{2}}{G_{F}}
    \left[ e^2 Q_q\left(A_{ND}^{\mu e}-A_{D}^{\mu e}\right)
    + \frac{g_{V}^q}{m_Z^2}F_Z^{\mu e} \right], \qquad q=u,d,
\end{equation}
where $A_{D}^{\mu e}$ and $A_{\mathrm{ND}}^{\mu e}$ denote respectively the reduced dipole and non-dipole photon-penguin coefficients, while $F_Z^{\mu e}$ is the left-handed $Z$-penguin form factor already entering the $\mu\to eee$ amplitude, and
\begin{equation}
    g_{V}^u=\frac{e}{s_Wc_W}\left(\frac12-\frac43s_W^2\right),
    \qquad
    g_{V}^d=\frac{e}{s_Wc_W}\left(-\frac12+\frac23s_W^2\right).
\end{equation}
The isoscalar and isovector combinations are then
\begin{equation}
    g_{LV}^{(0)}=\frac32\left(g_{LV}^{(u)}+g_{LV}^{(d)}\right),
    \qquad
    g_{LV}^{(1)}=\frac12\left(g_{LV}^{(u)}-g_{LV}^{(d)}\right).
\end{equation}
\subsection{Decay and collider properties of the $Z_2$-odd sector}
\label{sec:charged_decay_widths}
In the scalar-DM scenario considered here, $H_1^0$ is the lightest $Z_2$-odd state and is therefore stable. The heavier scalars $H_2^0$, $\omega_{2}$, and $H_{1,2}^{\pm}$ can decay through electroweak gauge interactions, provided that the corresponding transitions are kinematically open. Denoting the neutral scalar mass eigenstates by $S_{a}^{0} \in \{H_{1}^{0},H_{2}^{0},\omega_{2}\}$, the relevant scalar transitions are $H_{i}^{\pm}\to S_{a}^{0}W^{\pm(*)}$ and $S_{b}^{0}\to S_{a}^{0}Z^{(*)}$. The fermionic decays relevant to the benchmark analysis are $\psi^{\pm}\to N H_{i}^{\pm}$ and $\psi^{0}\to N S_{a}^{0}$. The gauge interactions that mediate these transitions
follow from the scalar kinetic terms after rotating to the mass basis
defined in Eq.~\eqref{eq:states}, and may be written as
\begin{equation}
    \mathcal{L}_{\mathrm{gauge}}
    \supset
    \sum_{i=1}^{2}\sum_{a=1}^{3}
    \left[
    g_{ia}^{W} W_{\mu}^{-}H_{i}^{+} \overleftrightarrow{\partial^{\mu}}S_{a}^{0} + \mathrm{h.c.}\right] + \sum_{k=1}^{2} g_{k}^{Z} Z_{\mu}H_{k}^{0} \overleftrightarrow{\partial^{\mu}}\omega_{2},
    \label{eq:scalar-gauge-interactions}
\end{equation}
where $A\overleftrightarrow{\partial^{\mu}}B \equiv A(\partial^{\mu}B)-(\partial^{\mu}A)B$. The coefficients $g_{ia}^{W}$ and $g_{k}^{Z}$ are fixed by the electroweak gauge coupling and the scalar mixing matrices. The charged-current coefficients contain the appropriate charged and neutral mixing factors, whereas $g_{k}^{Z}$ depends only on $\theta_N$. For the scalar transitions generated by Eq.~\eqref{eq:scalar-gauge-interactions}, we denote the initial and final scalar mass eigenstates by $S_I$ and $S_J$, respectively. When the two-body channel $S_I\to S_JV$ is kinematically open, the corresponding partial width is
\begin{equation}
    \Gamma\left(S_{I}\to S_{J}V\right) = \frac{\left|g_{IJ}^{V}\right|^{2}}
    {16\pi m_{S_{I}}^{3}m_{V}^{2}} \lambda_{\mathrm{K}}^{3/2}
    \left( m_{S_{I}}^{2}, m_{S_{J}}^{2}, m_{V}^{2} \right),
    \label{eq:scalar-two-body-width}
\end{equation}
where $V=W,Z$, $m_{S_{I}}>m_{S_{J}}+m_{V}$, and $\lambda_{\mathrm{K}}(x,y,z) = x^{2}+y^{2}+z^{2}-2xy-2xz-2yz$ is the K\"all\'en function. When $m_{S_I}-m_{S_J}<m_V$, the two-body channel is kinematically closed and the transition proceeds through an off-shell gauge boson, $S_I\to S_J V^{*}\to S_J f\bar f^{(\prime)}$. $S_I$ and $S_J$ denote the parent and daughter scalar mass eigenstates, respectively.

For the fermionic transitions, we write the relevant mass-basis interaction in the generic form
\begin{equation}
    \mathcal{L}_{\psi N S} \supset S^{\dagger} \overline{N}
    \left( c_{L}P_{L}+c_{R}P_{R}  \right)\psi +\mathrm{h.c.}.
    \label{eq:fermion-scalar-interaction}
    \end{equation}
Here, $\psi$ denotes a charged or neutral triplet fermion and $S$ the corresponding charged or neutral scalar mass eigenstate. For $M_{\psi} > M_{N}+m_{S}$, the partial width is
\begin{equation}
        \Gamma\left(\psi\to NS\right) =
        \frac{\lambda_{\mathrm{K}}^{1/2}
        \left( M_{\psi}^{2}, M_{N}^{2}, m_{S}^{2} \right)
        }{32\pi M_{\psi}^{3}}
        \times \left[ \left(|c_{L}|^{2}+|c_{R}|^{2} \right) \left( M_{\psi}^{2}+M_{N}^{2}-m_{S}^{2} \right) + 4 M_{\psi}M_{N} \operatorname{Re}\left(c_{L}c_{R}^{*}\right) \right].
        \label{eq:fermion-two-body-width}
\end{equation}
The chiral coefficients $c_L$ and $c_R$ follow from the Yukawa interactions after rotating the scalar fields to the mass basis. Eq.~\eqref{eq:fermion-two-body-width} applies to $\psi^{\pm} \to N H_i^{\pm}$ and $\psi^{0} \to N S_a^{0}$, with the corresponding channel-dependent couplings.

At proton--proton colliders, charged-scalar pairs $H_i^+H_j^-$ are produced through $s$-channel $\gamma/Z$ exchange, while charged-neutral associated production $H_i^\pm S_a^0$, with $S_a^0\in\{H_1^0,H_2^0,\omega_2\}$, proceeds through $s$-channel $W^\pm$ exchange. Neutral scalar combinations of the form $H_{1,2}^0\omega_2$ are produced through $Z$ exchange. The fermionic triplet states are produced through $pp\to\psi^+\psi^-$ and $pp\to\psi^\pm\psi^0$. Numerical production rates for the representative benchmarks are presented in Sec.~\ref{subsec:benchmark-phenomenology}.

\section{Dark matter}
\label{sec:dark_matter}
\textbf{Thermal relic abundance: } In the present work we restrict the numerical analysis to the scalar branch in which the lightest {\it CP}-even neutral state $H_1^0$ is the DM candidate.  With the convention used in Sec.~\ref{sec:model}, $H_1^0=c_Nh_2+s_N\eta^0$, so its phenomenology interpolates between an inert-doublet-like state and a real-triplet-like state.  The neutral and charged particles close to it in the $Z_2$-odd spectrum, namely $H_2^0$, $\omega_2$, $H_i^\pm$, and, if sufficiently degenerate, the triplet fermions $\psi^0,\psi^\pm$, can remain thermally populated at freeze-out.  The thermal abundance is computed in the standard freeze-out framework, following the coannihilation formalism used in radiative dark sectors~\cite{Griest:1990kh,Belanger:2018ccd}. For the total number density of the thermally accessible odd sector, $n=\sum_i n_i$, one has
\begin{equation}
    \frac{dn}{dt}+3Hn = -\langle \sigma_{\mathrm{eff}} v\rangle
    \left(n^2-n_{\mathrm{eq}}^2\right),
    \label{eq:dm_boltzmann_n}
\end{equation}
where $H$ is the Hubble parameter and $n_{\mathrm{eq}}=\sum_i n_i^{\mathrm{eq}}$.  Equivalently, for the yield $Y=n/s$, with $s$ the entropy density and $x=m_{H_1^0}/T$,
\begin{equation}
    \frac{dY}{dx} = -\frac{s}{Hx} \langle\sigma_{\mathrm{eff}}v\rangle \left(Y^2-Y_{\mathrm{eq}}^2\right).
    \label{eq:dm_boltzmann_y}
\end{equation}
The effective cross section contains all annihilation and coannihilation channels among the states that are not Boltzmann suppressed at freeze-out,
\begin{equation}
    \langle\sigma_{\mathrm{eff}}v\rangle(x) = \sum_{i,j} \langle\sigma_{ij}v_{ij}\rangle r_i(x)r_j(x), \qquad r_i(x) = \frac{g_i(1+\Delta_i)^{3/2}e^{-x\Delta_i}} {\sum_k g_k(1+\Delta_k)^{3/2}e^{-x\Delta_k}},
    \label{eq:dm_sigma_eff}
\end{equation}
where
\begin{equation}
    \Delta_i=\frac{m_i-m_{H_1^0}}{m_{H_1^0}},
    \qquad
    m_i=m_{H_1^0}(1+\Delta_i).
    \label{eq:dm_delta_i}
\end{equation}
At freeze-out one typically has $x_F\simeq20$--$30$.  Therefore particles lying within a few to ten percent of $m_{H_1^0}$ can have sizeable weights $r_i$, and their annihilation rates can dominate $\langle\sigma_{\mathrm{eff}}v\rangle$ even when the self-annihilation rate of $H_1^0$ is modest.  The relic abundance may be written schematically as
\begin{equation}
    \Omega h^2 \simeq \frac{1.07\times10^9~{\mathrm{GeV}}^{-1}} {M_{\mathrm{Pl}}\sqrt{g_*(x_F)}}
    \frac{1}{J(x_F)}, \qquad J(x_F)=\int_{x_F}^{\infty} \frac{\langle\sigma_{\mathrm{eff}}v\rangle}{x^2}\,dx,
    \label{eq:dm_relic_integral}
\end{equation}
which shows explicitly that increasing the effective annihilation rate lowers the final abundance.  In the scan this calculation is performed with \texttt{micrOMEGAs} using the full physical spectrum and the \texttt{CalcHEP} model files.

For $H_1^0$ DM, several mechanisms can be active.  Around the Higgs funnel, $m_{H_1^0}\simeq m_h/2$, the $s$-channel Higgs resonance can deplete the abundance even for a moderate portal coupling.  Away from the resonance, annihilation into SM fermions, electroweak gauge bosons, and Higgs final states is governed by the portal combination, by the doublet-triplet composition of $H_1^0$, and by the available phase space.  In the high-mass region, the correct abundance is usually not obtained from $H_1^0 H_1^0$ annihilation alone; it is instead controlled by coannihilation with $H_2^0$, $\omega_2$, and $H_i^\pm$, and occasionally by the electroweak-triplet fermions if they are close enough in mass.  Thus the relevant quantities are the splittings $m_{H_2^0}-m_{H_1^0}$, $m_{\omega_2}-m_{H_1^0}$, $m_{H_i^\pm}-m_{H_1^0}$, and $m_{\psi}-m_{H_1^0}$, together with the scalar mixing angles and Higgs-portal couplings.\\

\textbf{Direct detection: } Direct detection probes the nonrelativistic elastic scattering of $H_1^0$ on nuclei. The dominant spin-independent contribution is mediated by the SM-like Higgs boson.  In the neutral scalar basis, the Higgs-portal matrix is obtained by differentiating the {\it CP}-even inert mass matrix with respect to the electroweak VEV.  With the same sign convention used for the charged-scalar couplings in Eq.~\eqref{eq:hcharged-diagonal}, the diagonal portal coupling relevant for elastic scattering is
\begin{equation}
    g_{hH_1^0H_1^0} = \upsilon_H (\kappa_1 +\kappa_2 + \kappa_3) c_N^2 + \upsilon_H  \kappa_4s_N^2 + \sqrt{2}\,\mu_\phi s_Nc_N.
    \label{eq:dm_hH01H01}
\end{equation}
This expression makes the physical dependence transparent: the direct-detection rate is controlled by the doublet fraction, the triplet fraction, and the same dimensionful mixing parameter $\mu_\phi$ that connects the two neutral inert scalars.  A cancellation among the three terms in Eq.~\eqref{eq:dm_hH01H01} can suppress the elastic Higgs exchange, while the individual couplings still affect the scalar spectrum, coannihilation, and $h\to\gamma\gamma$. At the nucleon level we write the scalar Higgs coupling as
\begin{equation}
    g_{h\mathcal{N}\mathcal{N}} = \frac{m_{\mathcal{N}}}{v}
    \left[
    \sum_{q=u,d,s} f_{Tq}^{(\mathcal{N})}
    +\frac{2}{27}f_{TG}^{(\mathcal{N})}
    \right],
    \qquad
    f_{TG}^{(\mathcal{N})} = 1-\sum_{q=u,d,s} f_{Tq}^{(\mathcal{N})},
    \label{eq:dm_hnn}
\end{equation}
where $\mathcal{N}=p,n$.  For an interaction written as $\mathcal{L}\supset-\frac12g_{hH_1^0H_1^0}h(H_1^0)^2$, the Higgs-mediated spin-independent cross section scales as
\begin{equation}
    \sigma_{\mathrm{SI}}^{H_1^0\mathcal{N}} = \frac{\mu_{1\mathcal{N}}^2}{4\pi m_{H_1^0}^2}
    \left|
    \frac{g_{hH_1^0H_1^0}g_{h\mathcal{N}\mathcal{N}}}{m_h^2}
    \right|^2,
    \qquad \mu_{1\mathcal{N}} = \frac{m_{H_1^0}m_{\mathcal{N}}}{m_{H_1^0}+m_{\mathcal{N}}}.
    \label{eq:dm_si_scalar}
\end{equation}
The numerical scan uses the \texttt{micrOMEGAs} direct-detection likelihood, which evaluates the relevant effective nuclear matrix elements from the full model implementation~\cite{Belanger:2006is,Belanger:2018ccd}. Gauge interactions can in principle induce neutral-current scattering for inert multiplet DM.  In this model, the relevant $Z$-mediated neutral-scalar interactions are off-diagonal in the neutral inert sector, so they lead to inelastic transitions~\cite{Tucker-Smith:2001myb} rather than to a dominant elastic $H_1^0\mathcal{N}\to H_1^0\mathcal{N}$ signal. In our numerical scan, the mass splittings between $H_1^0$, $H_2^0$, and $\omega_2$ in the accepted sample are far above the nuclear recoil scale, making such inelastic channels kinematically ineffective in underground detectors. The phenomenologically relevant direct-detection observable is therefore the spin-independent Higgs-mediated cross section.

\section{Results}
\label{results}
\subsection{Numerical scan and constraints}
\label{subsec:numerical_scan}
We scanned the model parameter space by fitting the relevant experimental observables subject to theoretical constraints. For each sampled parameter vector $\boldsymbol{\theta}$, we constructed the physical spectrum and evaluated the neutrino observables, Higgs diphoton branching ratio, oblique parameters $\Delta S$ and $\Delta T$, and cLFV observables. During the scan, points satisfying the preliminary requirements were evaluated with \texttt{micrOMEGAs} to calculate the relic density and direct-detection observables~\cite{Belanger:2018ccd,Belanger:2006is}.  The total statistical fit is decomposed as 
\begin{equation}
    \label{eq:chi2_total}
    \chi^2_{\mathrm{tot}}(\boldsymbol{\theta}) = 
    \chi^2_{\Omega}(\boldsymbol{\theta})
    +\chi^2_{\nu}(\boldsymbol{\theta})
    +\chi^2_{ST}(\boldsymbol{\theta})
    +\chi^2_{h\gamma\gamma}(\boldsymbol{\theta}).
\end{equation}
where $\chi^2_{\Omega}$, $\chi^2_{\nu}$, $\chi^2_{ST}$, and $\chi^2_{h\gamma\gamma}$ denote the relic density, the six neutrino oscillation , the electroweak precision, and Higgs diphoton branching ratio chi-square terms, respectively. The charged lepton masses appearing in cLFV processes were fixed to their PDG values. For each neutrino mass ordering, we identified the best-fit parameter vector as the sampled vector with the smallest value of $\chi^2_{\rm{tot}}$ among those satisfying all hard constraints. The relic density $\Omega h^2$ was treated with a symmetric Gaussian pull,
\begin{equation}
    \label{eq:chi2_omega}
    \chi^2_{\Omega}(\boldsymbol{\theta})=
    \left[
    \frac{(\Omega h^2)_{\mathrm{th}}(\boldsymbol{\theta})
    -(\Omega h^2)_{\mathrm{exp}}}
    {\sigma_{\Omega h^2}}
    \right]^2.
\end{equation}
For the oblique parameters, the corresponding contribution to the fit statistic is
\begin{equation}
    \chi^2_{ST} = \left(\Delta\mathbf{X}\right)^{\mathrm T}
    C_{ST}^{-1}
    \left(\Delta\mathbf{X}\right),
\end{equation}
where $\Delta\mathbf{X}=\bigl(\Delta S-\Delta S_0,
\Delta T-\Delta T_0\bigr)^{\mathrm T}$. Using the correlation coefficient $\rho_{ST}=0.93$ from the electroweak fit \cite{ParticleDataGroup:2024cfk}, the covariance matrix is
\begin{equation}
    C_{ST} =
    \begin{pmatrix}
        \sigma_S^2 &
        \rho_{ST}\sigma_S\sigma_T \\
        \rho_{ST}\sigma_S\sigma_T &
        \sigma_T^2
    \end{pmatrix}.
\end{equation}

We constructed the neutrino oscillation contribution using the NuFIT~6.0 best-fit values and quoted $3\sigma$ intervals~\cite{Esteban:2024eli}. The six observables, treated separately for NO and IO, are $\sin^2\theta_{12}$, $\sin^2\theta_{13}$, $\sin^2\theta_{23}$, $\delta_{\rm{CP}}$, $\Delta m_{21}^2$, and $\Delta m_{3\ell}^2$. Each one of them was modeled by an asymmetric Gaussian pull, with the lower and upper $1\sigma$ widths set to one third of the distances from the best-fit value to the corresponding $3\sigma$ limits. The periodicity of $\delta_{\rm{CP}}$ was acocunted for and included the Higgs diphoton branching ratio $R_{\gamma\gamma}$ through an additional asymmetric Gaussian term. We further imposed hard constraints on the cosmological neutrino mass sum, current and projected cLFV limits, direct detection, quartic and Yukawa perturbativity, the implemented BFB conditions, the mass gap requirements, and the requirement that $H_1^0$ be the lightest $Z_2$-odd state selected by \texttt{micrOMEGAs} as the DM candidate. Table~\ref{tab:scan_ranges} summarizes the sampled parameters and their broad input ranges. Mass parameters and positive scalar couplings were sampled logarithmically where appropriate; the rest of the scalar couplings were sampled linearly, and the initial real and imaginary parts of the complex Yukawa couplings were sampled logarithmically.
\begin{table}[htbp]
\centering
\renewcommand{\arraystretch}{1.15}
\begin{ruledtabular}
\begin{tabular}{lcc}
            \textbf{Parameters}
            & \textbf{Range} & \textbf{Sampling} \\
            \colrule
            $m_{H_{1}^{0}}$ & $[1,2000]~\rm{GeV}$ & logarithmic \\
            $\Delta m_{H_{2}^{0}},\Delta m_{\omega_{2}}$ &
            $[0.03,250]~\rm{GeV}$ & logarithmic \\
            $\Delta m_{H_{1}^{\pm}},\Delta m_{H_{2}^{\pm}}$
            & $[0.03,300]~\rm{GeV}$ & logarithmic \\
            $\Delta M_{\psi}$ & $[1,1200]~\rm{GeV}$ & logarithmic \\
            $M_{N}$ & $[0.1,2000]~\rm{GeV}$ & logarithmic \\
            $\theta_{N},\theta_{C}$ & $[0.03,\pi/2-0.03]$ & linear \\
            $\lambda_{2},\lambda_{3}$ & $[0.005,1.2]$ & logarithmic \\
            $\kappa_{1}$ & $[-0.4,0.8]$ & linear \\
            $\kappa_{2}$ & $[-0.4,0.4]$ & linear \\
            $\kappa_{4},\kappa_{5}$ & $[-0.2,0.8]$ & linear \\
            $g_{hH_{1}^{0}H_{1}^{0}}$ & $[-10,10]~\rm{GeV}$ & linear \\
            $\operatorname{Re}(y_{\eta}),~ \operatorname{Im}(y_{\eta})$ &
            $\pm[10^{-4},1]$ & logarithmic \\
            $\operatorname{Re}(Y_{N})_{\alpha},~\operatorname{Im}(Y_{N})_{\alpha}$ &
            $\pm[10^{-8},5\times10^{-2}]$ & logarithmic \\
            $\operatorname{Re}(y_{\phi})_{\alpha},~\operatorname{Im}(y_{\phi})_{\alpha}$ & $\pm[10^{-8},2\times10^{-1}]$ & logarithmic \\
\end{tabular}    
\end{ruledtabular}  
\caption{Initial parameter ranges and sampling prescriptions for the NO and IO scans. Here, $\alpha=e,\mu,\tau$, $\Delta m_X\equiv m_X-m_{H_1^0}$, $\Delta M_\psi\equiv M_\psi-m_{H_1^0}$, and $\pm[a,b]\equiv[-b,-a]\cup[a,b]$. The $Y_N$ and $y_\phi$ ranges are pre-fit; $m_{H_1^0}$ is sampled linearly in the Higgs-funnel region and logarithmically elsewhere.}
\label{tab:scan_ranges}
\end{table}
Table~\ref{tab:scan_constraints} summarizes the main experimental inputs and hard constraints used in the scan.
\begin{table}[htbp]
\begin{ruledtabular}
\begin{tabular}{lcc}
            Observable or constraint
            & Value or bound
            & Treatment
            \\
            NuFIT~6.0 observables
            & Best fits and $3\sigma$ ranges
            & Asymmetric Gaussian $+\;3\sigma$ cut~\cite{Esteban:2024eli}
            \\
            $\sum_i m_i$
            & $\leq0.12~\mathrm{eV}$
            & Hard cut~\cite{Planck:2018vyg}
            \\
            $\Omega h^2$
            & $0.120\pm0.001$
            & Gaussian $+\;3\sigma$ cut~\cite{Planck:2018vyg}
            \\
            $R_{\gamma\gamma}$
            & $1.088^{+0.095}_{-0.089}$
            & Asymmetric Gaussian $+\;1\sigma$ cut~\cite{ATLAS:2022vkf}
            \\
            $(\Delta S,\Delta T)$
            & $(-0.05\pm0.07,\;0.00\pm0.06)$
            & {\scriptsize Correlated Gaussian,
            $\rho_{ST}=0.93$, $\chi^2_{ST}\leq11.83$~\cite{ParticleDataGroup:2024cfk}}
            \\
            $m_e,\;m_\mu,\;m_\tau$
            & $(0.00051099895,\;0.1056583755,\;1.77693)\,\mathrm{GeV}$
            & Fixed inputs
              $(\chi^2_{\ell}=0)$~\cite{ParticleDataGroup:2024cfk}
            \\
            cLFV observables
            & Current and projected limits
            & Hard cuts
            \\
            Direct detection
            & $p_{\mathrm{DD}}\geq0.10$
            & Hard cut
            \\
            Quartic and Yukawa couplings
            & $|\lambda_i|<4\pi,\;|y_i|<\sqrt{4\pi}$
            & Hard cuts
            \\
            Scalar potential
            & Implemented BFB conditions
            & Hard cuts
            \\
            Derived scalar parameter
            & $m_\phi^2>0$
            & Hard cuts
            \\
            Odd-sector spectrum
            & Mass-gap and $H_1^0$ DM requirements
            & Hard cuts
            \\
\end{tabular}
\end{ruledtabular}
\caption{Main inputs and constraints used in the NO and IO scans.}
\label{tab:scan_constraints}    
\end{table}

\subsection{Results of the fits and benchmark points}
\label{subsec:benchmark-phenomenology}
We now present the $H_1^0$ scalar-DM fit results. Separate scans were performed considering both neutrino mass orderings, each with $10^4$ sampled parameter vectors. After applying the likelihood requirements and hard cuts, 822 points remained for NO and 830 for IO. We selected ten accepted scan points, summarized in Tables~\ref{tab:selected-benchmark-masses} and~\ref{tab:selected-benchmark-couplings} as collider benchmarks. Among them, C01 and C10 are the best-fit points with $\chi^2_{\rm{tot}}=4.45$ and $\chi^2_{\rm{tot}}=3.41$ for NO and IO, respectively. The remaining eight were chosen to span the range of scalar masses and charged scalar decay lengths from low to high mass regions. All these benchmarks lie within the $2\sigma$ electroweak precision and neutrino projections as displayed later in this section using scatter plots.
\begin{table}[htbp]
    \centering
    \begin{ruledtabular}
        \begin{tabular}{lcccccccccc}
            Benchmark & Ordering & $M_N$ & $m_{H_1^0}$ & $m_{\omega_2}$ & $m_{H_2^0}$ & $m_{H_1^\pm}$ & $m_{H_2^\pm}$ & $M_{\psi}$ & $\Delta_1^\pm$ & $\Delta_2^\pm$ \\
            \colrule
            C01 & NO & $0.273$ & $855.0$ & $855.1$ & $855.1$ & $855.2$ & $886.7$ & $885.8$ & $0.257$ & $31.7$ \\
            C02 & IO & $2.89$ & $847.3$ & $851.8$ & $853.4$ & $847.3$ & $847.5$ & $889.3$ & $0.0551$ & $0.179$ \\
            C03 & NO & $0.435$ & $1017.0$ & $1017.0$ & $1017.1$ & $1019.7$ & $1047.0$ & $1036.3$ & $2.73$ & $30.0$ \\
            C04 & IO & $0.410$ & $26.4$ & $97.4$ & $65.0$ & $40.2$ & $31.1$ & $50.9$ & $13.8$ & $4.63$ \\
            C05 & NO & $0.127$ & $50.8$ & $62.0$ & $57.7$ & $62.0$ & $98.2$ & $70.3$ & $11.2$ & $47.4$ \\
            C06 & NO & $3.02$ & $993.4$ & $995.0$ & $997.2$ & $993.4$ & $1010.5$ & $1064.7$ & $0.0462$ & $17.1$ \\
            C07 & IO & $0.811$ & $1292.2$ & $1292.3$ & $1294.8$ & $1292.3$ & $1304.8$ & $1566.9$ & $0.0564$ & $12.6$ \\
            C08 & NO & $52.5$ & $1164.2$ & $1183.1$ & $1165.0$ & $1193.9$ & $1194.5$ & $1462.9$ & $29.8$ & $30.4$ \\
            C09 & IO & $0.119$ & $1104.2$ & $1131.3$ & $1104.2$ & $1118.8$ & $1121.6$ & $1420.8$ & $14.6$ & $17.4$ \\
            C10 & IO & $2.622$ & $899.9$ & $904.3$ & $910.6$ & $926.9$ & $917.4$ & $1038.7$ & $27.0$ & $17.6$ \\
        \end{tabular}
    \end{ruledtabular}
    \caption{Mass spectra of the ten benchmark points used in the collider analysis. All masses and mass splittings are in $\mathrm{GeV}$.}
    \label{tab:selected-benchmark-masses}
\end{table}
\begin{table}[htbp]
    \centering
    \begin{ruledtabular}
        \begin{tabular}{lccccccccc}
            Benchmark & $\theta_N$ & $\theta_C$ & $\lvert y_{\eta}\rvert$ & $\lVert Y_N\rVert$ & $\lambda_2$ & $\kappa_1$ & $\kappa_2$ & $\kappa_4$ & $\kappa_5$ \\
            \colrule
            C01 & $0.789$ & $0.739$ & $0.0375$ & $1.86\times10^{-5}$ & $0.0461$ & $0.132$ & $-0.282$ & $0.240$ & $0.375$ \\
            C02 & $0.637$ & $1.413$ & $0.00162$ & $5.58\times10^{-5}$ & $0.427$ & $-0.244$ & $0.0456$ & $0.260$ & $0.276$ \\
            C03 & $0.518$ & $1.344$ & $0.000441$ & $9.00\times10^{-4}$ & $0.140$ & $0.451$ & $-0.524$ & $0.158$ & $0.319$ \\
            C04 & $0.037$ & $1.032$ & $0.00265$ & $6.66\times10^{-5}$ & $0.464$ & $-0.0967$ & $0.241$ & $0.088$ & $0.161$ \\
            C05 & $0.282$ & $0.698$ & $0.000163$ & $9.77\times10^{-5}$ & $0.601$ & $-0.0477$ & $0.0574$ & $0.103$ & $0.164$ \\
            C06 & $0.254$ & $0.731$ & $0.890$ & $8.06\times10^{-6}$ & $0.0894$ & $-0.109$ & $0.0959$ & $0.509$ & $0.0664$ \\
            C07 & $1.407$ & $0.552$ & $0.112$ & $1.42\times10^{-5}$ & $0.0424$ & $0.0874$ & $3.01$ & $-0.0795$ & $-0.0113$ \\
            C08 & $0.826$ & $0.246$ & $0.00781$ & $2.55\times10^{-4}$ & $0.808$ & $0.0526$ & $0.0942$ & $0.455$ & $0.619$ \\
            C09 & $0.639$ & $0.813$ & $0.0301$ & $1.99\times10^{-5}$ & $0.101$ & $0.994$ & $-0.224$ & $0.479$ & $-0.156$ \\
            C10 & $0.579$ & $0.632$ & $0.0149$ & $2.80\times10^{-5}$ & $0.594$ & $0.478$ & $-0.792$ & $0.391$ & $0.00523$ \\
        \end{tabular}
    \end{ruledtabular}
    \caption{Values of the mixing and coupling parameters for the ten benchmarks.}
    \label{tab:selected-benchmark-couplings}
\end{table}

 \paragraph{\bf DM, neutrino mass, $R_{\gamma\gamma}$, oblique parameters and cLFV:}
To examine how the different observables constrain the parameter space, we organize the results by observable class, showing the DM, electroweak precision, Higgs diphoton, neutrino oscillation, and cLFV observables separately for each mass ordering. 
\begin{figure}[tb]
\centering
\includegraphics[width=0.38\textwidth]{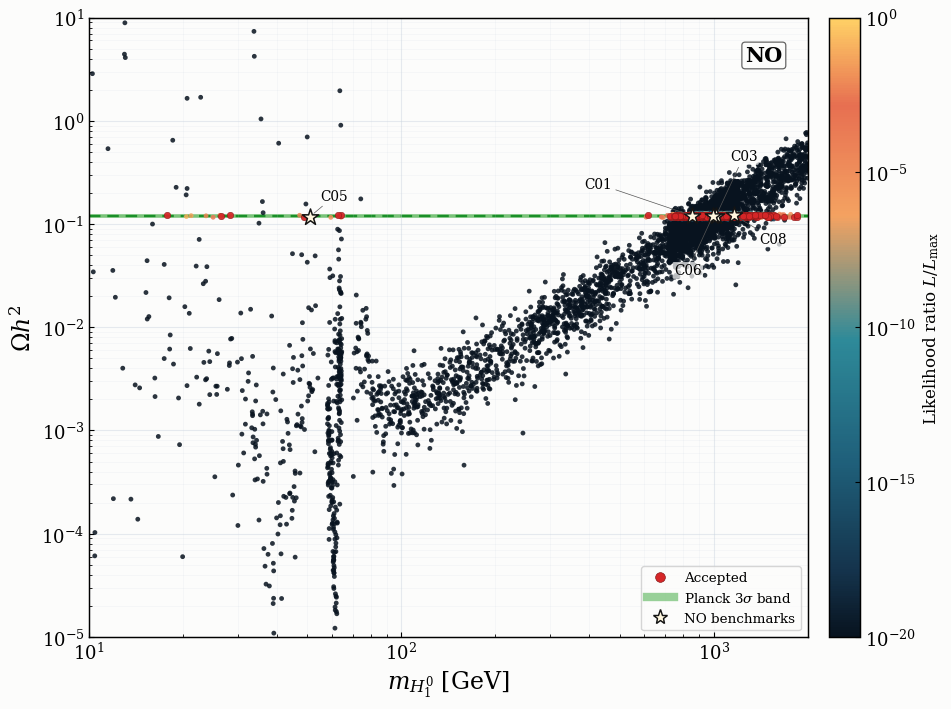}
\hspace{1cm}
\includegraphics[width=0.38\linewidth]{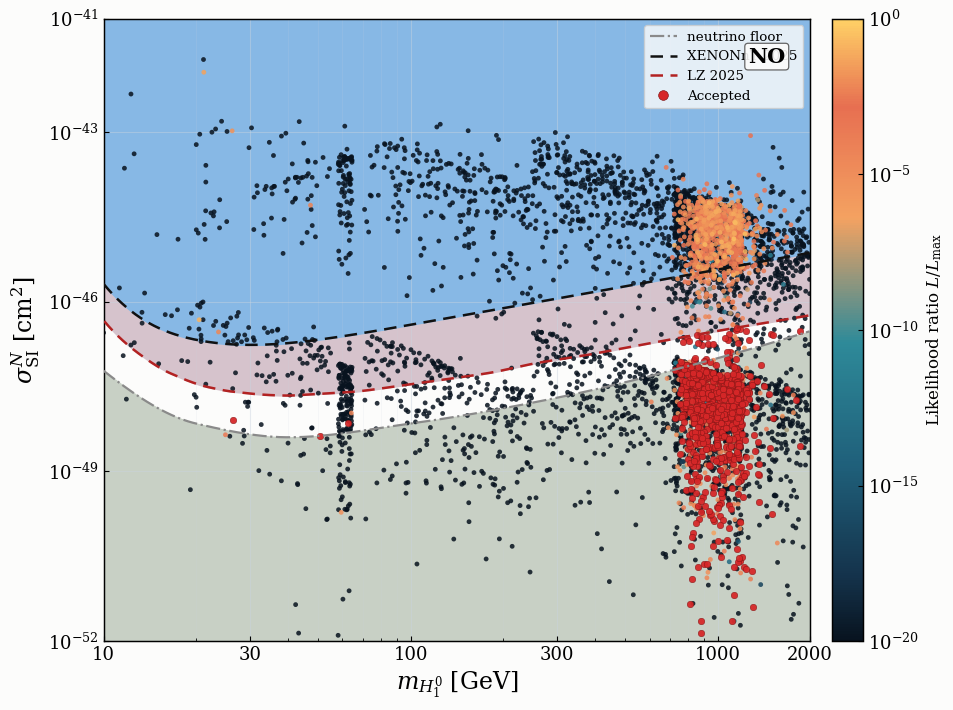}
\caption{Scalar DM in the NO case. Left panel: relic abundance as a function of $m_{H_1^0}$. Right panel: spin-independent cross section as a function of $m_{H_1^0}$. The Planck $3\sigma$ band, current LZ and XENONnT bounds, and the neutrino-floor reference are shown where relevant. The color scale gives the likelihood ratio relative to the best-fit point.}
\label{f2}
\end{figure}
Let us begin with the DM observables. In the left panels of Figs.~\ref{f2} and \ref{f6}, we plot the relic abundance as a function of the DM mass for NO and IO, respectively. The accepted $H_1^0$ masses, which lie within the imposed Planck $3\sigma$ band~\cite{Planck:2018vyg}, span from $6.46~\mathrm{GeV}$ to $\sim1.85~\mathrm{TeV}$. Most points lie above $600~\mathrm{GeV}$: 814 for NO and 823 for IO. The other accepted points lie below $80~\mathrm{GeV}$. Of these, two NO and three IO points lie within $2~\mathrm{GeV}$ of the Higgs-funnel condition $m_{H_1^0}=m_h/2$, while no accepted points occur between $80$ and $600~\mathrm{GeV}$. For every accepted point, at least one other $Z_2$-odd scalar lies within $10~\mathrm{GeV}$ of $H_1^0$. Such small mass splittings reduce the Boltzmann suppression at freeze-out, allowing coannihilation channels to contribute to the effective annihilation rate.
\begin{figure}[tb]
\centering
\includegraphics[width=0.30\textwidth]{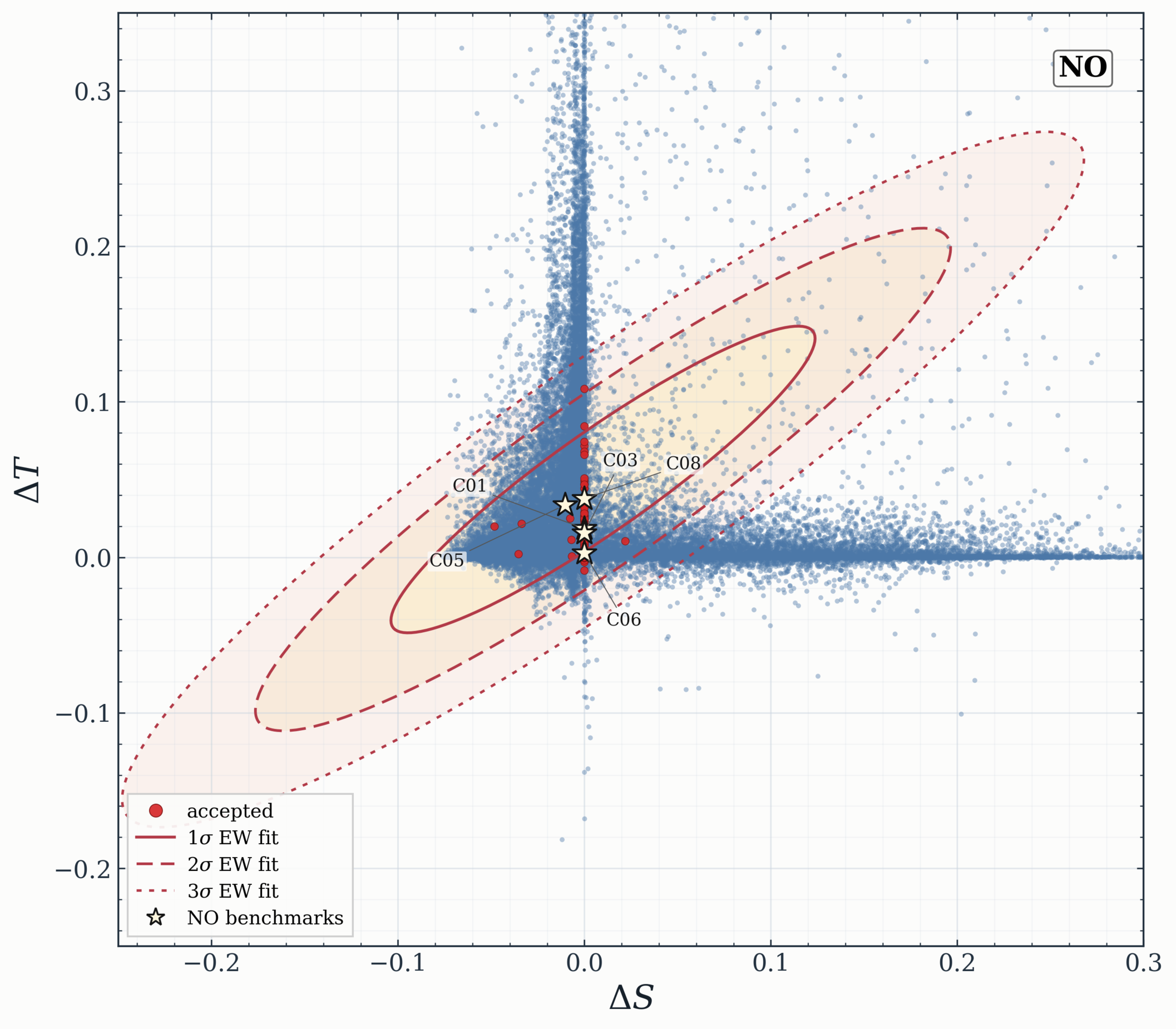}
\includegraphics[width=0.34\linewidth]{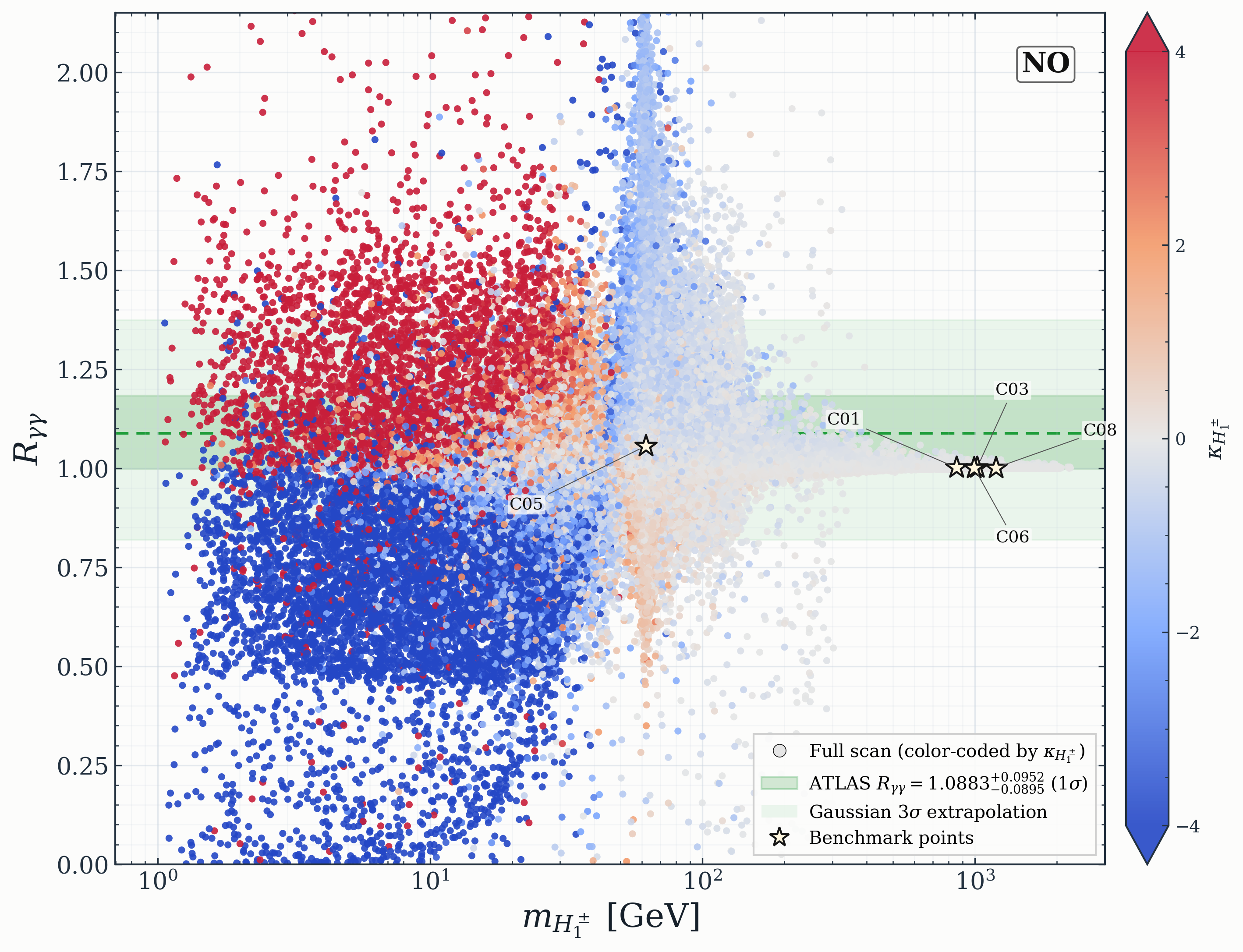}
\includegraphics[width=0.34\linewidth]{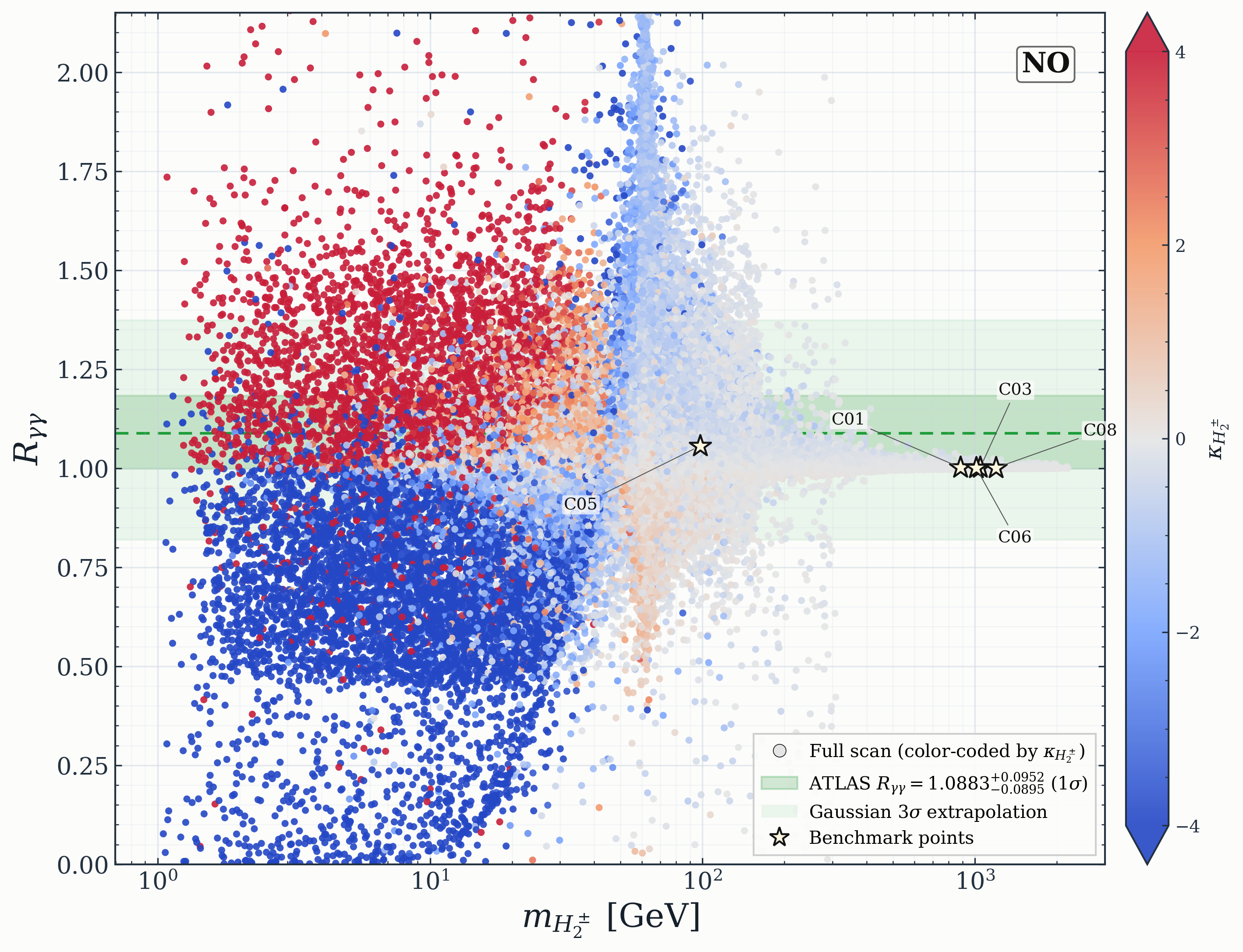}
\caption{Electroweak and Higgs diphoton constraints for scalar DM with NO. Left panel: allowed points in the $(\Delta S,\Delta T)$ plane with the electroweak-fit confidence ellipses. Middle and right panels: $R_{\gamma\gamma}$ as functions of $m_{H_1^\pm}$ and $m_{H_2^\pm}$, respectively. The color scales indicate $\kappa_{H_1^\pm}$ and $\kappa_{H_2^\pm}$. The ATLAS $1\sigma$ band and its Gaussian $3\sigma$ extrapolation are shown, while stars mark the NO benchmark points.}
\label{f3}
\end{figure}
\begin{figure}[tb]
\centering
\includegraphics[width=0.32\linewidth]{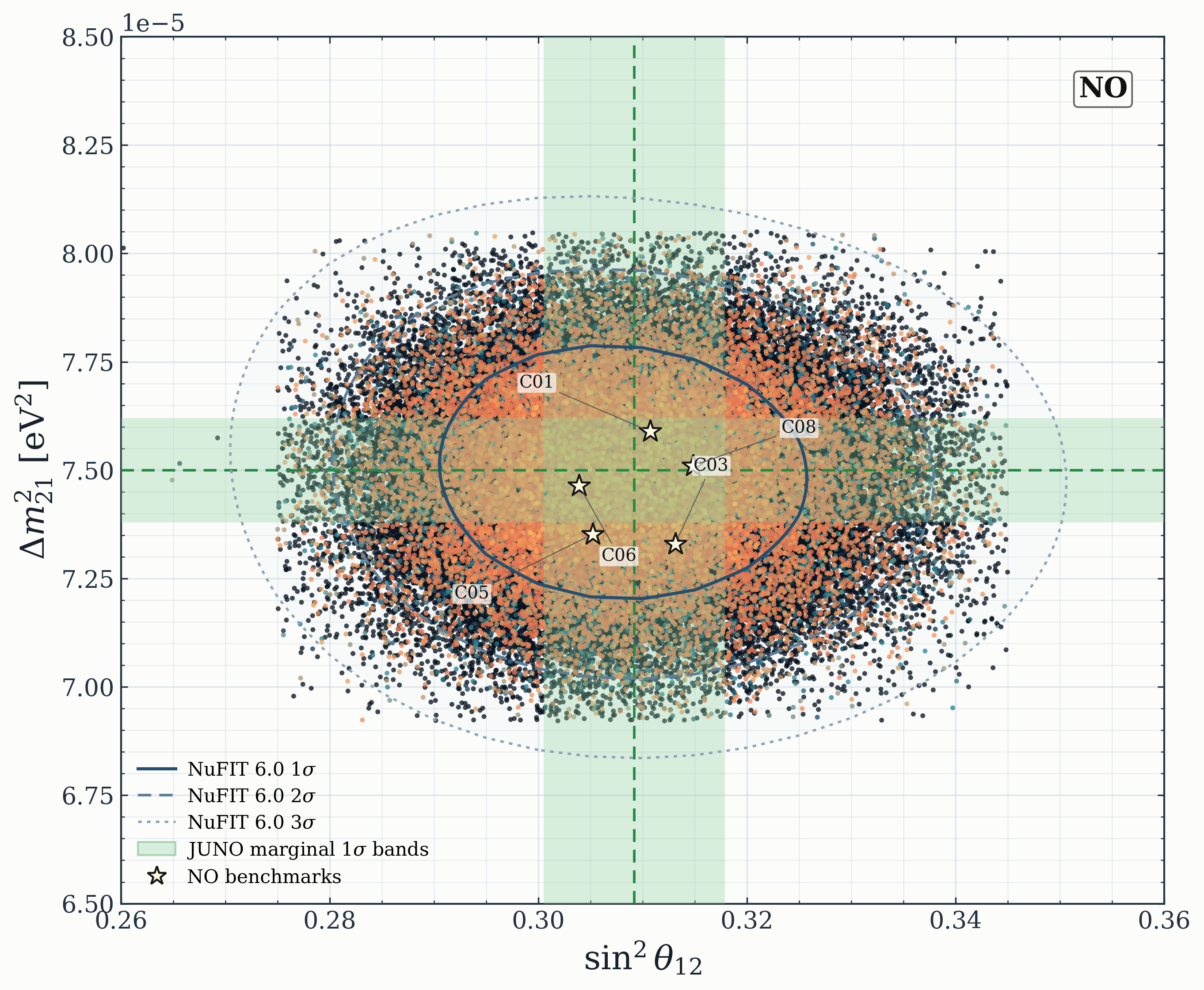}
\hfill
\includegraphics[width=0.32\linewidth]{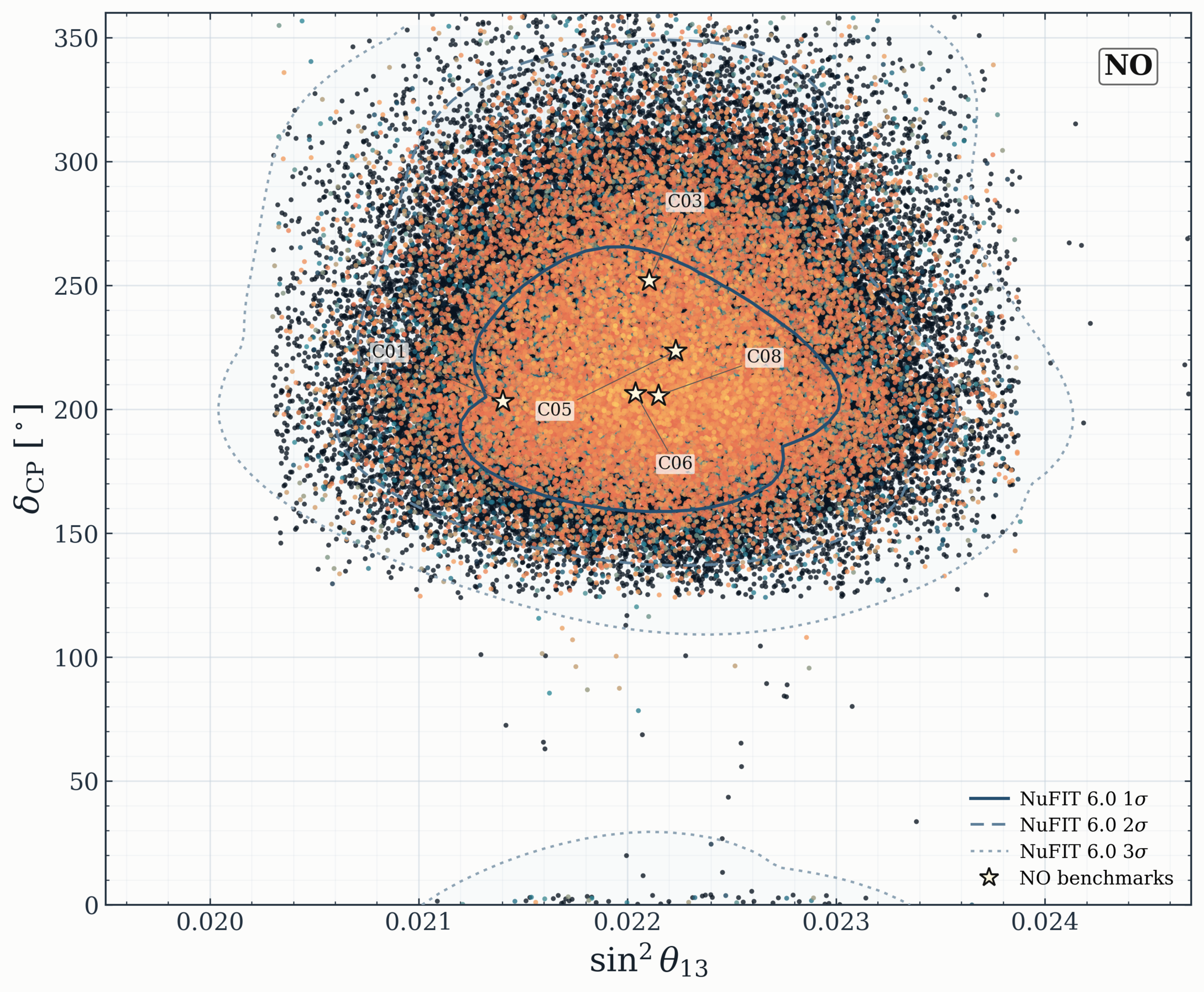}
\hfill
\includegraphics[width=0.32\linewidth]{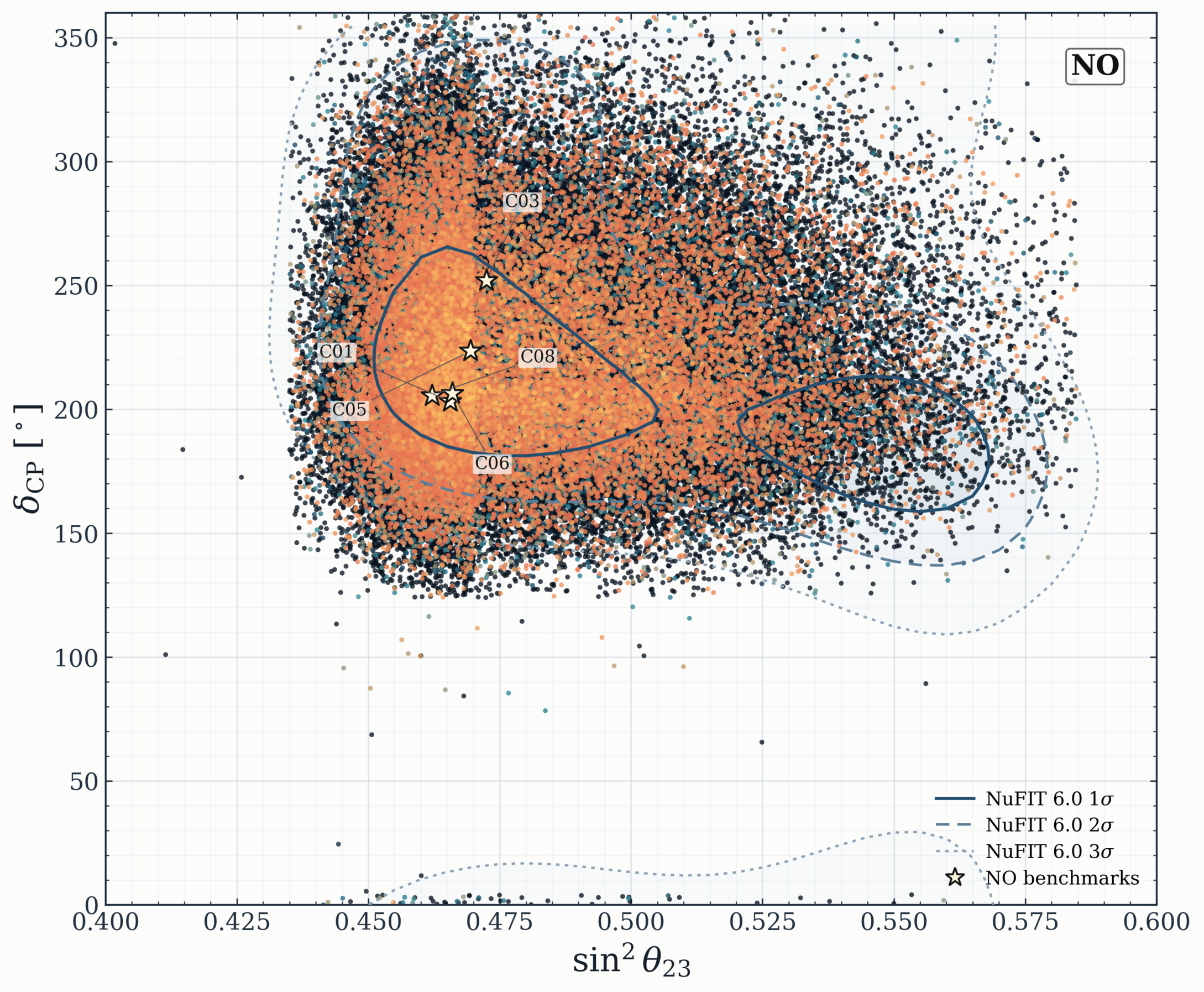}
\caption{Neutrino oscillation predictions for NO. Left panel: solar projection in the $(\Delta m_{21}^{2},\sin^{2}\theta_{12})$ plane. Middle and right panels: $\delta_{\mathrm{CP}}$ projections with $\sin^{2}\theta_{13}$ and $\sin^{2}\theta_{23}$, respectively. The NuFIT 6.0 preferred regions are shown in all panels, while the JUNO $1\sigma$ bands are included in the left panel.}
\label{f4}
\end{figure}

The spin-independent cross section as a function of $m_{H_1^0}$ is displayed in the right panels of Figs.~\ref{f2} and~\ref{f6} for NO and IO, respectively. The experimental curves indicate representative upper limits from current xenon-based direct-detection searches; points above a given curve are excluded by that bound, while points below it remain allowed by that particular limit~\cite{XENON:2023cxc,LZ:2024zvo}. 
\begin{figure}[tb]
\centering
\includegraphics[width=0.45\linewidth]{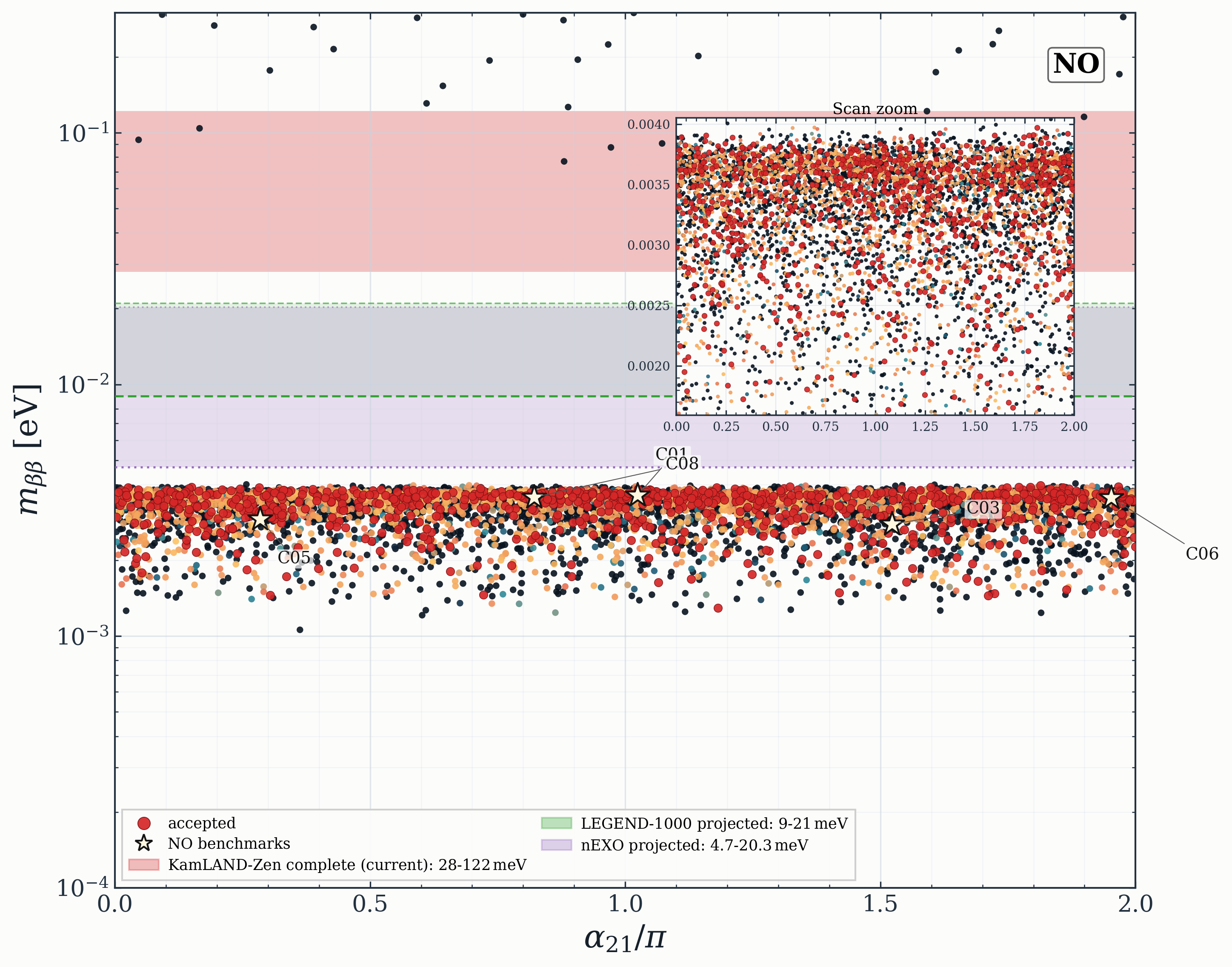}
\hfill
\includegraphics[width=0.45\linewidth]{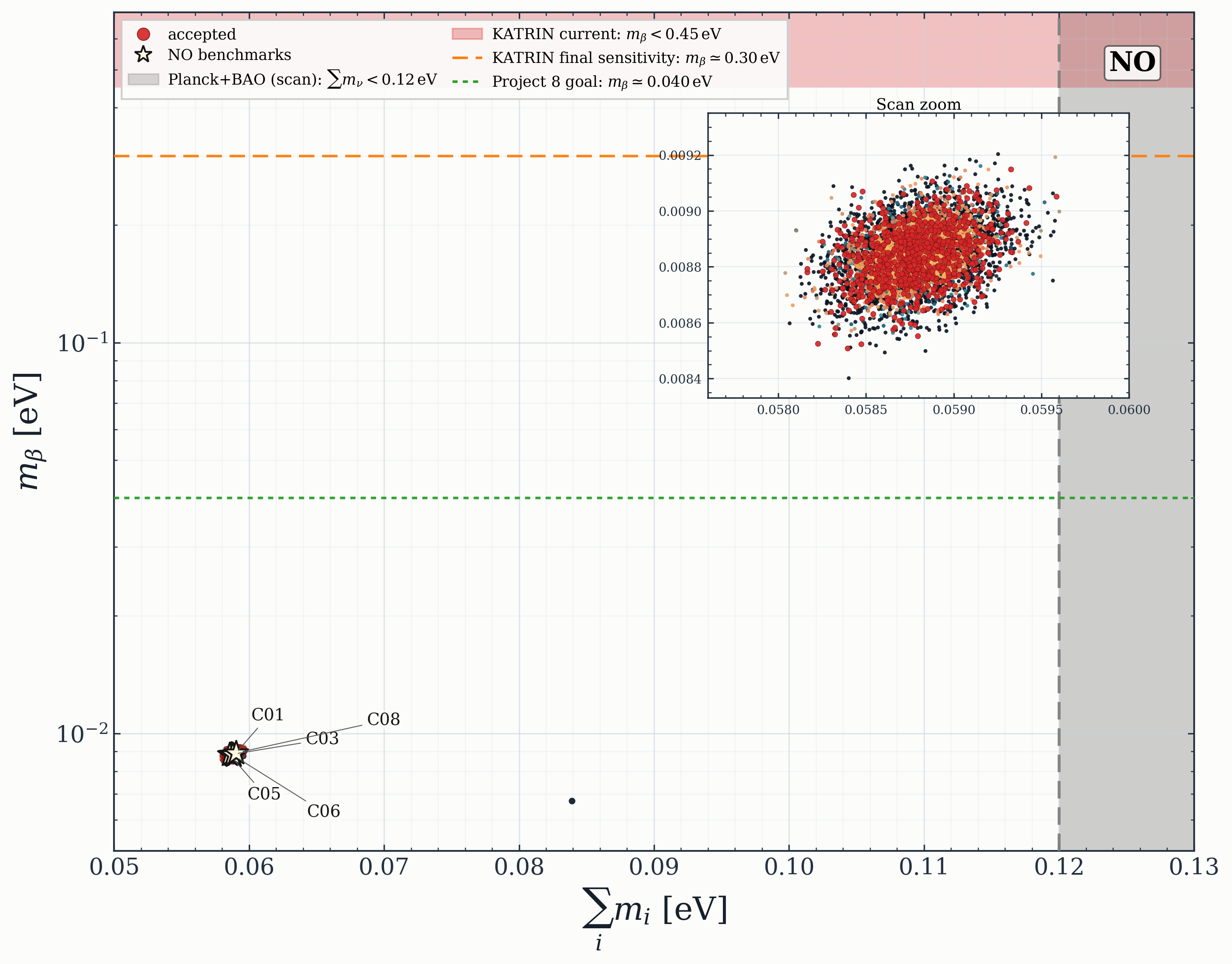}
\caption{Absolute neutrino mass predictions for NO. Left panel: $m_{\beta\beta}$ as a function of the Majorana phase $\alpha_{21}$. Right panel: $m_{\beta}$ as a function of $\sum_i m_i$. Current experimental limits and projected sensitivities are shown where applicable.}
\label{f4_mass}
\end{figure}
The accepted points are consistent with the current direct-detection constraints and span several orders of magnitude in $\sigma_{\rm SI}^{N}$, including values below the neutrino-floor reference~\cite{OHare:2021utq}. Their concentration near $1~\mathrm{TeV}$ is consistent with the mass scale reported by the latest LUX-ZEPLIN (LZ) analysis~\cite{LZ:2026axp}.
\begin{figure}[tb]
\centering
\includegraphics[width=0.32\textwidth]{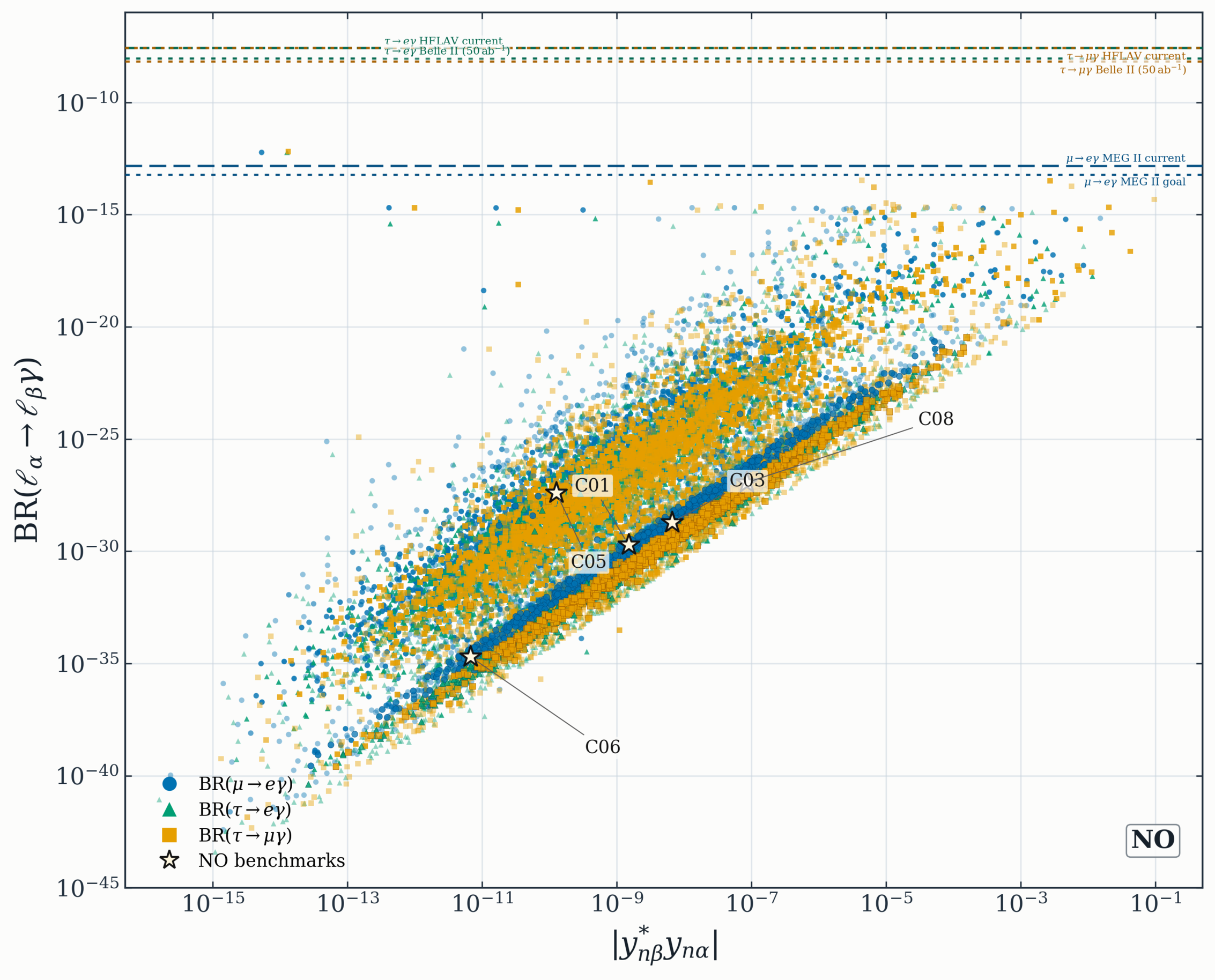}
\includegraphics[width=0.32\linewidth]{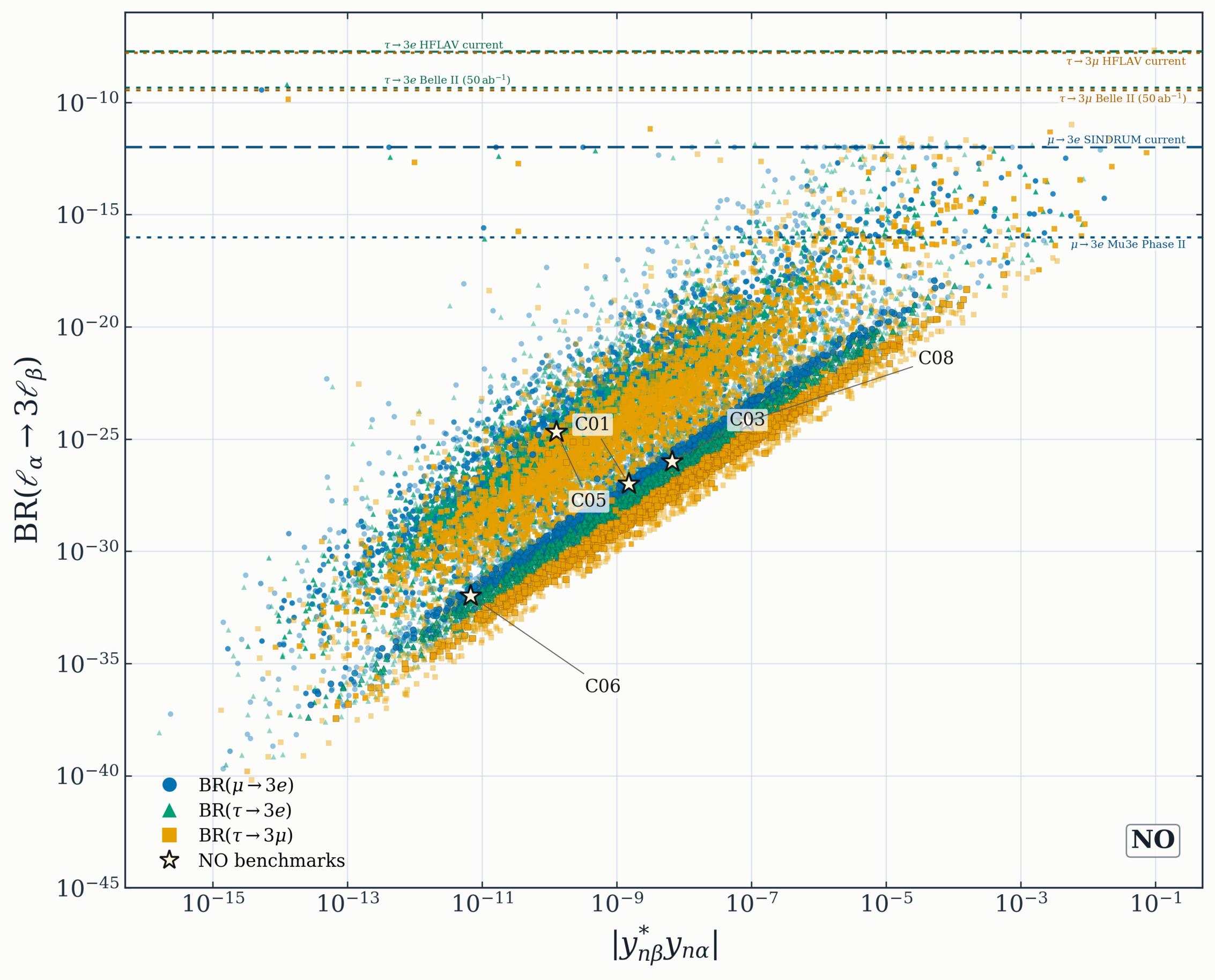}
\includegraphics[width=0.32\linewidth]{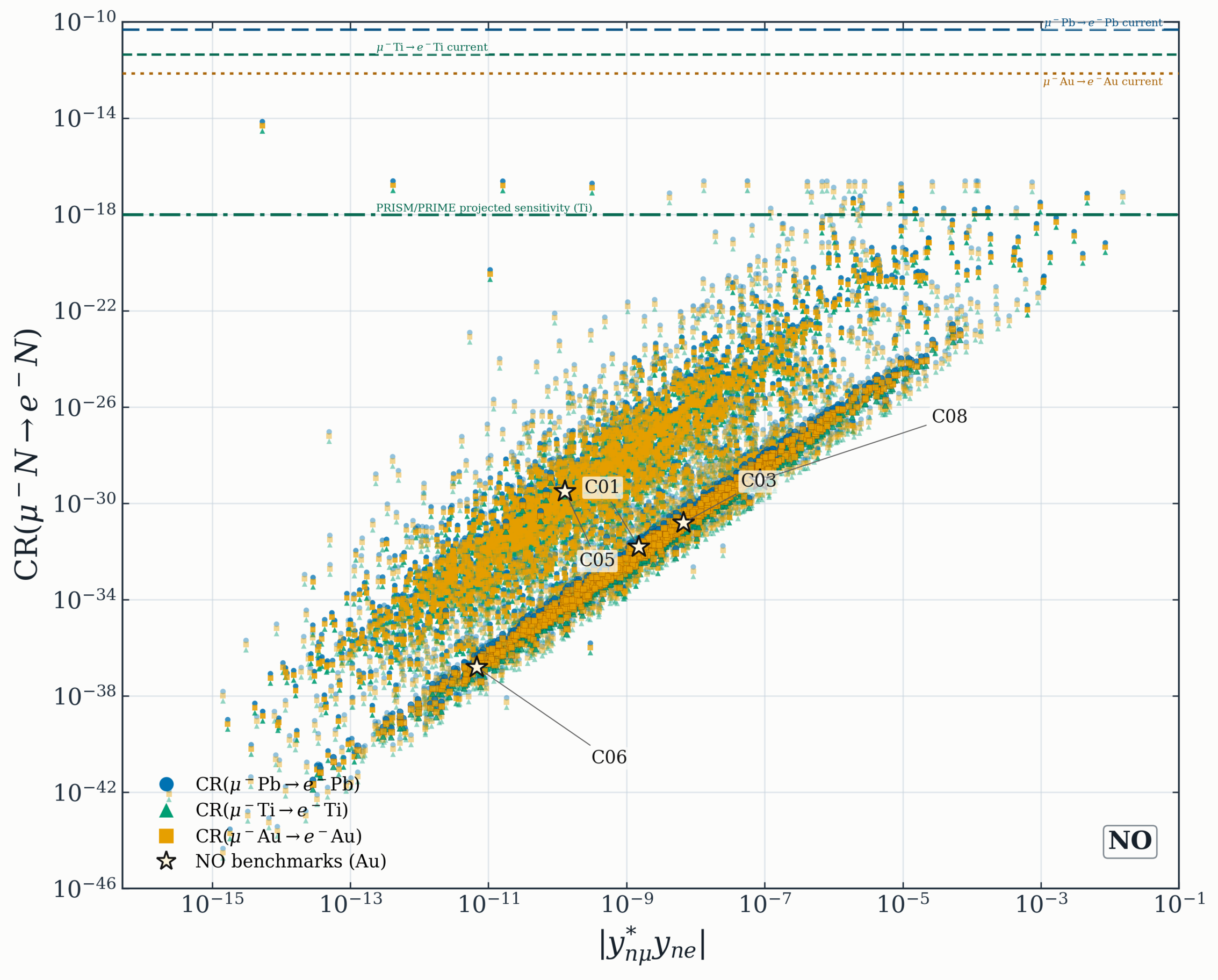}
\caption{cLFV predictions for scalar DM with NO. Left panel: radiative decays $\ell_{\alpha}\to\ell_{\beta}\gamma$. Middle panel: three-body decays $\ell_{\alpha}\to3\ell_{\beta}$. Right panel: coherent $\mu-e$ conversion in nuclei. Present limits and projected sensitivities are included.}
\label{f5}
\end{figure}

\begin{figure}[tb]
\centering
\includegraphics[width=0.30\textwidth]{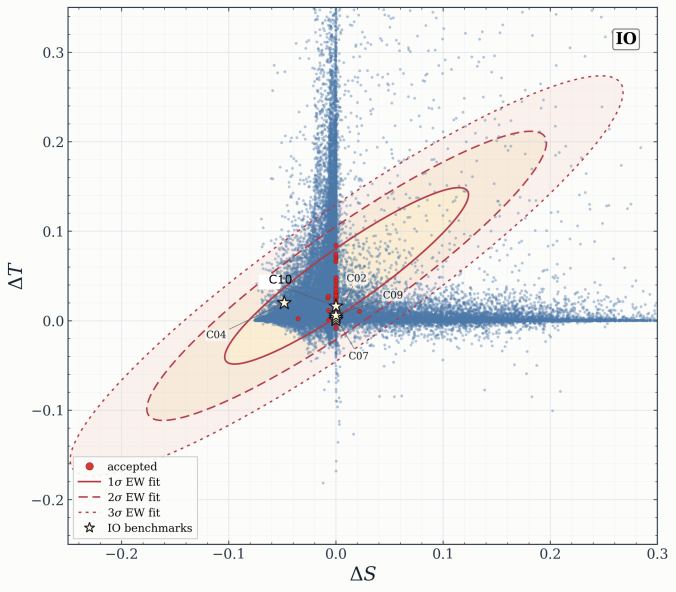}
\includegraphics[width=0.34\linewidth]{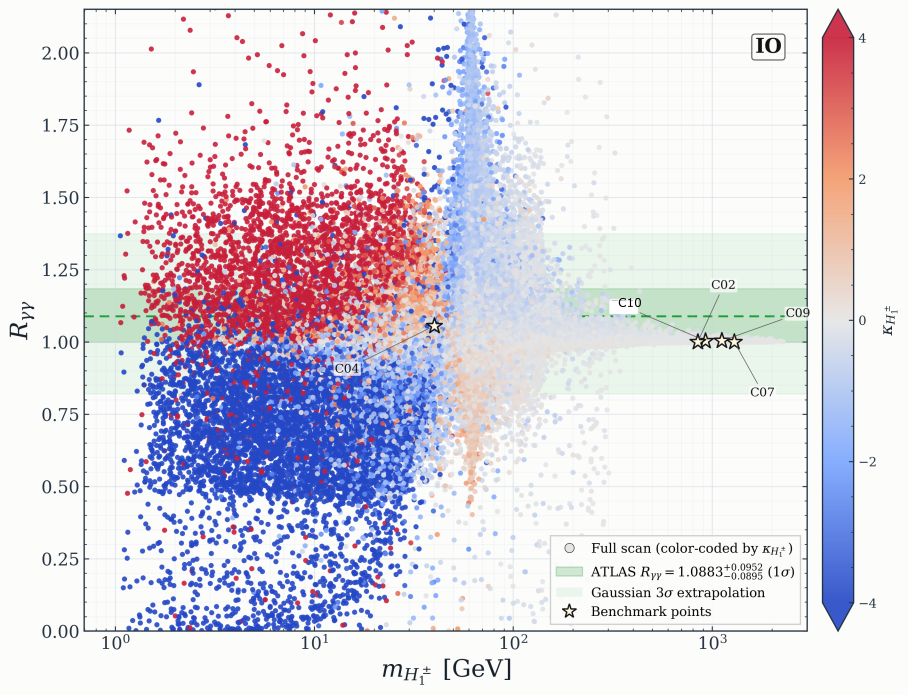}
\includegraphics[width=0.34\linewidth]{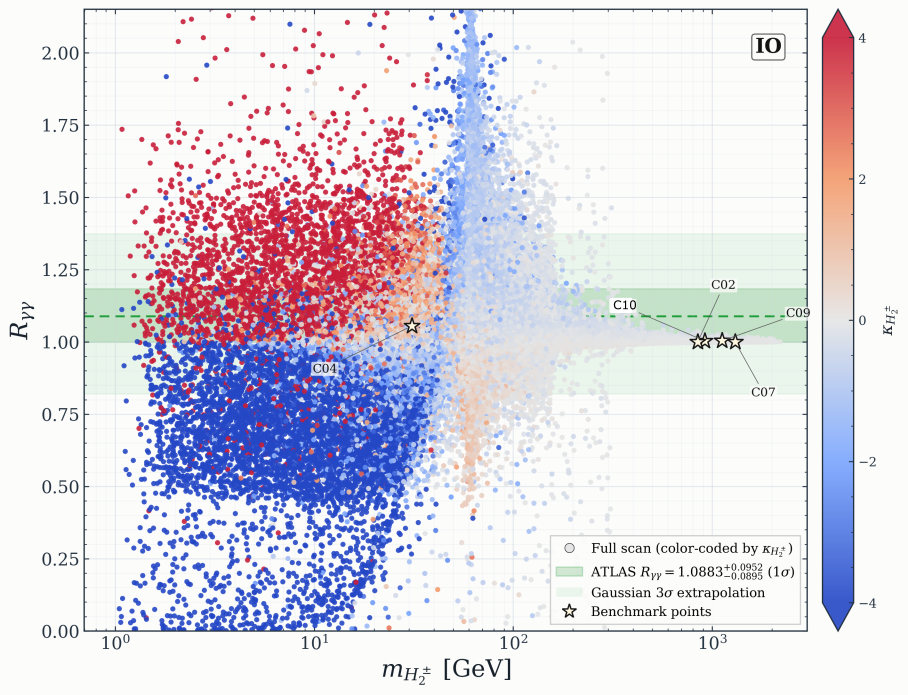}
\caption{Same as Fig.~\ref{f3} but for IO.}
\label{f7}
\end{figure}

\begin{figure}[tb]
\centering
\includegraphics[width=0.32\linewidth]{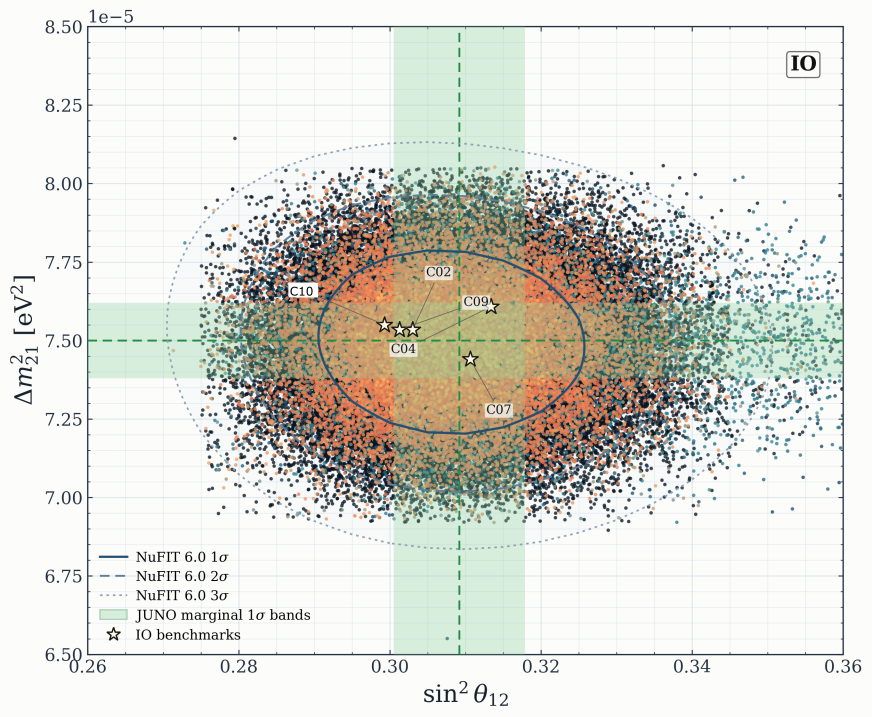}
\hfill
\includegraphics[width=0.32\linewidth]{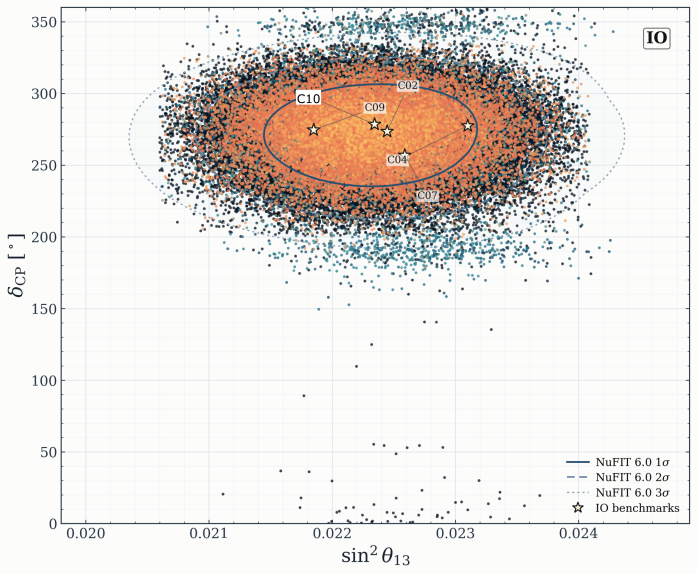}
\hfill
\includegraphics[width=0.32\linewidth]{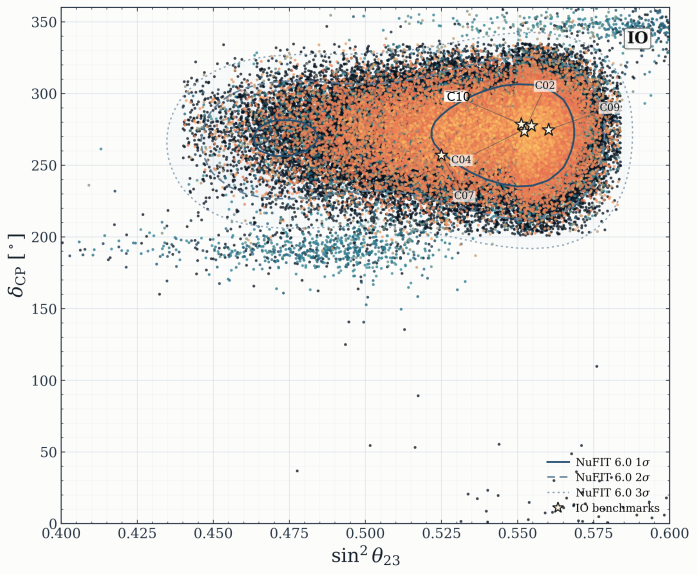}
\caption{Same as Fig.~\ref{f4}, but for IO.}
\label{f8}
\end{figure}

\begin{figure}[tb]
\centering
\includegraphics[width=0.45\linewidth]{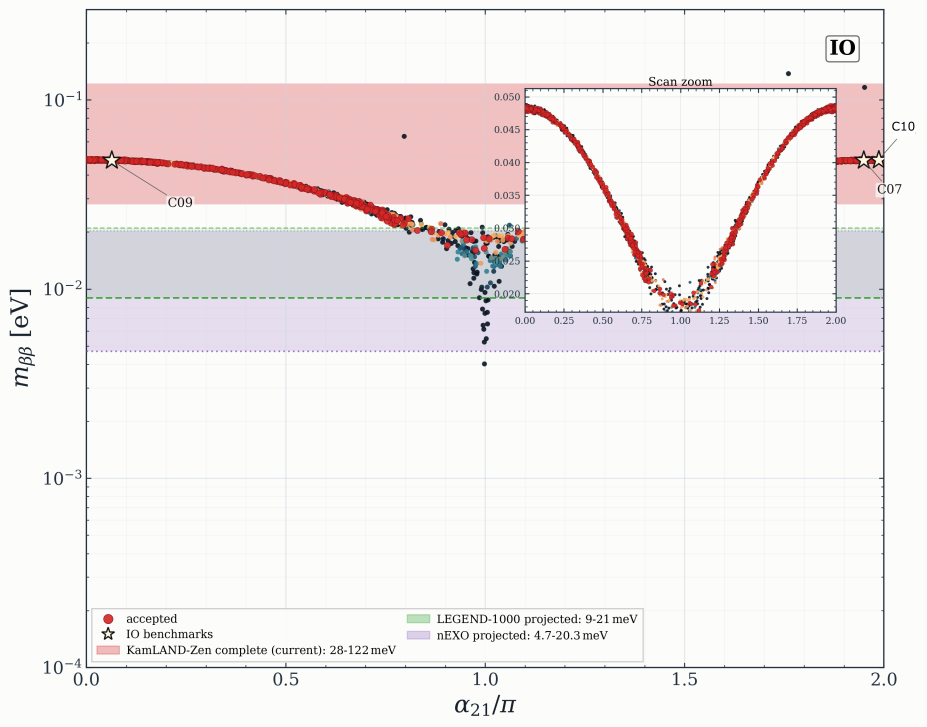}
\hfill
\includegraphics[width=0.45\linewidth]{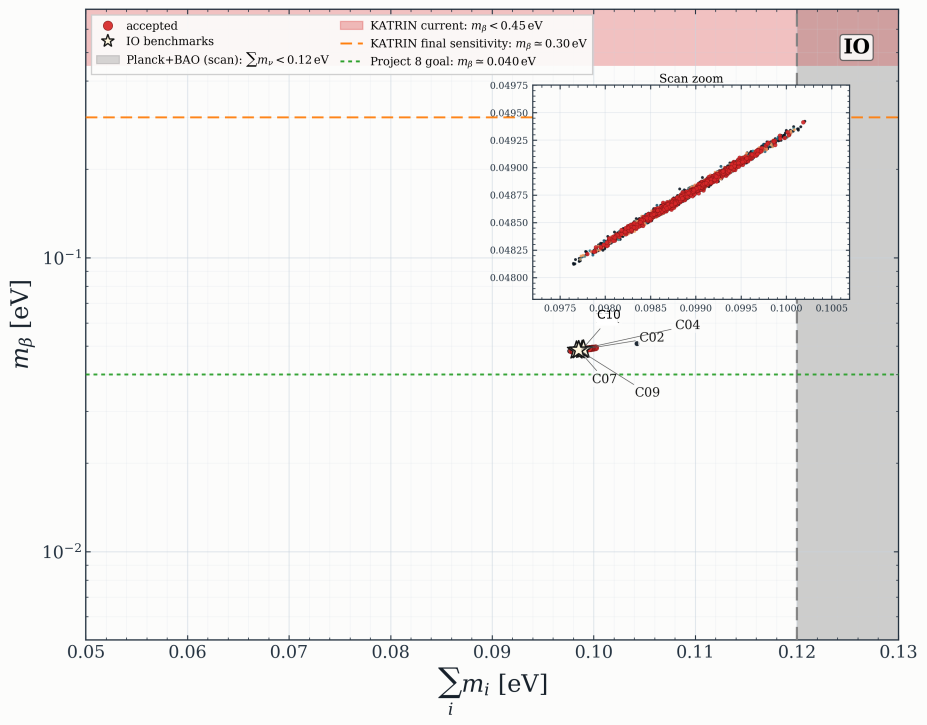}
\caption{Same as Fig.~\ref{f4_mass}, but for IO.}
\label{f8_mass}
\end{figure}

\begin{figure}[tb]
\centering
\includegraphics[width=0.32\textwidth]{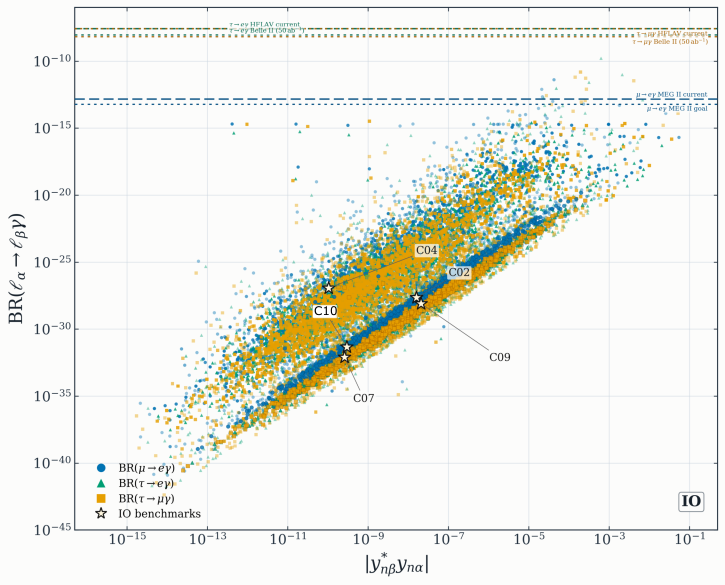}
\includegraphics[width=0.32\linewidth]{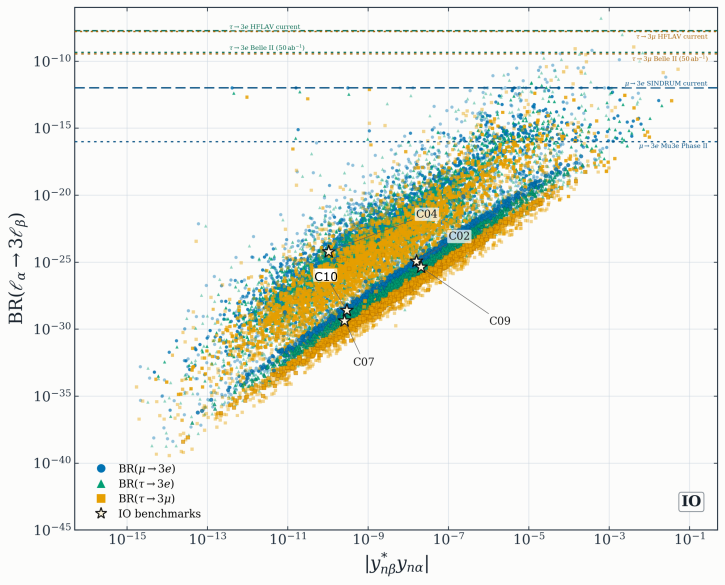}
\includegraphics[width=0.32\linewidth]{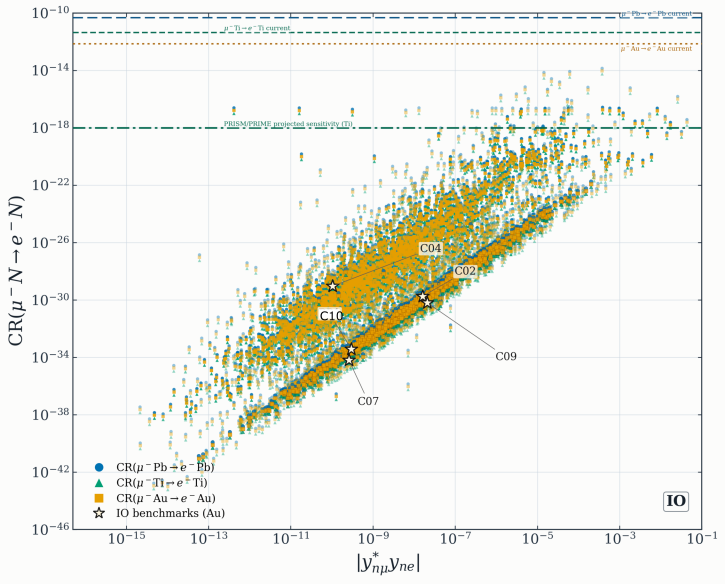}
\caption{Same as Fig.~\ref{f5} but for IO.}
\label{f9}
\end{figure}

The oblique parameters and Higgs diphoton branching ratio results for NO and IO are shown in Figs.~\ref{f3} and~\ref{f7}, respectively. In both orderings, the accepted points cluster around the region favored by the electroweak data in the ($\Delta S,\Delta T$) plane, while the diphoton panels show that the effect of the charged scalars becomes smaller as $m_{H_i^\pm}$ increases, causing $R_{\gamma\gamma}$ to approach its SM value. At lower masses, the sign and size of the deviation in $R_{\gamma\gamma}$ depend on $\kappa_{H_i^\pm}$.
\begin{figure}[tb]
\centering
\includegraphics[width=0.38\textwidth]{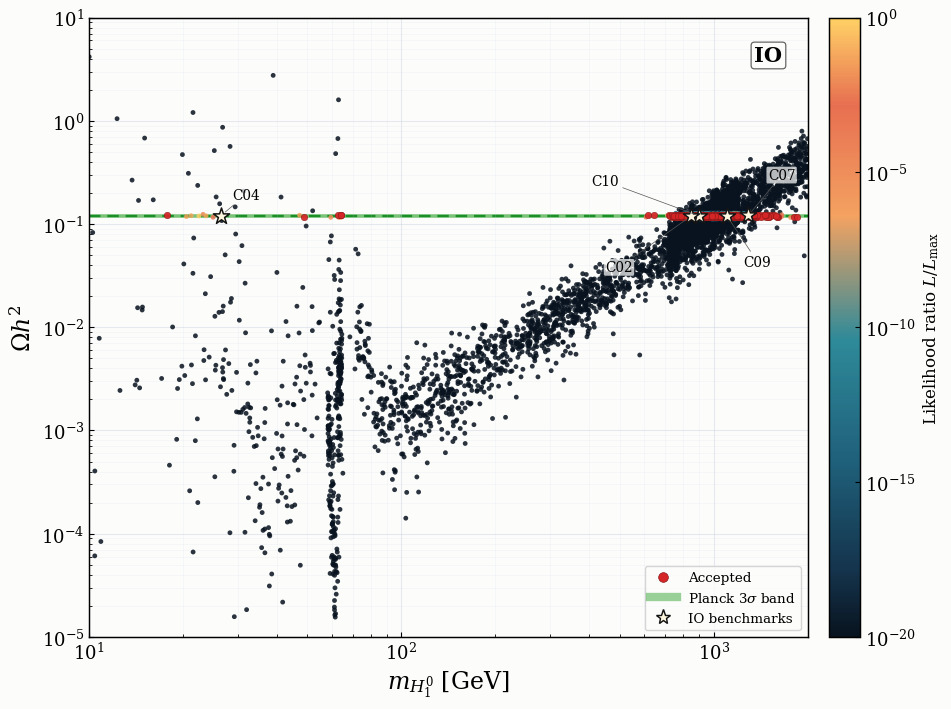}
\hspace{1cm}
\includegraphics[width=0.38\linewidth]{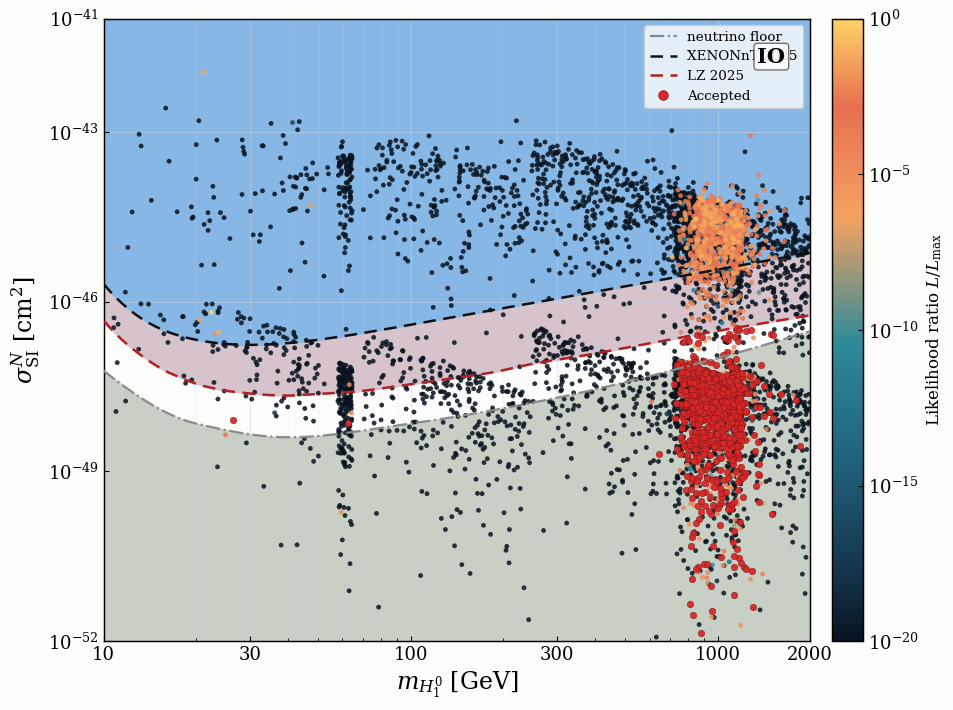}
\caption{Same as Fig.~\ref{f2} but for IO.}
\label{f6}
\end{figure}
Turning to the neutrino sector, Figs.~\ref{f4} and~\ref{f8} show that the oscillation observables are consistent with NuFIT 6.0's preferred regions for both orderings~\cite{Esteban:2024eli}. The solar panels further compare the $(\Delta m_{21}^2,\sin^2\theta_{12})$ plane with JUNO’s first reactor-antineutrino oscillation results~\cite{JUNO:2025gmd}. All benchmark points are compatible with the JUNO measurement at the $2\sigma$ level. Next, we have the mass observables in Figs.~\ref{f4_mass} and~\ref{f8_mass} for NO and IO, respectively. The left plots show the effective Majorana mass $m_{\beta\beta}$ as a function of the Majorana phase $\alpha_{21}$, while the right plots show the effective electron neutrino mass $m_\beta$ as a function of the sum of the three active neutrino masses $\sum_i m_i$. The results exhibit an ordering dependence: NO predicts smaller values of $\sum_i m_i$, $m_\beta$, and $m_{\beta\beta}$, while the larger IO values lie closer to the reach of KATRIN~\cite{KATRIN:2024cdt}, and Project~8~\cite{Project8:2022wqh} in direct neutrino mass searches and of KamLAND-Zen~\cite{KamLAND-Zen:2024eml}, LEGEND-1000~\cite{LEGEND:2021bnm}, and nEXO~\cite{nEXO:2021ujk} in neutrinoless double-beta decay searches. Finally, the cLFV results in Figs.~\ref{f5} and~\ref{f9} for NO and IO, respectively, show how radiative and three-body charged-lepton decays, together with coherent $\mu$--$e$ conversion in nuclei, constrain the Yukawa couplings. The figures include the adopted limits from MEG~II, HFLAV, SINDRUM, and SINDRUM~II, together with the projected sensitivities of MEG~II, Belle~II, Mu3e, and PRISM/PRIME~\cite{HeavyFlavorAveragingGroupHFLAV:2024ctg,SINDRUM:1987nra,MEGII:2018kmf,Belle-II:2018jsg,Mu3e:2020gyw,COMET:2025sdw}. Most model points and all the 10 benchmarks lie below the current experimental bounds and projected sensitivity thresholds imposed in the scan.
\newline


\paragraph{\bf Collider phenomenology:} To examine the collider phenomenology in detail, we focus on benchmark point C10. The point has the lowest $\chi^2_{\rm{tot}}$ and its predictions for $m_\beta$ and $m_{\beta\beta}$ lie within the projected sensitivities shown in Fig.~\ref{f8_mass}. At this point, the DM mass is $m_{H_1^0}=899.887~\mathrm{GeV}$. Since the charged scalar and neutral scalar mass splittings lie below $m_W$ and $m_Z$, respectively, the scalar transitions proceed through off-shell $W$ or $Z$ bosons, limiting the energy available to the visible decay products. By contrast, $M_\psi=1038.746~\mathrm{GeV}$ and $M_N=2.622~\mathrm{GeV}$ provide sufficient phase space for the two-body decays $\psi^\pm\to N H_i^\pm$ and $\psi^0\to N H_j^0$. Table~\ref{tab:bp1_widths_ctau} summarizes the total widths and proper decay lengths of the unstable scalars and fermionic triplet states, while Table~\ref{tab:charged_BRs_BP1} lists the dominant charged state branching fractions. We obtain the proper decay lengths from $c\tau_X=\hbar c/\Gamma_X$.
\begin{table}[htbp]
    \centering
    \begin{tabular}{c c c}
        \hline
        State & $\Gamma$ [GeV] & $c\tau$ [m] \\
        \hline
        $H_1^\pm$ & $3.904967\times 10^{-5}$ & $5.053230\times 10^{-12}$ \\
        $H_2^\pm$ & $5.573501\times 10^{-7}$ & $3.540449\times 10^{-10}$ \\
        $\psi^\pm$ & $1.049519\times 10^{-4}$ & $1.880166\times 10^{-12}$ \\
        $\psi^0$ & $2.519088\times 10^{-4}$ & $7.833271\times 10^{-13}$ \\
        $H_2^0$ & $2.462710\times 10^{-9}$ & $8.012595\times 10^{-8}$ \\
        $\omega_2$ & $9.269651\times 10^{-10}$ & $2.128742\times 10^{-7}$ \\
        \hline
    \end{tabular}
    \caption{Total widths, lifetimes, and proper decay lengths of the states entering the C10 decay chains. The scalar DM state $H_1^0$ is stable by the imposed $Z_2$ symmetry and is not listed.}
    \label{tab:bp1_widths_ctau}
\end{table}
\begin{table*}[htbp]
    \centering
    \small
    \setlength{\tabcolsep}{4pt}
    \begin{tabular}{c c c c c c}
        \hline
        \multicolumn{2}{c}{$H_1^+$}
        & \multicolumn{2}{c}{$H_2^+$}
        & \multicolumn{2}{c}{$\psi^+$} \\
        Decay channel & BR & Decay channel & BR & Decay channel & BR \\
        \hline
        $H_1^0 u\bar d$ & $2.722943\times10^{-1}$
        & $H_1^0 u\bar d$ & $1.898986\times10^{-1}$
        & $NH_2^+$ & $6.847421\times10^{-1}$ \\

        $H_1^0 c\bar s$ & $2.685042\times10^{-1}$
        & $H_1^0 c\bar s$ & $1.843007\times10^{-1}$
        & $NH_1^+$ & $3.151633\times10^{-1}$ \\

        $H_1^0\mu^+\nu_\mu$ & $9.660262\times10^{-2}$
        & $\omega_2 u\bar d$ & $1.015251\times10^{-1}$
        & Others & $9.460000\times10^{-5}$ \\

        $H_1^0 e^+\nu_e$ & $9.630812\times10^{-2}$
        & $\omega_2 c\bar s$ & $9.847850\times10^{-2}$
        & & \\

        $H_1^0\tau^+\nu_\tau$ & $9.383435\times10^{-2}$
        & $H_1^0\mu^+\nu_\mu$ & $6.671570\times10^{-2}$
        & & \\

        $\omega_2 u\bar d$ & $4.191841\times10^{-2}$
        & $H_1^0 e^+\nu_e$ & $6.670135\times10^{-2}$
        & & \\

        $\omega_2 c\bar s$ & $4.060726\times10^{-2}$
        & $H_1^0 \tau^+\nu_\tau$ & $6.312909\times10^{-2}$
        & & \\

        Others & $8.993074\times10^{-2}$
        & Others & $2.292510\times10^{-1}$
        & & \\
        \hline
    \end{tabular}
    \caption{Dominant branching fractions of the charged states in the C10 cascade.}
    \label{tab:charged_BRs_BP1}
\end{table*}
The charged scalars decay exclusively through the three-body channels $H_i^\pm\to S^0W^{\pm *}\to S^0f\bar f'$, with $S^0=H_1^0,H_2^0,\omega_2$. Therefore, the charged scalar entries in Table~\ref{tab:charged_BRs_BP1} list the fermion-pair final states produced through the virtual $W$ boson. For $H_1^+$, the largest modes are $H_1^0u\bar d$ and $H_1^0c\bar s$, followed by $H_1^0\ell^+\nu_\ell$. The $H_2^\pm$ state has a smaller total width because of the reduced available phase space, with dominant modes involving $H_1^0$, $\omega_2$, or $H_2^0$ and a fermion pair. These differences in decay modes and available phase space lead to a range of proper decay lengths. The longest proper decay length among the unstable states is $c\tau(\omega_{2})\simeq2.1\times10^{-7}~\mathrm{m}$, well below the millimeter scale. 
\begin{figure}[htbp]
    \centering
    \includegraphics[width=0.65\textwidth]
    {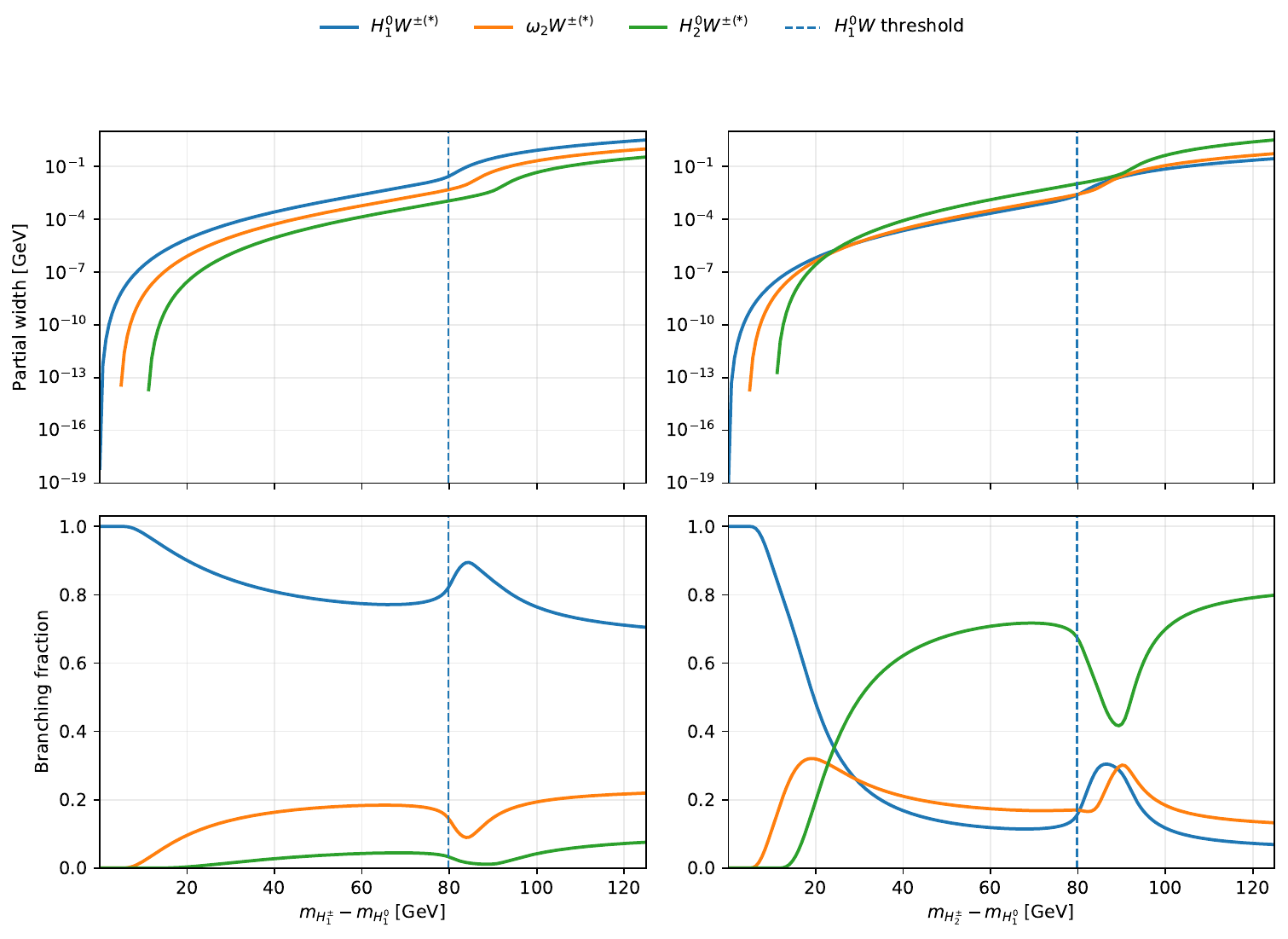}
    \caption{Charged-scalar partial widths (upper panels) and branching fractions (lower panels) as functions of the charged-neutral mass gap for $H_1^\pm$ (left) and $H_2^\pm$ (right), with all other parameters fixed to C10. The dashed line marks the $H_1^0W^\pm$ on-shell threshold.}
    \label{fig:bp1-charged-scalar-gap-scan}
\end{figure}

\begin{figure*}[t]
    \centering
    \includegraphics[width=0.36\textwidth]{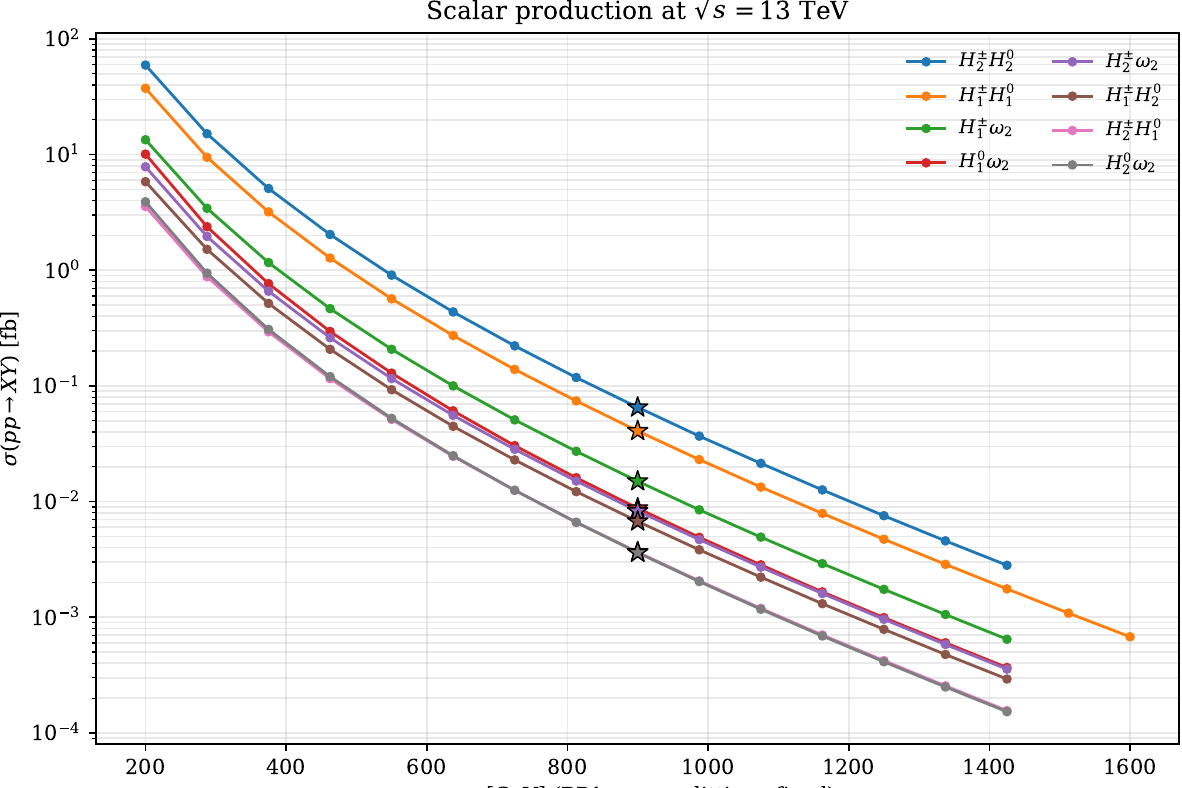}
    \hspace{0.03\textwidth}
    \includegraphics[width=0.35\textwidth]{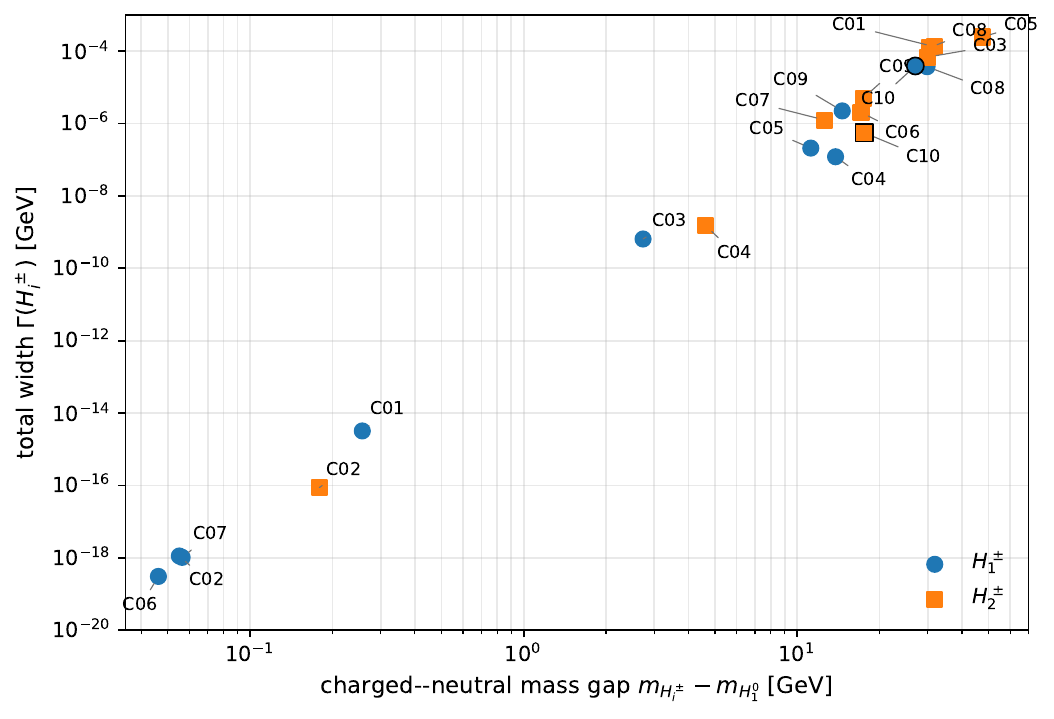}
    \caption{Leading-order associated scalar-production cross sections at $\sqrt{s}=13~\mathrm{TeV}$ along a mass trajectory through C10 (left), and charged-scalar total widths for the benchmarks as functions of the charged--neutral mass splitting (right).}
    \label{fig:scalar-production-widths}
\end{figure*}

To examine the parameter dependence of these results, Fig.~\ref{fig:bp1-charged-scalar-gap-scan} extends this comparison by showing how the charged scalar partial widths and branching fractions vary with the charged neutral mass gap. The partial widths increase with the mass gap as more phase space becomes available. The dashed line marks the on-shell $H_1^0W^\pm$ threshold; the $\omega_2W^\pm$ and $H_2^0W^\pm$ channels open at slightly larger gaps, producing the changes in the branching fractions.
\begin{figure}[htbp]
    \centering
    \includegraphics[width=0.70\textwidth]
    {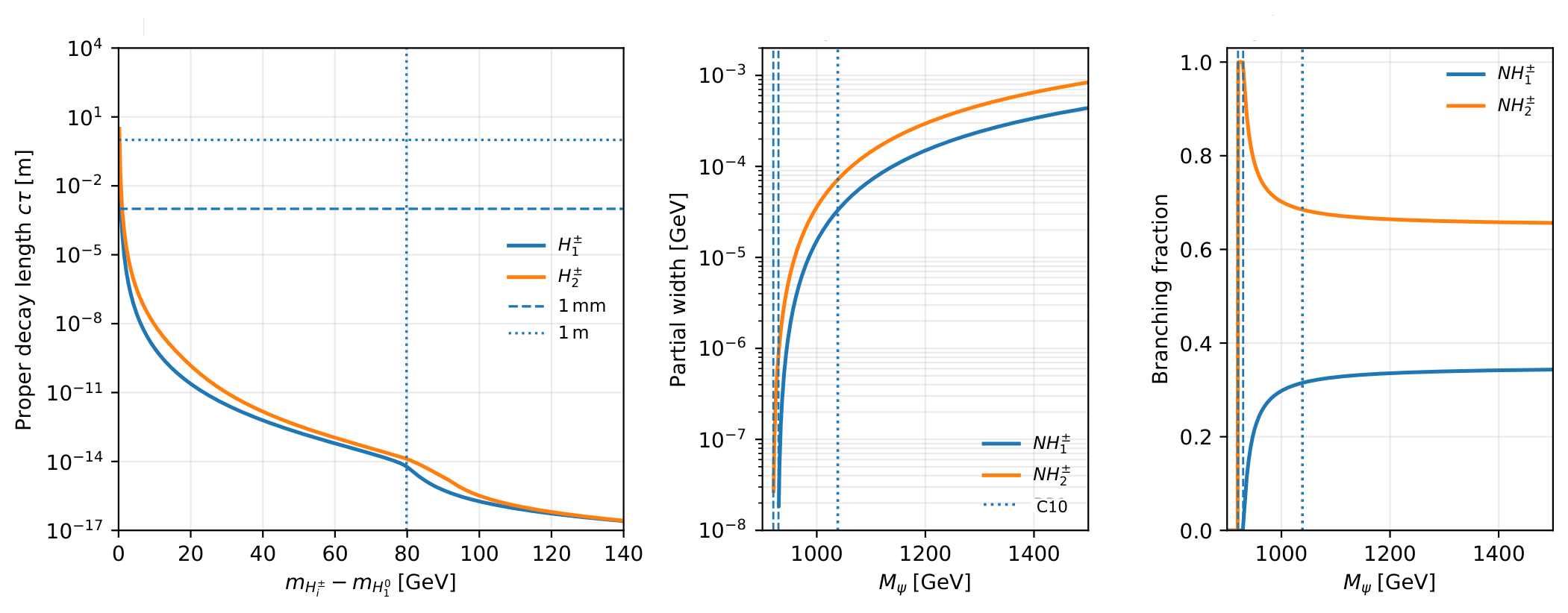}
    \caption{Left: proper decay lengths of $H_1^\pm$ and $H_2^\pm$ as functions of the charged-neutral mass gap. Center and right: partial widths and branching fractions for $\psi^\pm\to NH_{1,2}^\pm$ as functions of $M_\psi$. All other parameters are fixed to C10; the horizontal lines indicate $c\tau=1~\mathrm{mm}$ and $1~\mathrm{m}$, while the vertical line marks the C10 value of $M_\psi$.}
    \label{fig:bp1-lifetime-fermion-scan}
\end{figure}
Similarly, Fig.~\ref{fig:bp1-lifetime-fermion-scan} shows that the charged scalar decay lengths decrease rapidly as the charged neutral mass splitting increases. In the $M_\psi$ scan, the channels $\psi^\pm\to N H_2^\pm$ and $\psi^\pm\to N H_1^\pm$ have thresholds at $M_\psi\simeq920.0~\mathrm{GeV}$ and $929.5~\mathrm{GeV}$, respectively. Above these thresholds, both partial widths increase with $M_\psi$, and $\psi^\pm\to N H_2^\pm$ remains dominant. At C10, where \(\theta_C=0.632\), the combined effects of the charged-scalar mixing and phase space give
\(\mathrm{Br}(\psi^\pm\to NH_2^\pm)=0.685\) and
\(\mathrm{Br}(\psi^\pm\to NH_1^\pm)=0.315\), while \(c\tau(\psi^\pm)=1.88\times10^{-12}\,\mathrm{m}\). To illustrate scalar production, the left panel of Fig.~\ref{fig:scalar-production-widths} shows that the associated scalar production cross sections decrease with increasing scalar mass, with $pp\to H_2^\pm H_2^0$ giving the largest cross section. The right panel of the same figure compares the charged scalar widths for all ten benchmarks as functions of the charged-DM mass splittings $\Delta_i^\pm=m_{H_i^\pm}-m_{H_1^0}$. The corresponding scalar masses and mass splittings are listed in Table~\ref{tab:selected-benchmark-masses}.

When the fermionic sector is included, the electroweak triplet-fermion channels $pp\to\psi^\pm\psi^0$ and $pp\to\psi^+\psi^-$ dominate the production rate, while the scalar associated channels are subleading. The compressed scalar cascades produce soft leptons or jets accompanied by missing transverse momentum, as in inert-doublet and scotogenic models~\cite{Kalinowski:2020rmb,Baumholzer:2019twf}. Here, doublet-triplet scalar mixing and the additional electroweak fermion triplet provide further production and cascade channels. The mixing angles and couplings governing these production and cascade channels are listed in Table~\ref{tab:selected-benchmark-couplings}.
\begin{figure}[!h]
\centering
\includegraphics[width=0.34\textwidth]{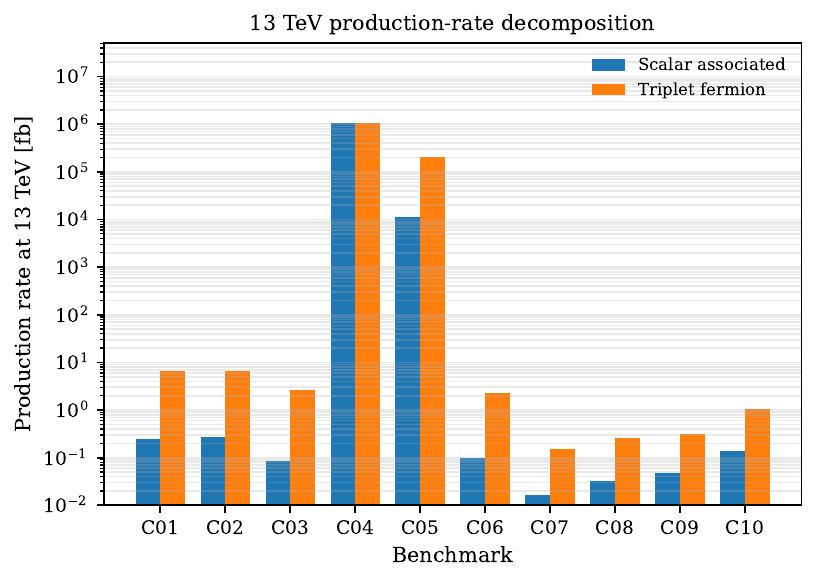}
\hspace{0.03\textwidth}
\includegraphics[width=0.38\textwidth]{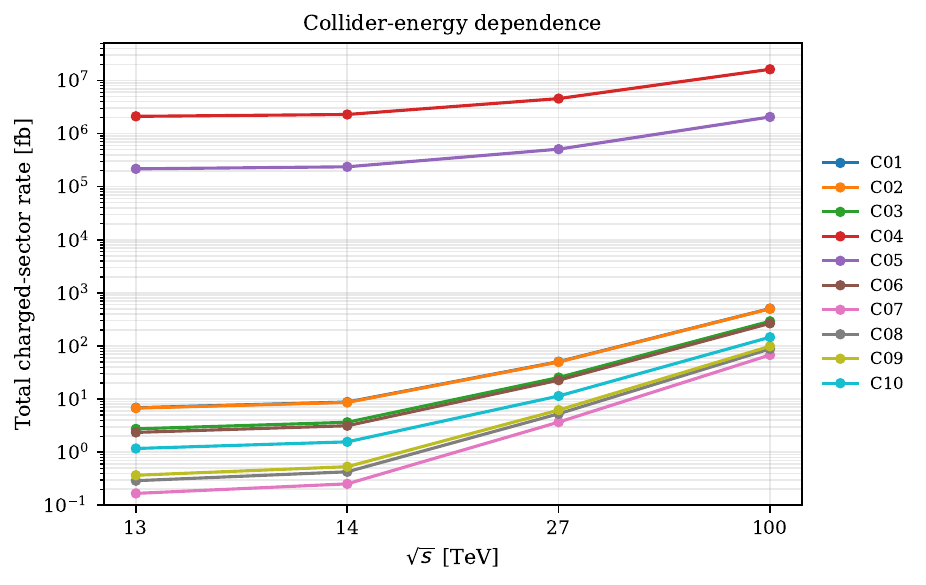}
\caption{ Left: Charged-sector production rates at $\sqrt{s}=13~\mathrm{TeV}$ for the ten benchmarks, separated into charged-neutral scalar associated production and electroweak triplet-fermion production. Right: total reliable
charged-sector production rate at $\sqrt{s}=13$, $14$, $27$, and $100~\mathrm{TeV}$. Charge-conjugate channels are combined.
}
\label{fig:selected-production-rates}
\end{figure}
To compare these production channels across benchmark points, the left panel of Fig.~\ref{fig:selected-production-rates} shows that triplet-fermion production dominates scalar-associated production. For the collider-energy dependence, the right panel shows that the total production rates increase with collider energy, particularly for the heavier spectra. Since C04 and C05 contain charged states below $100~\mathrm{GeV}$, they require a dedicated comparison with LEP and LHC direct-search limits which is beyond the scope of the present study. 
\begin{figure}[!h]
\centering
\includegraphics[width=0.35\textwidth]
{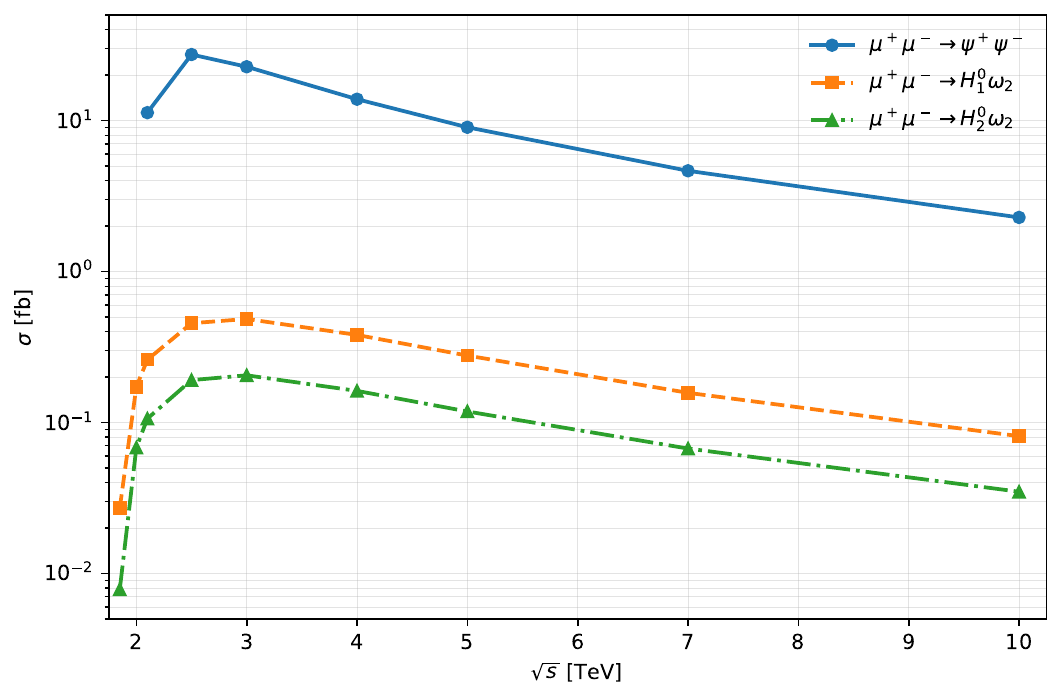}
\caption{ Leading-order C10 production cross sections at a muon collider as functions of the centre-of-mass energy. The channels shown are $\mu^+\mu^-\to\psi^+\psi^-$, $\mu^+\mu^-\to H_1^0\omega_2$, and $\mu^+\mu^-\to H_2^0\omega_2$.}
\label{fig:bp1-muon-collider-energy-scan}
\end{figure}

To close this section, let us use the benchmark point C10 to comment on the possible production channels at a future muon collider. We obtained the cross sections using \texttt{MadGraph5\_aMC@NLO}~\cite{Alwall:2014hca}. Figure~\ref{fig:bp1-muon-collider-energy-scan} shows that $\mu^+\mu^-\to\psi^+\psi^-$ is the dominant channel above its threshold at $\sqrt{s} \simeq2.08~\mathrm{TeV}$, reaching approximately $27~\mathrm{fb}$ near $\sqrt{s}=2.5~\mathrm{TeV}$, while the neutral associated-production channels are smaller. A quantitative sensitivity assessment would require the inclusion of beam-spectrum effects, detector acceptance, and SM backgrounds.
\section{Conclusion}
\label{sec:concl}
We have investigated the phenomenology of the \texttt{T4-3-i-C1} realization of the \texttt{T4-3-i} topology, a geniune one-loop model linking radiative neutrino mass generation to scalar DM, in which the resulting neutrino mass matrix has rank two and therefore leaves one neutrino massless. Both NO and IO remains compatible with the imposed constraints, with IO yielding modestly better fit and larger absolute neutrino mass observables, with $m_{\beta\beta}\simeq 16-49~{\rm meV}$, $m_\beta\simeq 48-49~{\rm meV}$, and $\sum_i m_i\simeq 0.098-0.100~{\rm eV}$, compared with $m_{\beta\beta}\simeq 1.3-4.0~{\rm meV}$, $m_\beta\simeq 8.5-9.1~{\rm meV}$, and $\sum_i m_i\simeq 0.058-0.060~{\rm eV}$ for NO. The DM analysis reveals a restricted low-mass region governed by Higgs-resonant annihilation and threshold effects, together with a dominant TeV-scale region extending to approximately $1.85~{\rm TeV}$, where coannihilation with nearby states determines the relic abundance. The viable points remain below current direct-detection limits through suppressed effective Higgs couplings or cancellations among scalar contributions.

The compressed spectra favored by the relic-density constraint also produce distinctive collider signatures. Depending on the charged-neutral mass splittings, the charged states may decay promptly, produce displaced signatures, or become long-lived on detector scales. Triplet-fermion production generally provides the largest charged-state rates at hadron colliders, whereas charged--neutral scalar associated production probes the scalar mixing structure more directly. These channels offer collider tests complementary to direct detection and low-energy observables. Since our results are based on leading-order production-level predictions, detector-level analyses incorporating acceptances, backgrounds, event selection, and realistic beam effects are required to establish the sensitivity of future hadron and lepton colliders. Related realizations containing doubly charged scalar and fermion states (corresponding to $\alpha=\pm2$) also warrant a separate study.
\appendix
\section{Perturbative Unitary Constraints}
\label{sec:unitarity}
Perturbative unitary constraints are derived from all possible $2 \to 2$ scalar amplitudes in the high-energy limit, where only quartic interactions contribute. The analysis is performed in the gauge eigenstates basis defined in Section~\ref{sec:model}. We consider the quartic scalar interactions given in Eq.~\eqref{scal.pot}, and construct the scattering matrix by organizing the two-particle states according to the conserved quantum numbers: electric charge $Q$, CP, and the $Z_2$ symmetry. Accordingly, the full scattering matrix decomposes into independent blocks. The neutral sector $(Q=0)$ splits into: \{{\it CP}-even, $Z_2$-even\}, \{{\it CP}-odd, $Z_2$-even\}, \{{\it CP}-even, $Z_2$-odd\}, and \{{\it CP}-odd, $Z_2$-odd\} blocks. The singly charged $(Q=\pm1)$ and doubly charged $(Q=\pm2)$ scattering sectors each decompose into $Z_2$-even and $Z_2$-odd subspaces. For each subspace, we construct a complete basis of linearly independent two-particle states, evaluate the corresponding scattering matrix, and diagonalize it to obtain the eigenvalues entering the perturbative-unitarity bounds.

To illustrate the procedure, we consider the neutral, $CP$-even, $Z_2$-even scattering sector. We use consistently normalized two-particle states; in particular, the conventional factor $1/\sqrt{2}$ for identical-particle channels is understood in the basis definitions. In the ordered basis
\begin{equation}
    \mathcal{B}_{0,+}^{(+)} = 
    \left\{h_1 h_1,\; G^0 G^0,\; G^+ G^-,\; h_2 h_2,\; \omega_2 \omega_2,\;
    \phi^+ \phi^-,\; \eta^0 \eta^0,\; \eta^+ \eta^-,\; h_2 \eta^0,\;
    \dfrac{\phi^+ \eta^- + \eta^+ \phi^-}{\sqrt{2}} \right\},
\end{equation}
The corresponding scattering matrix takes the form 
\begin{equation}
    \mathcal{M}_{0,+}^{(+)} =
    \begin{pmatrix}
        6\lambda_1 & 2\lambda_1 & 2\lambda_1 & \kappa_{123}^{(+)} & \kappa_{123}^{(-)} & \kappa_1 & \kappa_4 & \kappa_4 & 0 & 0 \\
        2\lambda_1 & 6\lambda_1 & 2\lambda_1 & \kappa_{123}^{(-)} & \kappa_{123}^{(+)} & \kappa_1 & \kappa_4 & \kappa_4 & 0 & 0 \\
        2\lambda_1 & 2\lambda_1 & 4\lambda_1 & \kappa_1 & \kappa_1 & \kappa_1 + \kappa_2 & \kappa_4 & \kappa_4 & 0 & 0 \\
        \kappa_{123}^{(+)} & \kappa_{123}^{(-)} & \kappa_1 & 6\lambda_2 & 2\lambda_2 & 2\lambda_2 & \kappa_5 & \kappa_5 & 0 & 0 \\
        \kappa_{123}^{(-)} & \kappa_{123}^{(+)} & \kappa_1 & 2\lambda_2 & 6\lambda_2 & 2\lambda_2 & \kappa_5 & \kappa_5 & 0 & 0 \\
        \kappa_1 & \kappa_1 & \kappa_1 + \kappa_2 & 2\lambda_2 & 2\lambda_2 & 4\lambda_2 & \kappa_5 & \kappa_5 & 0 & 0 \\
        \kappa_4 & \kappa_4 & \kappa_4 & \kappa_5 & \kappa_5 & \kappa_5 & 24\lambda_3 & 8\lambda_3 & 0 & 0 \\
        \kappa_4 & \kappa_4 & \kappa_4 & \kappa_5 & \kappa_5 & \kappa_5 & 8\lambda_3 & 16\lambda_3 & 0 & 0 \\
        0 & 0 & 0 & 0 & 0 & 0 & 0 & 0 & \kappa_5 & 0 \\
        0 & 0 & 0 & 0 & 0 & 0 & 0 & 0 & 0 & \kappa_5
    \end{pmatrix},
\end{equation}
where
\begin{equation}
\kappa_{123}^{(\pm)} \equiv \kappa_1+\kappa_2\pm\kappa_3.
\end{equation}
The final two basis states are decoupled from all remaining channels and therefore yield two degenerate eigenvalues, each equal to $\kappa_5$. The remaining $8\times8$ block can be further block-diagonalized through an orthogonal transformation into an analytically tractable $2\times2$ block and a residual $6\times6$ block. Although this decomposition is useful for identifying part of the spectrum analytically, in the numerical analysis we implemented the complete scattering matrices for all sectors classified by electric charge, {\it CP}, and $Z_2$ parity, and diagonalized them numerically at each parameter point.

\bibliographystyle{JHEP}
\bibliography{bibliography.bib}

\end{document}